%% file: XSGRB_sample_arxiv.tex
\documentclass[longauth]{aa}    

\usepackage{graphicx,url,twoopt,natbib}
\usepackage[varg]{txfonts}           

\usepackage{pdfcomment}              
\usepackage{acronym}                 

\usepackage{color}
\definecolor{linkcolor}{rgb}{0,0.3,0.7}
\hypersetup{colorlinks=true,
	linkcolor=linkcolor, 
	citecolor=linkcolor,
	filecolor=linkcolor, 
	urlcolor=linkcolor}

\usepackage{url}

\bibpunct{(}{)}{;}{a}{}{,}    

\input{definitions.tex}

\usepackage{xcolor}

\usepackage{float}

\begin{document}
	
\title{The X-shooter GRB afterglow legacy sample (XS-GRB)\thanks{Based on
		observations collected at the European Southern Observatory, Paranal, Chile,
		Program IDs: 084.A-0260, 084.D-0265, 085.A-0009, 086.A-0073, 087.A-0055,
		088.A-0051, 089.A-0067, 090.A-0088, 091.A-0877, 0091.C-0934, 092.A-0124,
		092.D-0056, 092.D-0633, 093.A-0069, 094.A-0134, 095.A-0045, 095.B-0811,
		096.A-0079, 097.A-0036, 098.A-0136, and 098.A-0055.}}

\titlerunning{The X-shooter GRB afterglow legacy sample}

\input{authorlist.tex}

\date{Received/ accepted}

\authorrunning{Selsing et al.}

\abstract{In this work we present spectra of all $\gamma$-ray burst (GRB)
	afterglows that have been promptly observed with the X-shooter spectrograph
	until \termdate. In total, we obtained spectroscopic observations of 103 individual GRBs
	observed within 48 hours of the GRB trigger. Redshifts have been measured for 97
	per cent of these, covering a redshift range from 0.059 to 7.84. Based on a set
	of observational selection criteria that minimize biases with regards to
	intrinsic properties of the GRBs, the follow-up effort has been focused on
	producing a homogeneous sample of 93 afterglow spectra for GRBs discovered by
	the {\em Swift} satellite. We here provide a public release of all the reduced
	spectra, including continuum estimates and telluric absorption corrections. For
	completeness, we also provide reductions for the 18 late-time observations of
	the underlying host galaxies. We provide an assessment of the degree of
	completeness with respect to the parent GRB population, in terms of the X-ray
	properties of the bursts in the sample and find that the sample presented here
	is representative of the full \textit{Swift} sample. We constrain the fraction
	of dark bursts to be < 28 per cent and we confirm previous results that higher
	optical darkness is correlated with increased X-ray absorption. For the 42
	bursts for which it is possible, we provide a measurement of the neutral
	hydrogen column density, increasing the total number of published H{\sc I}
	column density measurements by $\sim$ 33 per cent. This dataset provides a
	unique resource to study the ISM across cosmic time, from the local progenitor
	surroundings to the intervening universe.}


\maketitle

\section{Introduction}

Gamma-ray bursts (GRBs) are bright glimpses of electromagnetic radiation that
pierce through the universe, all the way from the edges of the observable
universe. They provide constraints on a very wide range of topics in
astrophysics. Examples range from small-scale phenomena relating to magnetars,
properties of highly relativistic jets, hyper/supernova explosions, the
interstellar medium, dust extinction curves, starburst galaxies, chemical and
molecular abundances, escape of ionizing radiation, the ionization state of the
intergalactic medium, intervening absorption systems to standard candles in
cosmology \citep[e.g.,][]{Wijers1998, Savaglio2006, Ghirlanda2007, Molinari2007,
	Amati2008, Vergani2009, Prochaska2009, HjorthBloom2012, Rowlinson2017,
	Christensen2017}.

The \textit{Neil Gehrels Swift Observatory} ({\it Swift}) satellite
\citep{Gehrels2004, Gehrels2009}, which was launched in 2004, has made it
possible to harvest much of the great potential in using GRBs as probes of the
intergalactic medium, which was already hinted at by results from earlier
missions \citep[e.g.,][]{Paradijs2000, Ricker2004}. With the three on-board
instruments, the Burst Alert Telescope \citep[BAT;][]{Barthelmy2005}, the X-Ray
Telescope \citep[XRT;][]{Burrows2005}, and the UltraViolet and Optical Telescope
\citep[UVOT;][]{Roming2005}, \swift~is an ideal observatory for GRB hunting. A
crucial aspect of the success of the {\it Swift} mission has been the extensive
ground-based follow-up observations of the afterglows and of the host galaxies
of the GRBs, involving a large community of researchers. This fruitful
collaboration has been facilitated by the open data access policy of the {\it
	Swift} mission. The close collaboration between detection facilities and
electromagnetic  follow-up campaigns continue to be immensely rewarding, as
recently highlighted by the simultaneous detection of gravitational waves and
light from the neutron star merger in the shape of GW170817/GRB170817A/AT2017gfo
\citep{LIGOScientificCollaboration2017a, LIGOScientificCollaboration2017}.

In the beginning of the {\it Swift} era most of the follow-up afterglow
spectroscopy was secured using low-resolution spectrographs \citep[typically
$R=\lambda/\Delta\lambda$$<$1000, e.g.,][]{Fynbo2009}. Spectroscopy is powerful
as it allows us to secure information even for very faint targets
\citep{Kruhler2012}. This allows the measurement of a number of important
parameters such as redshifts, spectral slopes, and extinction. For a handful of
very bright afterglows high-resolution (typically R$>$20000) spectra have been
secured, and for these events much more information about conditions inside the
host galaxies were extracted \citep[e.g.,][]{Fiore2005, Thone2007,
	Prochaska2007, Vreeswijk2007, DElia2009, Castro-Tirado2010}.

The X-shooter spectrograph \citep{Vernet2011} was the first of the second
generation instruments at the ESO Very Large Telescope (VLT). It was designed
very much with transient follow-up in mind as the fading luminosities of
such sources makes it urgent to secure as extensive wavelength coverage as possible in the
shortest possible time. At the same time, the resolution was designed to be in
the range 4000--9000 in order to be able to get a large useful spectral range
between the many sky-background emission lines in the red and near-IR spectral
ranges. The near-IR spectral coverage allows for spectroscopic observations of
the highest redshift GRBs.

In this paper, we present the results of a dedicated effort over the years 2009
-- 2017 to use the X-shooter spectrograph to secure spectroscopic observations
of afterglows and host galaxies of GRBs detected by {\it Swift}. We here make
all the data resulting from the survey publicly available in reduced form (see Sect
\ref{products}).

The paper is organized in the following way: In Sect.~\ref{sample} we describe
the sample including the sample selection criteria and the observational
strategy. In Sect.~\ref{obs}, we describe the observations and the instrumental
setups, and in Sect.~\ref{proc} we detail the methodological strategies adopted
in the data reduction process and auxiliary material. In Sect.~\ref{results} we
describe the results of the survey, i.e. the efficiency of the follow-up effort
and the characteristics of the observed bursts. We also assess the
completeness of the realized sample. Finally, we offer our conclusions
in Sect.~\ref{conclusions}. We use the $\Lambda$CDM cosmology parameters
provided by the \citet{Planck2015} in which the universe is flat with $H_0 =
67.7$\,km\,s$^{-1}$ Mpc$^{-1}$ and $\Omega_\mathrm{m} = 0.307$.

\section{Sample selection criteria and observations}\label{sample}


\subsection{Sample selection criteria} \label{samplecrit}

Being of transient nature, it is difficult to impose strong sample selection
criteria on GRBs without hampering the follow-up effort. Many natural follow-up
restrictions exist already, being it weather conditions, pointing restrictions
of the telescope or unconstrained burst localizations as reported by the alerting
facility. To maximize the return of the follow-up campaign, we have chosen a
few selection criteria that facilitate an unbiased selection of bursts, while
at the same time allowing for a high follow-up success rate. The importance of
defining unbiased selection criteria has been highlighted previously
\citep{Jakobsson2006b, Salvaterra2012, Hjorth2012, Vergani2015, Perley2016a},
when trying to address the intrinsic underlying distribution functions such as
the redshift distribution, host metallicity distribution, or afterglow
brightness distribution. When investigating a specific distribution function, a
high degree of \textit{completeness} is desired \citep[e.g.,][]{Perley2016b}.

In defining the selection criteria, we simultaneously aim to minimize any biases
against intrinsic astrophysical conditions while at the same time maximizing the
likelihood of successful observations, hence allowing us to obtain a higher
degree of completeness. By restricting the selection criteria to conditions
local to the Milky Way and therefore independent of intrinsic GRB properties,
the aim is that the collected sample represents the underlying distribution of
GRBs in a fair way. The selection criteria used here are based on previous,
similar studies \citep{Jakobsson2006b, Fynbo2009, Hjorth2012}. We characterize
the sample completeness in Sect. \ref{completeness}.

The selection criteria that define a GRB as part of our initial \textit{statistical}
	sample are:

\begin{enumerate}
	
	\item GRB trigger by BAT onboard the \swift~satellite

	\item XRT started observing within 10 minutes after the GRB; an XRT position
	must be distributed within 12 hr.
	
	\item The target must be visible from Cerro Paranal for at least 60 minutes, 30
degrees above the horizon, with the Sun below $-12$ degrees\footnote{Note that
	in the P84 proposal the criteria have been stated differently, the visibility
	constraint being replaced by a declination + Sun angle constraint. The criteria
	listed above are, however, those defining the sample.}.

	\item Galactic $A_V \leq 0.5$ mag according to the maps of \citet{Schlegel1998}.
	
	\item No bright, nearby stars within $ 1.8 + 0.4 \times \exp[(20 - R)/2.05$]
arcsec, where $R$ is the USNO magnitude of the star).
	
\end{enumerate}

Our ability to observe GRB afterglows is strongly dependent on the timing and
the precision of the target positions delivered by the triggering facilities. By
selecting only bursts that have been triggered on board the \textit{Swift} space
telescope \citep{Gehrels2004}, based on the BAT, we start out with a sample
where burst characteristics are delivered immediately, allowing for an informed
follow-up strategy. However, the BAT sensitivity varies across its field of
view, so selection is not entirely homogeneous. Despite the complexity of the
triggering mechanism on board \textit{Swift} \citep{Band2006, Coward2013a},
attempts at inferring properties of the underlying GRB population based on the
detection thresholds and triggering algorithms have been made \citep{Lien2014,
	Graff2016}. Restricting the follow-up effort to bursts detected by
\textit{Swift}, we therefore ensure that the limitations of the parent sample
are well studied.

Because the localization accuracy of BAT is 1 -- 4 arcminutes
\citep{Barthelmy2005}, an afterglow identification based on BAT alone would be
harder, and  host association impossible. We therefore additionally require an
X-ray position from XRT to be distributed to the GCN network
\citep{Barthelmy2000} within 12 hours and to account for observing constraints
on \textit{Swift} that XRT began observations within 10 minutes. The additional
timing requirement of the XRT follow-up means that all bursts in our sample have
detected X-ray afterglows. Because the XRT completeness is very high for
promptly-repointed GRBs \citep{Burrows2007}, this cut should not alter the
parent sample significantly.

To ensure a minimum of observability, we require that the GRB is visible from
the telescope site at Cerro Paranal, Chile, for a least 1 hour after the trigger
with the sun below $-12$ degrees. This secures time for the spectroscopic
observations to be completed. Since the GRB population is isotropically
distributed on the sky as seen from Earth, and because the GRB properties do not
depend on position on the sky \citep{Meegan1992, Briggs1996, Ukwatta2016}, this
cut does not influence our ability to fairly sample the underlying GRB
population. The same arguments apply to the requirement that there are no nearby
foreground bright stars. We additionally require that the Galactic extinction is
below $A_V \lesssim 0.5$ mag, based on the extinction maps by
\citet{Schlegel1998}\footnote{We use the updated values of \citet{Schlafly2011}
	to correct for foreground extinction, but the sample criterion is based on the
	old \citet{Schlegel1998} values for consistency with the first semesters of our
	program.}. Choosing low-extinction sightlines also reduces the problem of field
crowding and the contamination from Galactic outbursts posing as GRB impostors.
These additional cuts should not influence the optical properties of the bursts
themselves, only our ability to successfully secure the observations that allow
us to investigate the spectroscopic properties of GRBs.

Bursts that fulfill these selection criteria are what we define as our initial
\textit{statistical} sample. To reach a higher degree of sample completeness, we
impose additional cuts that increase our ability to infer population properties.
These are presented in Sect. \ref{badbursts}. Using this sample we will be able
to address population properties of \swift-detected bursts. We further discuss
the effect of these selection criteria and their implication for the
completeness of the sample in Sect. \ref{completeness}.

\subsection{Follow-up procedure}

Our collaboration has set up a procedure to promptly react to alerts from
\swift~and other satellites (such as \textit{Fermi}). We process automatically
GCN notices in order to flag those events belonging to our statistical sample.
Two people are permanently on alert to manually supervise each event, for
example to recognize events of special interest beyond their inclusion in the
sample. Members of our collaboration have access to numerous facilities spread
throughout the world, among which the William Herschel Telescope (WHT), Nordic
Optical Telescope (NOT), the Gran Telescopio Canarias (GTC), the Telescopio
Nazionale Galileo (TNG), the Gamma-Ray Burst Optical Near-infrared Detector
(GROND \citealt{Greiner2008}), and the Xinlong observatory. Thanks to this
network, we can often conduct searches for optical/NIR afterglows ahead of the
time of the X-shooter spectroscopy, thus helping to plan the spectroscopic
observations. If this is not possible, we use the built-in acquisition camera of
X-shooter to search for counterparts, and adapt the observing strategy on the
fly. In many cases we could interact directly with the staff at the telescope to
aid the observation. Raw data are usually available within minutes in the
\href{http://archive.eso.org/wdb/wdb/eso/xshooter/form}{ESO archive} and are
promptly reduced by members of our collaboration, often using archival
calibration data which are readily available. This allows us in most cases to
report the preliminary results (redshift, identification of the most prominent
emission and absorption feature, etc.) within a few hours after the beginning of
the observations and plan additional X-shooter observations if deemed necessary.

\subsection{Rapid response mode (RRM) observations} \label{RRM}

Under rare circumstances, the use of the ESO rapid response mode (RRM;
\citealt{Vreeswijk2010}) has been possible. The RRM is a system to automatically
override ongoing observations at the telescope. This allows the shortest
possible delay between the GRB trigger and the initiation of observations, where
spectroscopic integration of the rapidly fading, optical transient can commence
within minutes of the burst. One limitation is that no instrument change is
allowed in RRM, which lowers the number of successful triggers (X-shooter shares the telescope with two other instruments). In case of a promptly visible GRB, a robotic
trigger is sent to the telescope if at the time of the GCN notice the GRB
fulfills the following criteria:

\begin{enumerate}
	\item The GRB triggered \textit{Swift} onboard.
	\item The X-ray position must be available less than 1 hr after the GRB.
	\item The elevation of the source in the sky is $> 22^\circ$ (both at trigger time and 15 min after).
	\item The Sun elevation from Paranl is $< -12^\circ$ (both at trigger time and 15 min after).
	\item Galactic $A_V \leq 1.0$~mag according to the maps of \citet{Schlegel1998}.
	\item The \swift-circulated tags: KNOWN\_SOURCE, COSMIC\_RAY, DEF\_NOT\_GRB are set to false (PROB\_NOT\_GRB can be true)\footnote{These tags are distributed as part of the GCN notices. Explanations for all tags are \href{https://gcn.gsfc.nasa.gov/sock_pkt_def_doc.html}{available online}.}.
\end{enumerate}

The criteria to trigger RRM are looser than those that define the statistical
sample. This is both because of the expected larger brightness of GRB
counterparts soon after the explosion, and because of the rarity of RRM
triggers.

The use of RRM is unique as it allows to sample a long logarithmic time span in
the GRB lifetime, and it exploits the extreme brightness of early afterglows. A
pivotal exampl is the study of temporal variability of GRB afterglow absorption
systems due to effect of the GRB itself on the sorrounding medium \citep[e.g.,
see][]{Dessauges-Zavadsky2006, Vreeswijk2007, DElia2009, Vreeswijk2013}.

There are nine GRBs that have been observed with X-shooter in RRM mode. One of
the RRM triggers is outside the statistical sample, and two of the RRM triggers
are on short GRBs. Our fastest response (between the \swift~GRB trigger time and
the beginning of spectroscopic integration) was for GRB~160410A, for which the
delay was only 8.4 minutes.

In many cases, unfortunately, we could not use RRM even for promptly visible
events, e.g.{} because of the unavailability of X-shooter. Potentially, the
proximity of the telescope (Cerro Paranal, Chile) to the South Atlantic Anomaly
(SAA) affects the rate of RRM triggers \citep{Greiner2011}. BAT switches off
when the satellite goes through the SSA and that means we have fewer events
immediately observable, compared to the rest of the \swift~orbit.

\subsection{Observations} \label{obs}

\begin{figure*}
	\centerline{\includegraphics[width=\linewidth]{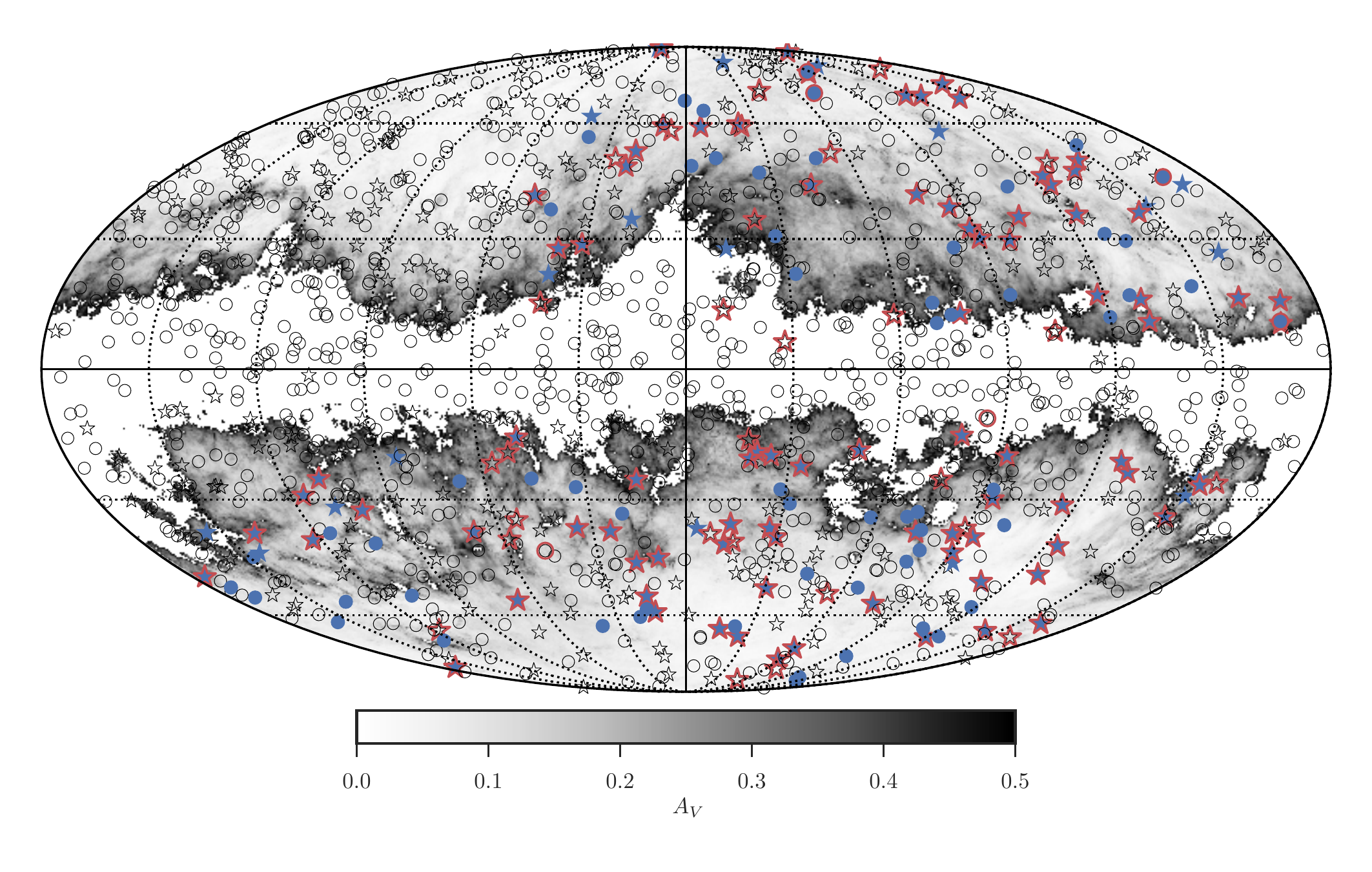}} \caption{Mollweide projection
	in Galactic coordinates of the full sky showing the positions on the sky of the
	bursts presented in this work. The equator is the Galactic plane. The empty
	stars/circles are the positions of all the 1266 \textit{Swift} bursts detected
	until \termdate. Stars indicate bursts with measured redshifts and circles
	indicate those without. Blue stars indicate the position of the 104 bursts
	fulfilling the sample criteria specified in Sect. \ref{samplecrit} that have a
	measured redshift. Red outlines are added to the 84 GRBs that enter our
	statistical sample with both X-shooter spectroscopy and a measured redshift.
	Red outlines of empty stars represent bursts which have been observed with
	X-shooter and has a measured redshift, but is not a part of our statistical
	sample. The blue dots show the positions of the 61 GRBs of our statistical
	sample that lack redshift measurements. A red outline is added around the six
	bursts in our statistical sample which were observed with X-shooter, but did
	not yield a redshift measurement. Two empty circles with red outlines indicate
	bursts outside the statistical sample that were followed up with X-shooter, but
	without a redshift measurement. The different samples are compared in Sect.
	\ref{results}. The background shows the dust maps presented in
	\citet{Schlegel1998}. Please note that we removed the background where the
	sample criterion is violated ($A_V > 0.5$ mag) and replaced it with a white
	background. The grey scale bar below indicates the value of $A_V$ on the plot.
	The dotted lines indicate intervals of 30$^\circ$ in longitude and latitude.}
\label{fig:skymap}
\end{figure*}

The observations obtained for this sample have been secured with the
cross-dispersed echelle spectrograph, X-shooter \citep{Vernet2011}, mounted on
one of two Unit Telescopes at ESO/VLT, UT2 (Kueyen) and UT3 (Melipal) during
the duration of this follow-up campaign. The observations have been taken during
a period of eight years corresponding to the ESO observing periods P84 through
P98 under the following programme IDs: 084.A-0260, 085.A-0009, 086.A-0073,
087.A-0055, 088.A-0051, 089.A-0067, 090.A-0088, 092.A-0124, 093.A-0069,
094.A-0134, 095.A-0045, 096.A-0079, 097.A-0036, and 098.A-0055 (PI: Fynbo) and
0091.C-0934 (PI: Kaper). These proposals were initiated on Guaranteed Time. We
have included a few additional bursts, from the programmes 084.D-0265 (PI:
Benetti), 091.A-0877 (PI: Schady), 092.D-0056 (PI: Rau), 092.D-0633, 098.A-0136 (PI:
Greiner), and 095.B-0811 (PI: Levan). The total collection of spectra represents
\textit{all} GRB afterglows that have been followed up by X-shooter up to
\termdate, which marks the end of the XS-GRB legacy follow-up program.

The first GRB followed up was GRB~090313 \citep{DeUgartePostigo2010}, observed
on the 15th of March, 2009, during the commissioning of X-shooter on the second
Unit Telescope (UT2) of the VLT . The bursts observed during the commissioning
or the science verification process (GRB~090313, GRB~090530, GRB~090809,
GRB~090926A) are not a part of the sample we use to investigate the statistical
properties of GRB afterglows, due to the different criteria for their selection. The
first burst observed after science verification and the mounting of X-shooter on UT2,
was GRB~091018, which is the first burst entering our statistically homogeneous
sample. For all bursts that fulfill our sample selection criteria, described in
Sect. \ref{samplecrit}, spectroscopic follow-up have been attempted with
X-shooter. Various conditions can affect our ability to follow up a given burst,
and a discussion of these conditions and their consequences for the sample is
included in Sect. \ref{badbursts}.

X-shooter covers the spectral wavelength region from 300 nm to 2480 nm in a
single exposure, by splitting the light into three separate spectroscopic arms
through the use of two dichroics. The ultraviolet blue (UVB) arm covers 300 -
550 nm, the visual (VIS) arm covers 550 - 1020 nm, and the near-infrared (NIR)
arm covers 1020 - 2480 nm. For some of the observations, we have applied a
$K$-band blocking filter, cutting the coverage of the NIR arm at 2100 nm. The
$K$-band blocking filter is only used after 2012, where it was installed. This
is done to reduce the amount of scattered background light from the thermal
infrared. For the majority of observations, a nodding observing scheme
has been employed, with a nodding throw of 5\arcsec. Each nodding observation
has typically been carried out in a standard ABBA pattern. For some cases,
conditions during the follow-up (either technical or weather), have necessitated
alterations to this scheme as described in App. \ref{notes}. For RRM triggers, a
slightly different observing strategy was employed. Starting as rapidly as
possible, a simple stare mode sequence was started, with 5 spectroscopic
integrations with increasing exposure times.

For the majority of the bursts, we have observed with a slit width of 1\farc0,
0\farc9, and 0\farc9 for the UVB, VIS, and NIR-arm respectively. This sets a
lower limit on the delivered resolving power of the spectra based on the
tabulated values of the delivered resolutions, which is 4350, 7450, and 5300 for
the UVB, VIS and NIR-arm respectively\footnote{See
	\href{https://www.eso.org/sci/facilities/paranal/instruments/xshooter/inst.html}{this URL} for the nominal spectral resolutions.}. 
For accurate measurements involving line profiles, knowledge of the precise
instrumental resolution is required. The spectral resolution becomes better than
the nominal one, when the delivered seeing is smaller than the projected width
of the slit on the sky. We discuss how we determine the effective instrumental
resolution in Sect. \ref{resolution}.

Due to a mechanical failure, the atmospheric dispersion corrector (ADC) was
disabled from 1st of August 2012 until the end of this program. Only GRB~100728B
was affected by the failing ADC prior to disablement, resulting in a
lower-than-nominal throughput. To avoid chromatic slit losses due to atmospheric
dispersion, nearly all subsequent observations have been carried out at
parallactic angle. A consequence of this is that for all observations following
1st of August 2012, the centroid of the trace of the source changes position
across the spatial direction of the slit as a function of wavelength. This
effect has been modeled in the extraction procedure, as described in Sect.
\ref{extract}.

\input{tables/burst_overview.tex}

We provide an overview of all the observations in Table
\ref{tab:sample_overview} and plot the positions of all the bursts on the
celestial sphere in Galactic coordinates in Fig. \ref{fig:skymap}. Away from the
central zone of avoidance, due to high Galactic extinction cutoff (marked in
white), the GRB positions have an isotropic distribution, except in the upper
left quadrant which cannot be observed from Paranal due to the declination
constraints of the telescope.

Thirty percent of the spectra presented here, primarily host observations, have
already been published in \citet{Kruhler2015}. Single bursts have additionally
been published, based on unusual properties in their afterglows. We present
individual notes on all the X-shooter spectra and their previous use in App.
\ref{notes}. We include independent reductions of them here for completeness.

\section{Data processing} \label{proc}

In this section we describe how the final data products are produced and
subsequently post-processed. 
All post-processing scripts developed for this dataset are made publicly
available at \url{https://github.com/jselsing/XSGRB_reduction_scripts}.

Before any reductions are initiated, the raw object images are run through the
cosmic-ray removal algorithm \citep{VanDokkum2001} implementation,
\textit{Astro-SCRAPPY}\footnote{\url{https://github.com/astropy/astroscrappy}},
where a wide clipping radius was used around detected cosmic ray hits to ensure
that edge residuals are robustly rejected.

The basis for the reductions is the VLT/X-shooter pipeline, version
\texttt{2.7.1} or newer \citep{Goldoni2006, Modigliani2010}. The pipeline is
managed with the Reflex interface \citep{Freudling2013} and is used for
subtraction of bias level, flat-fielding, tracing of the echelle orders,
wavelength calibrations with the use of arc-line lamps, flux calibration using
spectrophotometric standards \citep{Vernet2010, Hamuy1994}, mirror flexure
compensation(see Sect. \ref{wavecal}), sky-subtraction and lastly the
rectification and merging of the orders. Errors and bad pixel maps are
propagated throughout the extraction. For the initial sky-subtraction, the
background has been estimated by a running median in regions adjacent to the
object trace clear of contaminating sources. Due to the broken ADC, for some
objects there is curvature in the object trace along the dispersion axis of the
slit(see Sect. \ref{extract}). This means that for these bursts, the initial
sky-estimate was made from a limited number of pixels in the spatial direction.
The subtraction of the sky background on the un-rectified image ensures that the
bulk of the sky background is not redistributed by the rectification process.

X-shooter is an echelle spectrograph, and therefore the individual echelle
orders are curved across image space. The individual orders therefore need to be
rectified. In order to transform the image space (pixels) into a physical
(wavelength-slit) space, the image pixels are resampled onto a physical grid,
while propagating the pixel uncertainties derived by the pipeline. This
rectification  process correlates neighboring pixels and in order to minimize
the degree of correlation, we need to choose a physical sampling that matches
the pixel sampling. We rectified the image onto an equidistant grid with a
dispersion sampling of 0.02 nm/pixel and a 0\farcs16 per pixel spatial sampling
for the UVB and VIS arm and 0.06 nm/pixel with a 0\farcs21 per pixel in the NIR
arm.  Because the tabulated resolution is a lower limit to the delivered
resolution, we choose a sampling of 0.02 nm/pixel to ensure that the lowest
wavelength part of neither of the arms have a sampling lower than the Nyquist
sampling rate of 2 pixels per resolution FWHM.

\subsection{Post-processing} \label{postproc}

For a typical observation, each of the exposures in the nodding sequence have
been reduced as a single observation and then subsequently combined to form a
single image. We employ this strategy so that we can reject outliers in the
stack and weight by an averaged measure of the inverse variance of the
background. When weighting images, where the noise in each pixel is dominated by
Poisson noise, it is important to estimate the background variance in a large
enough region, so that any correlation between the signal and the weights are
removed. To this end, the weight map is generated by a running median window
over the variance map produced by the pipeline, where the trace has been masked
and the width of the window is chosen to be wide enough for the median variance to
be calculated on the basis of several hundred pixels. This weighting scheme
automatically also optimally combines images of different exposure times or
images where the background is varying, which is often the case when a burst has
been observed close to twilight.

An additional sky-subtraction procedure is run on all rectified 2D spectra.
This is done to remove residual sky, still present. At each pixel in the
dispersion direction, the spatial dimension is fit with a low-order polynomial,
after all sources are masked out. This low order polynomial is then convolved
with a few pixel wide Gaussian filter in the dispersion direction and subtracted
from the entire 2D image.

In the NIR arm, where the background is very bright and there are a high number
of bright sky-lines, an alternative approach to sky subtraction has been
employed. When there are no contaminating sources in the slit, the sky has been
put back on the images and the images are combined in pairs of two before
extraction, subtracting the two from each other while keeping the WCS static.
Due to the nodding offsets used between observations, this conserves the
source flux while removing the sky at the expense of a decrease in
signal-to-noise by a factor of $\sqrt{2}$. This amounts to the regular nodding
reduction, only we can reject outliers and weight by the averaged inverse
variance map.

Reducing the images as single observations for all exposures, we additionally
get a spectrum of the sky which we can use to recalibrate the wavelength
solution in the post-processing steps.

\subsection{Correction for offsets in the wavelength calibration}    \label{wavecal}

Since X-shooter is installed at the VLT Cassegrain focus, it is prone to
flexures. The flexures modify the projection of the slit on the detector with
respect to the one obtained in daytime calibration. This requires a modification
of the wavelength solution in order to correctly process the night-time data.
Part of this correction is performed by the pipeline using the frames taken
during the X-shooter Active Flexure Compensation procedure\footnote{X-shooter
	User Manual available at
	\href{https://www.eso.org/sci/facilities/paranal/instruments/xshooter/doc.html}{ESO website}.}. The remaining offset is corrected by cross-correlating the observed sky spectrum with a synthetic sky spectrum \citep{Noll2012, Jones2013} after the continuum, estimated as the mode of all flux values, has been subtracted. To get the correct seeing point-spread-function (PSF) with which to convolve the synthetic sky, an initial refinement of the wavelength solution has been obtained by cross-correlating the observed sky with an unconvolved synthetic sky. This preliminary wavelength calibration is applied to the observed sky. The synthetic spectrum is then convolved with an increasing seeing PSF and the width that minimizes $\chi^2$ with the updated observed sky is chosen to be the effective sky-PSF. Using the synthetic sky with the matched resolution, a final wavelength calibration can then be calculated by cross-correlating the observed sky with the correctly broadened sky spectrum as a function of a velocity offset. Both a multiplicative and an additive offset to the wavelength calibration has been tested, but in terms of $\chi^2$, the model with only a multiplicative offset is preferred. The resulting offsets, which were smaller than 0.01 nm in the UVB and VIS data and smaller than 0.05 nm in the NIR spectra, but changing over short periods of time were applied to the corresponding spectra\footnote{The wavelength shifts have been extensively studied by ESO staff \href{https://www.eso.org/sci/facilities/paranal/instruments/xshooter/doc/XS_wlc_shift_150615.pdf}{in this document}.}. Using the convolved synthetic sky, the pixels containing the brightest sky lines have been flagged as such in the bad pixel map.

\subsection{Spectral resolution} \label{resolution}


\begin{figure}[!t]
	\centerline{\includegraphics[width=\columnwidth]{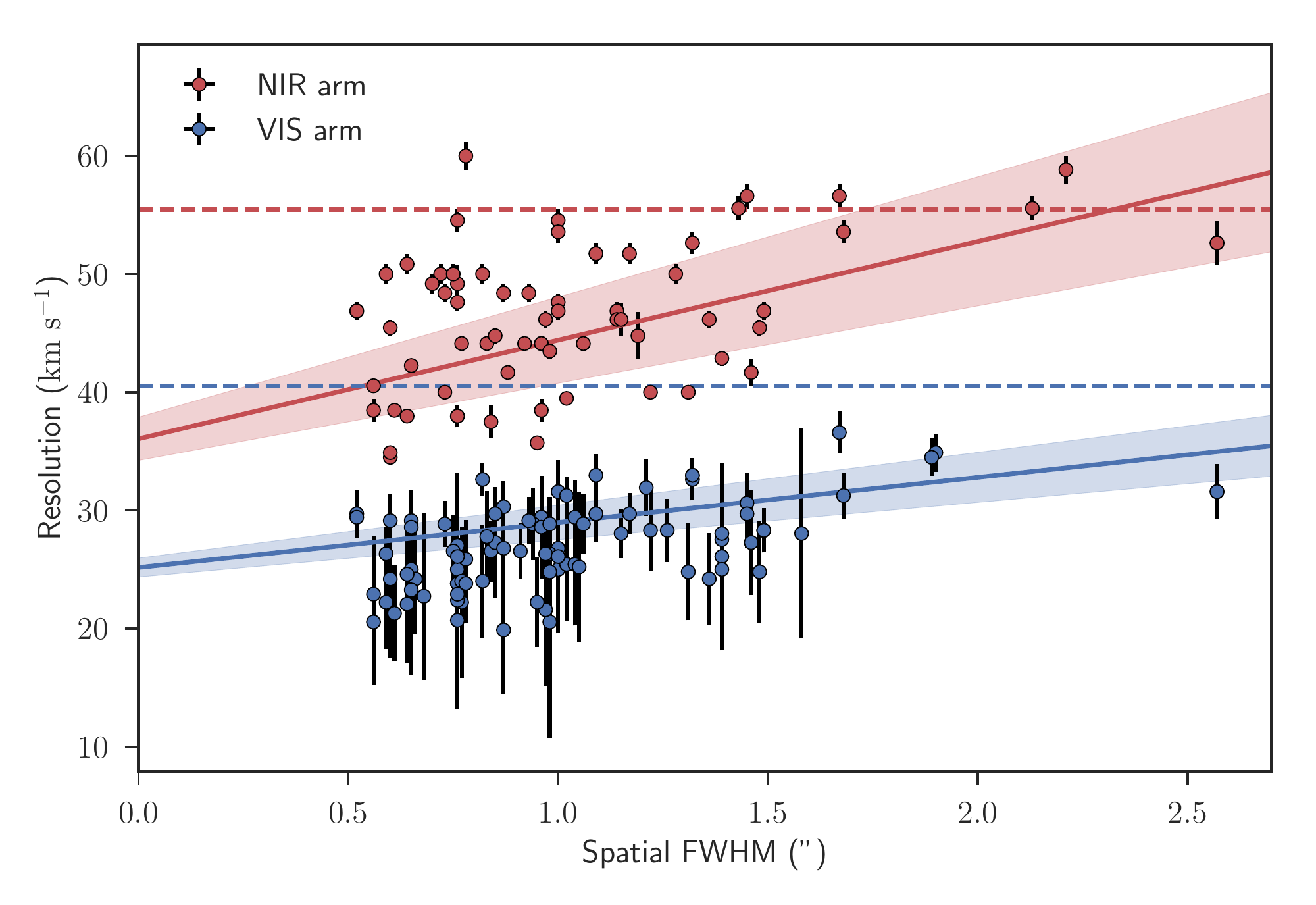}} \caption{Blue
		data points show the FWHM (km/s) of Gaussian fits to unresolved telluric
		absorption lines in the VIS spectra, as a function of the airmass corrected
		DIMM seeing. The red data points show a subsample of measurements obtained for
		NIR spectra. The colored lines show the corresponding linear fits to the
		data points. The colored bounds contain 68 per cent of the best-fit probability
		mass. The red and blue dashed lines indicate the nominal resolution for the
		slit-widths used for the telluric standard observations. As can be seen, the
		effective resolution is in many cases superior to the nominal one.}
	\label{fig:res}
\end{figure}

The afterglow spectra described in this paper are obtained in
Target-of-Opportunity mode. In most cases, there is little possibility to tweak
slit widths to the seeing at the time of observations (i.e. to optimize spectral
resolution and signal-to-noise), and almost all our data are therefore taken
with a fixed set of slit widths and binning, described above. In a fair number
of cases, the seeing full width at half maximum (FWHM) is considerably smaller
than the slit width, and the delivered spectral resolution will then be
determined by the seeing rather than by the slit width, as afterglows are point
sources (this is evidently not the case for extended sources, e.g for host
galaxies). The delivered resolution for slit-width dominated spectra
post-reduction and extraction can easily be determined from the bright sky
emission lines. For afterglow spectra with very high signal-to-noise, the
delivered spectral resolution can at times be determined from the science data
themselves. However, in the presence of multiple velocity components in
absorption, other forms of line broadening, and a lack of lines at some
redshifts, this is difficult to do at lower signal-to-noise ratios (the majority
of spectra in our sample). A broad starting value for the expected resolution
will help fitting of these spectra, and can be important in upper limit
determination, and for this reason we construct a crude relation between the
seeing and the delivered resolution at our slit width, binning, and reduction
pipeline settings.

To this end we use observations of telluric standard stars that are taken with
identical instrument settings as our afterglow spectra, usually just after the
science data, as part of the ESO X-shooter calibration plan. These spectra have
been reduced together with the afterglow spectra, using identical pipeline
settings with the same version of the pipeline. To get the resolution for each
observation, we select a series of atmospheric transitions that are resolved
multiples which should be intrinsically unresolved, and are in areas with well
defined continuum flux. We then fit the lines with Voigt-profiles and calculate
a corresponding FWHM, which we can then convert to a delivered resolution using
the wavelength of the chosen transitions. To get the resolution as a function of
ambient conditions, for each observations we also calculate the
airmass-corrected \href{http://www.eso.org/asm/ui/publicLog}{DIMM} seeing, which
is measured at 500 nm.

The resulting distribution of spectral FWHM (km/s) as a function of spatial FWHM
at 500 nm is fairly well described in the VIS arm by a linear relation $a +
b*x$, with $x$ the spatial FWHM in arcseconds, $a= 25.2 \pm 0.8$ km/s, $b=3.8
\pm 0.7$ (see Fig. \ref{fig:res}). For the NIR arm the corresponding relation is
$a= 36.0 \pm 1.8$ km/s, $b=8.4 \pm 1.8$. We use these linear relations as a way
to estimate the spectral resolution for medium to low signal-to-noise afterglow
spectra. To extend this to the UVB arm, we calculate the ratio between the VIS
and the NIR arm resolutions and find that the resulting distribution is
consistent with a simple scaling of the VIS arm relation by the ratio of
resolutions of the NIR and VIS arm for unresolved, slit filling, sources as
given on the
\href{https://www.eso.org/sci/facilities/paranal/instruments/xshooter/inst.html}{ESO instrument website}. The UVB arm contains no suitable absorption lines to use, and we therefore use a scaled value. This simple analysis gives a sufficiently accurate estimate for the analysis of the low signal-to-noise science spectra. In all cases the determined resolution is written to the header with the "RESOLUTION" keyword.

\subsection{Spectral extraction}    \label{extract}

To extract the afterglow spectrum from the rectified 2D-image, several
techniques have been employed based on the brightness of the afterglow and the
complexity of the objects entering the slit. Due to the malfunctioning ADC (see
Sect. \ref{obs}), the spectral trace changes position across the slit in the
spatial direction as a function of wavelength. For a large fraction of the
observed bursts, using a single aperture for the spectral extraction is
inadequate due to the large amount of background that would then enter the slit.
To optimally select the extraction regions we therefore need to model the trace
position.

To get the shape and the position of the spectral PSF as a function of location
on the image, we need to chose a model which can represent how the light falls
on the slit. We know from \citet{Trujillo2001} that a Moffat function
\citep{Moffat1969} with an index of $\beta = 4.765$ adequately describes an
imaging PSF due to atmospheric turbulence, but because of aberrations in the
optical dispersion elements and the rectification process, the PSF we are trying to
model is different from this profile. To allow for flexibility in the model, we
have chosen the Voigt function as a model for the spectral PSF and we describe
how this is evaluated in App. \ref{voigt}. Since the host galaxy could also give
a contribution to the image profile, this choice allows for the required freedom
if additional flux is in the wings of the profile.

To guide the estimated position of the trace on the slit as a function of
wavelength, we have used the analytic prescription for the trace position
described in \citet{Filippenko1982}, where the header keywords of the
observations have been queried for the ambient conditions which controls the
degree to which the trace changes position in the spatial direction. This
analytic approach is only valid for a plane-parallel atmosphere, but because the
final position is refined in the fit, it is adequate for our purposes.

Based on the signal-to-noise of the afterglow continuum, the 2D-image has been
binned in the spectral direction to a number of elements that allows for an
accurate tracing of the PSF, typically 200 bins for moderate signal-to-noise.
For each of the bins, using the analytically guided guess position, the spectral
PSF has been fit using the unweighted chi-squared minimization algorithm
implemented in \texttt{scipy.optimize.curve\_fit} \citep{scipy}. Since we know
that the trace varies slowly as a function of wavelength, we have then fitted a
low-order polynomial to the fit parameters as a function of wavelength, which
allows us to evaluate the spectral PSF at all wavelengths and in this way
accurately model the entire spectral PSF.

Equipped with a model for how the light is distributed across the entire
dispersion direction, we can  employ the optimal extraction algorithm
\citep{Horne1986}, which weights the extraction aperture by the spectral profile,
or alternatively sum all pixels within 1 FWHM of the modeled profile. Where
possible, we have used the optimal extraction. In cases where the trace is very
weak, even in the binned images, an aperture has been selected manually which
covers all emission lines, if present, and when nothing is immediately visible, the
entire nodding window. The error- and bad pixel maps are in all cases propagated
throughout the extraction.

In cases where multiple traces are visible in the slit, additional components
for the profile are used in the optimal extraction. The additional components do
not share the PSF parameters and in cases where the additional component is an
extended object, the fits have been inspected to ensure that the additional
component does not skew the fit towards a different PSF. The additional
components are not used for the weights in the extractions.

The spectra are corrected for Galactic extinction using the $E(B-V)$ value from
the dust maps of \citet{Schlegel1998} with the update in
\citet{Schlafly2011}\footnote{Queried from
	http://irsa.ipac.caltech.edu/applications/DUST/index.html using
	\texttt{astroquery} \citep{astroquery}.}, and the extinction curve by
\cite{Cardelli1989} with a total to selective extinction $R_V = 3.1$. The
wavelengths of the extracted 1D-spectra are wavelength recalibrated (described
in Sect. \ref{wavecal}), moved to vacuum, and corrected for barycentric motion.
Pixels with pixel-to-pixel variation larger than $50 \sigma$ are additionally
added to the bad pixel map.

\subsection{Telluric correction} \label{tell_corr}

For all Earth-based telescopes, the light first has to pass through Earth's
atmosphere, where the atmospheric content and conditions make an imprint on the
received spectrum. These telluric features can be corrected for in a multitude
of ways. We employ a prioritized list of methods here, depending on the
availability of the chosen methods. Since the observations are often taken at
odd times under varying conditions, this prioritized list ensures that we are
always doing the best possible correction.

The highest priority method is using the GRB afterglow continuum itself, where
the atmospheric conditions have directly been imprinted on the spectrum. The
telluric features can directly be fit with an atmospheric model \citep[\texttt{Molecfit};][]{Smette2015,
	Kausch2015}\footnote{\url{http://www.eso.org/sci/software/pipelines/skytools/molecfit}}, which can then be used to correct for the absorption. The accuracy of the correction depends on the signal-to-noise per pixel of the target spectrum, where we have chosen the requirement that the afterglow continuum spectrum has a median signal-to-noise higher than a value of 10.

If the afterglow is not sufficiently bright, telluric standard stars observed
close in time to the GRB can be used as a proxy for the atmospheric
condition during the GRB observation. Here we employ the telluric correction
method that has been developed in \citet{Selsing2015}, where a library of
synthetic templates is fit to the observed telluric standard.

In the last case, where the object is neither bright enough, or there for some
reason a telluric standard have not been observed, we rely on a synthetic sky
model \citep{Noll2012, Jones2013} for which we generate a
synthetic transmission spectrum, where the ambient parameters for the
observations have been used.

\begin{figure*}
	\centerline{\includegraphics[width=16cm]{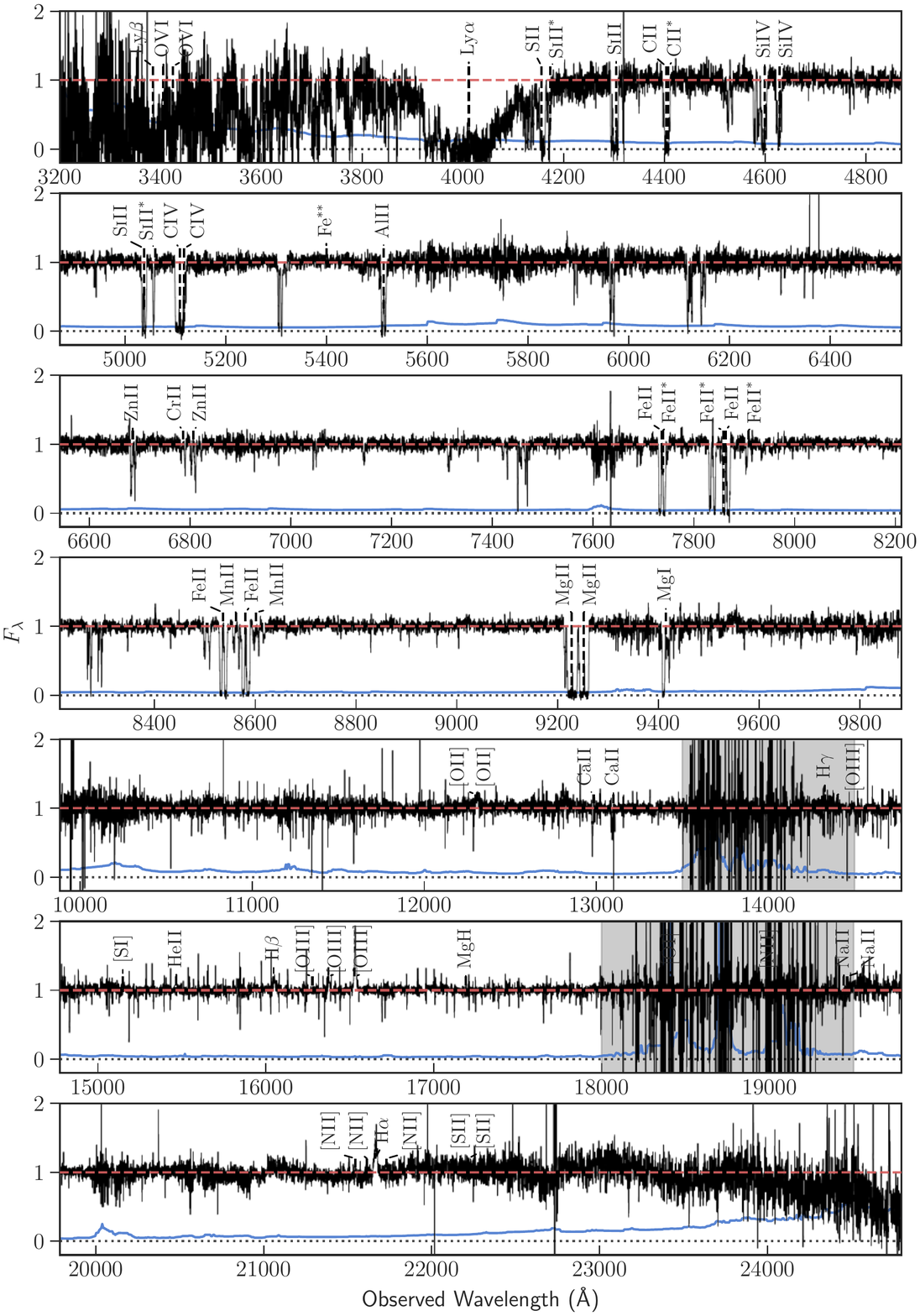}}
\caption{Telluric corrected, normalized spectrum of GRB~121024A at $z = 2.300$
	that illustrates the typical data quality. The continuum estimate is shown in
	dashed red and the error spectrum in solid blue. The acquisition magnitude is
	$R$ = 20, meaning it is in the brighter end of the sample presented here, but
	not the brightest. The spectrum is rich in absorption and emission lines, including
	absorption from molecular $\mathrm{H_2}$. The absorption trough visible at
	$\sim$ 4000 $\AA$ is due to \lya~in the host. We have marked the most prominent
	lines seen in GRB afterglows from \citet{Christensen2011a}. The regions of most
	severe telluric absorption are highlighted by grey-shading the background.
	Additionally, three intervening systems are seen in this sightline. This
	spectrum has been analyzed in detail in \citet{Friis2015}.}
\label{fig:spectrum}
\end{figure*}

\subsection{Continuum estimate} \label{continuum}

We additionally provide an estimate of the continuum for all the spectra
presented here. For this, we have developed an algorithmic approach that
attempts to automatically estimate the continuum placement along with the error
on the continuum estimate through an iterative procedure. The method is entirely
data-driven and does not rely on any physical assumptions. The method is applied
on each arm separately for each spectrum, to allow the widest possible
wavelength range of the spectral shape to guide the normalization.

To estimate the continuum, a number of points (typically on the order of 100)
are inserted at random positions along the wavelength direction, and the
flux-density of each point is determined by the median value of the spectrum in
a small region ($\sim 1~\mathrm{\AA}$) surrounding each point. The points are
fitted with a low order polynomial (we use \texttt{numpy.polynomial.chebyshev})
and iteratively, the point furthest away is removed until the polynomial fit
differed from the points by less than $\sim$ 5 per cent. This filtering is used
because the intrinsic afterglow continuum of GRBs can be modeled by power laws
\citep{Piran2005}, and removing points that differ significantly from a smooth
continuum shape will guide the continuum estimate to a shape more reasonable for
GRB afterglows. The reason for the non-physical model for the normalization is
that it has the flexibility to capture instrumental variations of the continuum
level that are not easily contained in a more physically motivated model.
Additionally, points spaced closer than 1 per cent of the total spectral
coverage are pruned. The remaining filtered and pruned points are then spline
interpolated using \texttt{scipy.interpolate.splrep.splev}, which serves as a
first estimate of the continuum placement. To prevent the spline from diverging
at the edges, the spline-based continuum is tapered with a low order polynomial.
An attempt to identify absorption and emission regions is then carried out,
where they are marked as such if the difference between the estimated continuum
and the observed spectrum is larger than 3 - 5 times the associated error
spectrum. All regions marked as affected are then masked.

Using the masked spectrum, the entire process is then repeated 500 times where
the final continuum estimate is the mean of the continuum realizations and the
associated error estimate is the standard deviation. This error reflects the
stability of the algorithm across the spectrum. An example of the performance is
shown in Fig. \ref{fig:spectrum}. In the \lya~forest, where there is very little
flux at the continuum level, the performance of the normalization algorithm
depends on the continuum coverage redwards of the \lya~line. In some cases, very
little continuum is contained in a single arm and a manual continuum estimate is
provided, similar to what is done in \citet{Lopez2016}. In these cases the
continuum error is set to 10 per cent. The code for the continuum estimate is
released along with the paper at
\url{https://github.com/jselsing/XSGRB-sample-paper}.

\subsection{Science data products} \label{products}


All the spectra are made available as a single
\href{https://sid.erda.dk/cgi-sid/ls.py?share_id=DBuNORk1lI}{ZIP file}, through
\url{http://grbspec.iaa.es} \citep{deUgartePostigo2014f}, and additionally through the ESO archive in the
form of \href{http://archive.eso.org/wdb/wdb/adp/phase3_main/form}{phase 3
	material}. This release includes both prompt afterglow observations as well as
late time observations of the associated hosts, and represents \textit{all}
afterglow spectra of GRBs carried out by the X-shooter spectrograph since the
commissioning of the instrument, \startdate, and until the end of the last
period of the program 098.A-0055, \termdate~and thus constitutes eight years of GRB
afterglow observations with X-shooter. An overview of all the spectra and their
observational setups is given in Table \ref{tab:sample_overview}. For each
burst, each individual observation is provided in a separate reduction, and in
cases where observations have been repeated for an increased signal-to-noise or
to follow the temporal evolution, a combined spectrum is also provided. No
attempt has been made to join the spectroscopic arms, so for each observation,
three spectra are provided in separate files.

All spectra are released in the ESO Science Data Product (SDP) format
\citep{Micol2016}, and formatted as binary FITS files. The naming convention is
based on the GRB name and the observation number, and follow the scheme
\texttt{GRBxxxxxxx\_OBxarm.fits}. For example, the visual arm of the third
observation of GRB~151021A, observed in RRM mode (see Sect. \ref{RRM}), is named
\texttt{GRB151021A\_OB3VIS.fits}.

Each file contains 7 columns with the following contents and descriptions:
\begin{itemize}
	
	\item WAVE - Observed wavelength in vacuum, corrected for barycentric motion
	and drifts in the wavelength solution ($\mathrm{\AA}$).
	
	\item FLUX - Observed flux density ($\mathrm{erg} \mathrm{s}^{-1}
	\mathrm{cm}^{-2} \mathrm{\AA}^{-1}$).
	
	\item ERR - Associated flux density error ($\mathrm{erg} \mathrm{s}^{-1}
	\mathrm{cm}^{-2} \mathrm{\AA}^{-1}$).
	
	\item QUAL - Bad pixel map, where a value different from zero indicates a bad
	pixel.
	
	\item CONTINUUM - Continuum estimate based on Sect. \ref{continuum}
	($\mathrm{erg} \mathrm{s}^{-1} \mathrm{cm}^{-2} \mathrm{\AA}^{-1}$).
	
	\item CONTINUUM\_ERR - Relative error on the continuum estimate.
	
	\item TELL\_CORR - Inverse transmission spectrum. Multiply FLUX and ERR column
	with this column to correct for telluric absorption.
	 
\end{itemize}

\section{Results} \label{results}

In this section, we describe the efficiency of the follow-up effort and the
characteristics of the observed bursts. We also assess the degree to which the
obtained sample is representative for the full \textit{Swift} sample. An
important note is that here we provide a release for \textit{all} GRBs after
\startdate, that have been observed with X-shooter, while only a subset of these
constitutes our \textit{statistical} sample. The statistical sample is based on
the selection criteria described in Sect. \ref{samplecrit}. Some bursts not
fulfilling the sample criteria have been followed up due to interesting
characteristics, e.g., curios properties of their light curves, their
brightness, etc. These bursts are not discussed as part of the investigation of
the statistical properties of the GRB population. A prime example of a
spectacular burst outside the statistical sample is the bright \textit{INTEGRAL}
burst GRB~161023A (de Ugarte Postigo et al., in prep), that contains at least 15
intervening absorption systems (See \ref{161023}).

\subsection{Follow-up timing and afterglow brightness} \label{timing}

\begin{figure}[!ht]
	\centerline{\includegraphics[width=\columnwidth]{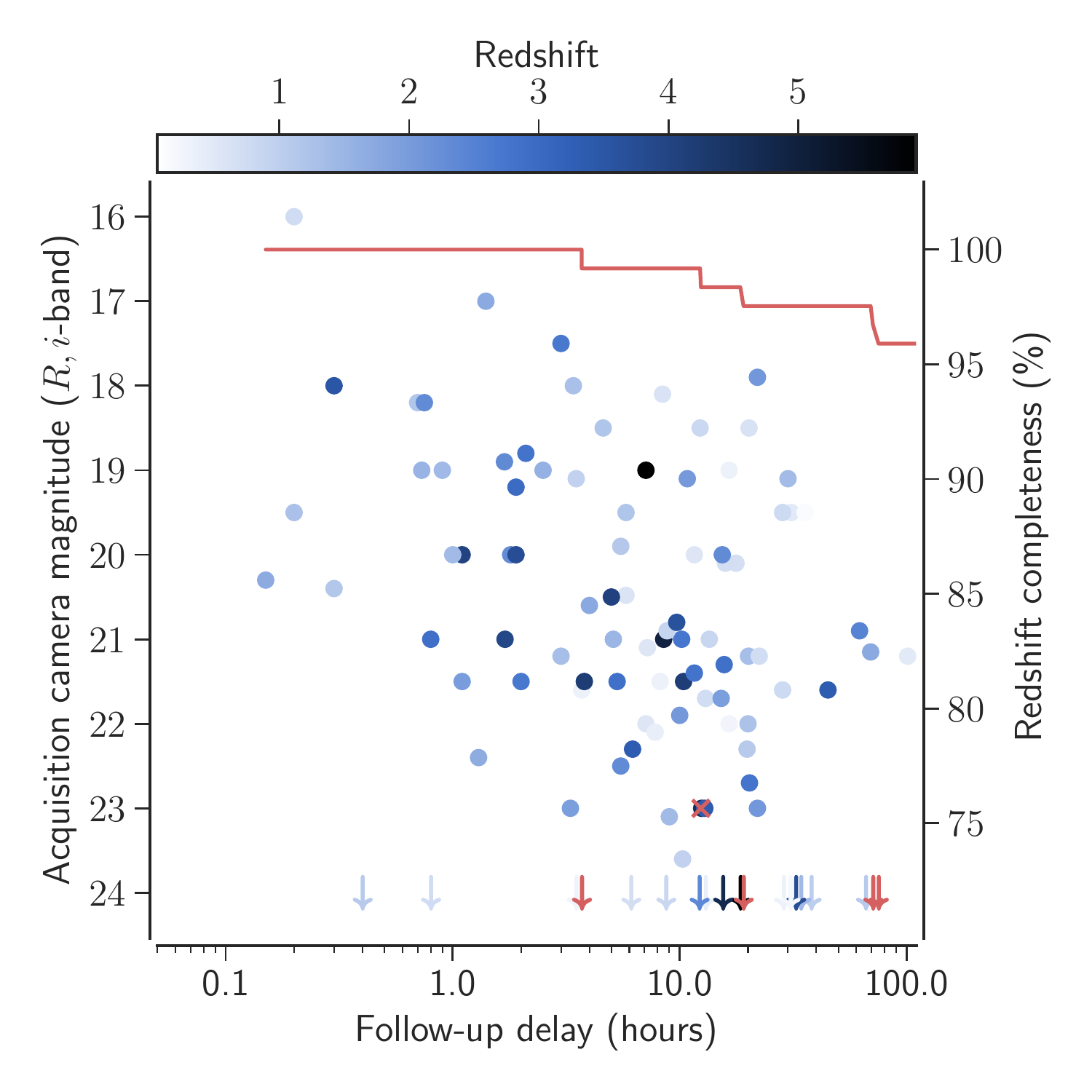}} \caption{Afterglow
	magnitude at the start of observation and redshift completeness as a function
	of follow-up delay for all the afterglows that have been followed up. The
	points have been colored based on the redshift of the corresponding burst. Red
	symbols indicate GRBs without a measured redshift and arrows indicate bursts
	for which the afterglow was not detected in the acquisition image. In red is
	shown the redshift completeness as a function of follow-up delay.}
\label{fig:timing}
\end{figure}

Redshift determination of bursts for which the host is too faint for a
spectroscopic redshift measurement relies on the detection of absorption lines
imprinted on the GRB afterglow continuum. Because the optical afterglow rapidly
fades (typically as $\sim t^{-1}$) a rapid follow-up is essential. In Fig.
\ref{fig:timing} we plot the delay from the BAT trigger to the start of the
spectroscopic observation. The shortest delays are observed in RRM-mode. The
fastest follow-up between BAT trigger and start of spectroscopic observations
for any observation is for the short, $z = 1.717$, GRB~160410A for which
spectroscopic integration was initiated only 8.4 minutes after the BAT trigger.
To illustrate the importance of the follow-up delay for the redshift
completeness, we plot the redshift completeness as a function of delay time in
Fig. \ref{fig:timing} for all the bursts we have followed up, including the ones
outside the statistical sample. As can be seen from the figure, the fraction of
GRBs with a redshift determination decreases with follow-up delay. The redshift
completeness for bursts that we have followed up is 94 per cent. This degree of
completeness in the followed bursts, illustrates the efficiency of VLT/X-shooter
in redshift determination. Not shown in the figure are an additional 12 bursts
that have redshift determinations based on late-time host observations with
delay times longer than $\sim$10 days.

\subsection{Sample completeness} \label{completeness}

\begin{figure*}
	\centerline{\includegraphics[width=\linewidth]{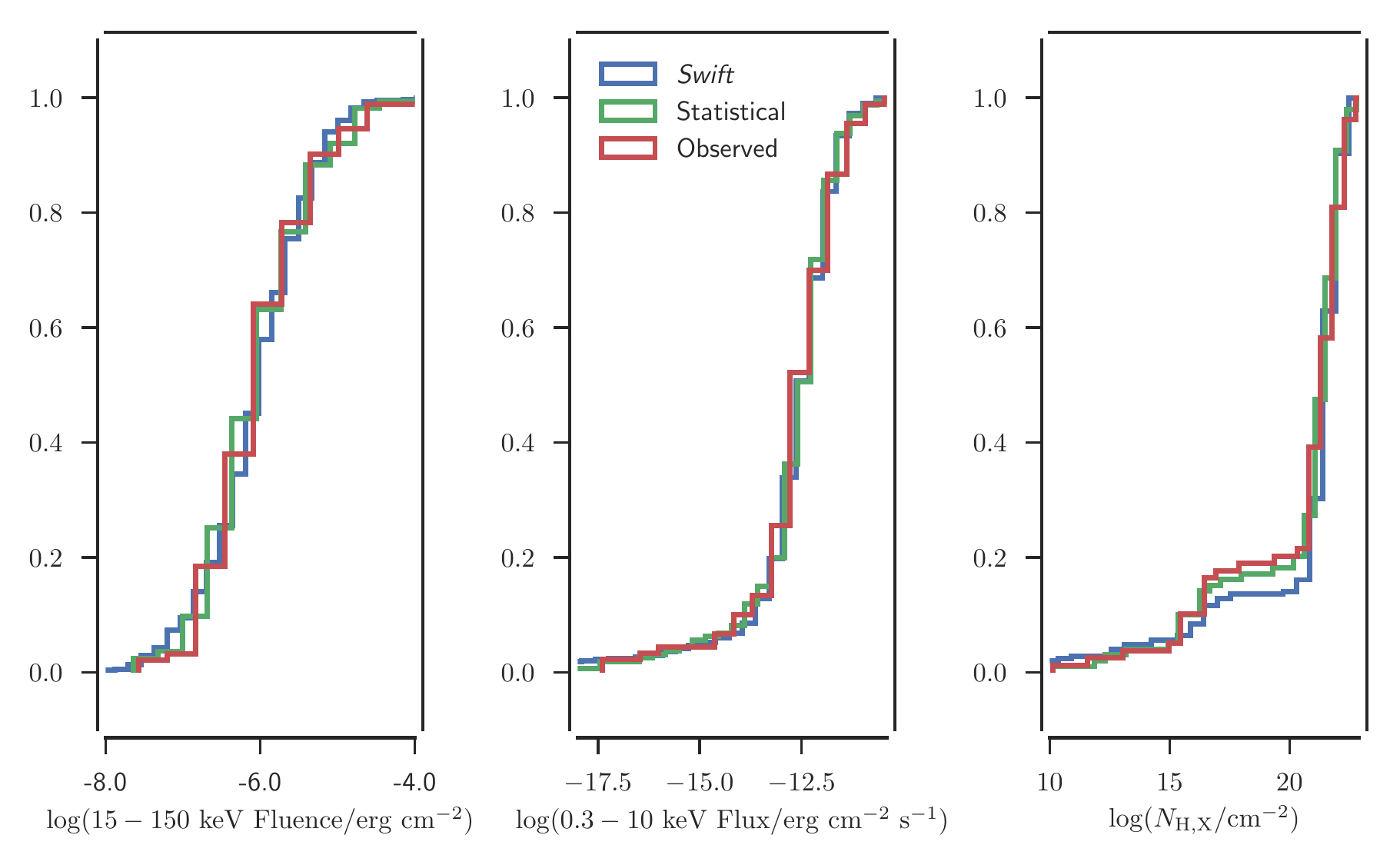}}
\caption{Comparison between the burst properties of all bursts observed with
	\textit{Swift} (blue), the statistical sample that fullfill the criteria
	specified in Sect. \ref{samplecrit} (green), and the subset that has been
	observed as part of the statistical sample (red). The left plot shows the
	fluence in the 15-150 keV band as observed by BAT. The middle panel shows the 0.3 - 10
	keV flux, 11 hours after the bursts as measured by XRT. In the right most
	panel, we show $N_{\mathrm{HI,X}}$ based on the XRT spectrum
	\citep{Evans2009}.} \label{fig:swift_complete}
\end{figure*}

Of all the BAT-triggered bursts, a total of 165 bursts fulfill the sample
criteria specified in Sect. \ref{samplecrit}, since the commissioning of
VLT/X-shooter. This sample constitutes the \textit{statistical sample} from
which we will derive statistical properties of the GRB population. The redshift
completeness of the full statistical sample is 61 per cent. We return to the
question of redshift completeness in Sect. \ref{redshift}. From this sample, 93
GRBs have been spectroscopically followed up with X-shooter. In order to assess
whether the subset of bursts followed up are representative of the underlying
GRB parent population, we compare intrinsic properties of GRBs in our sample to
GRBs in the full sample followed up by \textit{Swift}. We show the comparison
between the BAT (15 -- 150 keV) fluence, the XRT flux (0.3 -- 10 keV) at 11
hours, and the intrinsic X-ray derived equivalent hydrogen column density at the
redshift of the GRB, in excess of the Galactic X-ray absorption column,
$N_{\mathrm{HI,X}}$, in Fig. \ref{fig:swift_complete}. For the latter, we can
only use values of $N_{\mathrm{HI,X}}$ derived for bursts with a measured
redshift, excluding $\sim 75$ per cent of the full \swift~sample. We return to
the last point in Sect. \ref{darkness}.

Using the observational characteristics of the 1266 bursts observed until
\termdate~by \textit{Swift}, and the derived $N_{\mathrm{HI,X}}$
\citep{Evans2009}, we can quantify the degree to which our sample is biased
relative to the overall \textit{Swift} sample. The values are queried from the
online \textit{Swift}
database\footnote{\url{http://swift.gsfc.nasa.gov/archive/grb\_table/}}$^,$\footnote{\url{http://www.swift.ac.uk/xrt\_live\_cat/}}. Three samples are of interest in order to assess the completeness of the follow-up campaign (Fig.~\ref{fig:swift_complete}); the full \textit{Swift} sample consisting of all the bursts observed by \textit{Swift} (blue), all the bursts that fulfill the selection criteria imposed in Sect. \ref{samplecrit} (green), and the bursts actually followed up with X-shooter (red).

\input{tables/sample_properties.tex}

For each of the samples, we calculate the median, 16th, and 84th percentiles of
each of the distributions, which can be used as point estimates for the
population distribution. These are provided in Table
\ref{tab:sample_properties}. It can be seen from the values that the three
samples have very similar distributions in terms of the point estimates chosen.
This suggests that our selection criteria are unbiased compared to the
\textit{Swift}-sample and that additionally, the follow-up effort conserves the
distributions of the intrinsic GRB properties (except perhaps for
$N_{\mathrm{HI,X}}$, see Sect. \ref{darkness}).

\begin{figure}[!ht]
	\centerline{\includegraphics[width=\columnwidth]{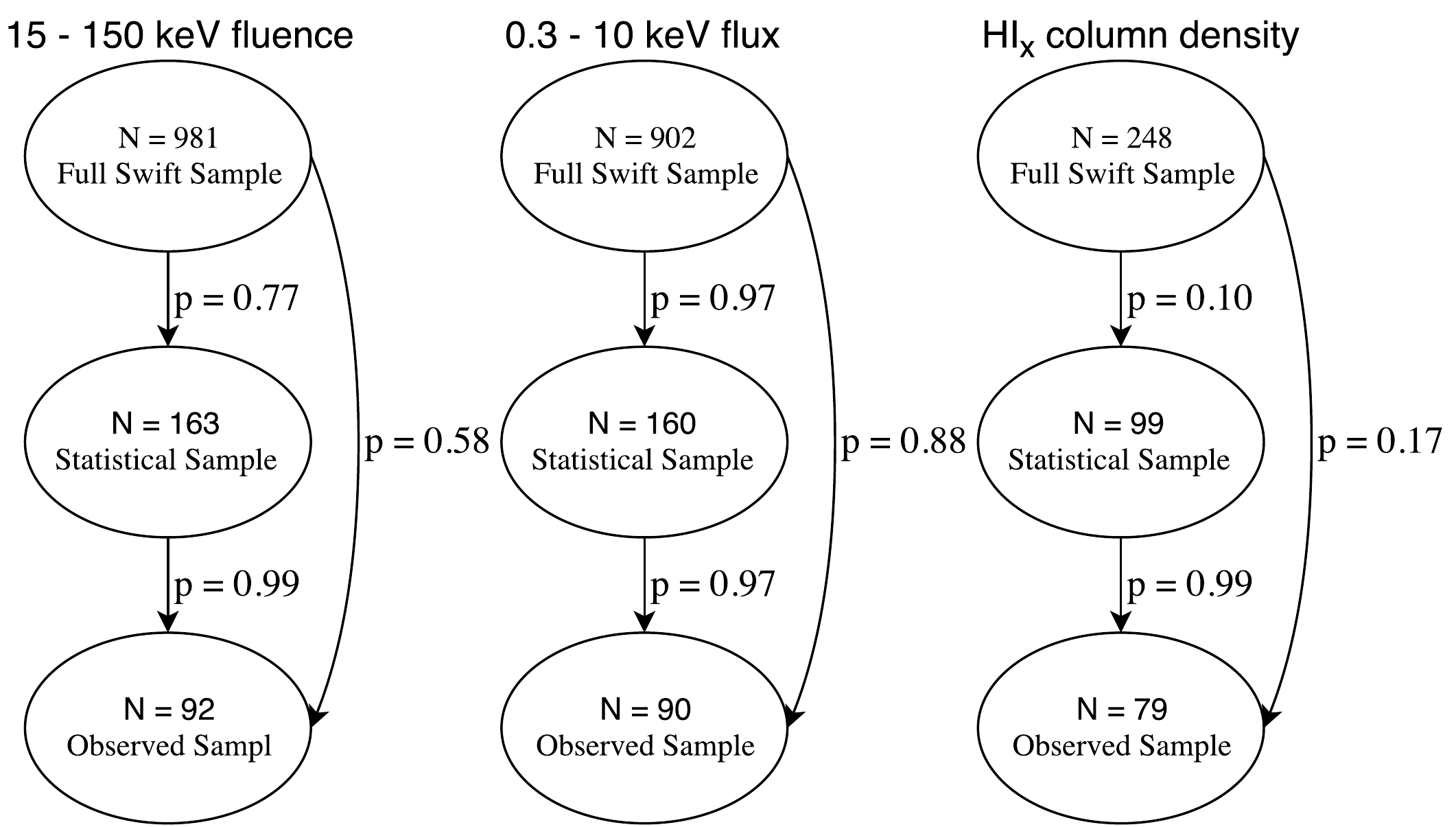}}
\caption{Relational graph showing the respective $p$-values. They all represent
	the degree that the different samples are drawn from the same underlying
	distribution. The arrows represent the comparison direction with each of the
	$p$-values they are listed next to. Only in the H{\sc I} column density distribution is
	there mild evidence against the null hypothesis, but the discrepancy is mainly
	driven by a relatively larger fractional contribution from low-column hosts in
	the statistical sample.} \label{fig:p_values}
\end{figure}

Additionally, using a 2-sided Kolmogorov-Smirnov test (KS-test), we can assess
the degree to which the null hypothesis, that the two distributions are drawn
from the same parent distribution, can be rejected. We show a graphic
representation of the test statistics in Fig. \ref{fig:p_values}. A high $p$-value
indicates little evidence against the null hypothesis . The distribution of
$N_{\mathrm{HI,X}}$ exhibits the highest degree of dissimilarity, but the two
distribution are still consistent with being drawn from the same underlying
distribution.

We therefore conclude that the statistics of the sample presented here,
conserves the intrinsic properties of the GRBs in the full \textit{Swift} sample
- at least in terms of BAT fluence and X-ray flux at 11 hours.


\subsubsection{Properties of rejected triggers} \label{badbursts}

Out of the 165 bursts meeting our initial selection criteria for the statistical
sample, 36 (22 per cent) were not observed due to reasons unrelated to the GRB
or afterglow properties. The reasons include unavailability of the telescope due
to technical maintenance (e.g., mirror re-coating), a visiting observer
rejecting the ToO trigger, or bad weather. Because this cut is unrelated to the
GRB properties, it will not change the statistical properties of the full
sample. Removing these bursts from the statistical sample, dramatically improves
the redshift completeness from 61 per cent to 88 per cent. The remaining burst
not followed up already had a redshift from other instruments or were very faint
and without a host association, thus observations were unlikely to yield a redshift
measurement. In the remainder of the text, we consider the 129 bursts our
statistical sample.

\subsection{On the redshift distribution of GRBs} \label{redshift}

One of the objectives of our follow-up campaign is to measure the redshift
distribution for a well-defined, observationally unbiased and statistically useful
sample of GRBs. The imposed selection criteria (see Sect. \ref{samplecrit})
ensure that the GRBs entering our homogeneous sample, fairly represent the
underlying population. The redshift distribution of such a sample holds valuable
information about the occurrence of GRBs through cosmic time
\citep{Jakobsson2012, Perley2016a}.

\input{tables/redshift_comparison}

\begin{figure*}[!ht]
	\centering \includegraphics[width=\linewidth]{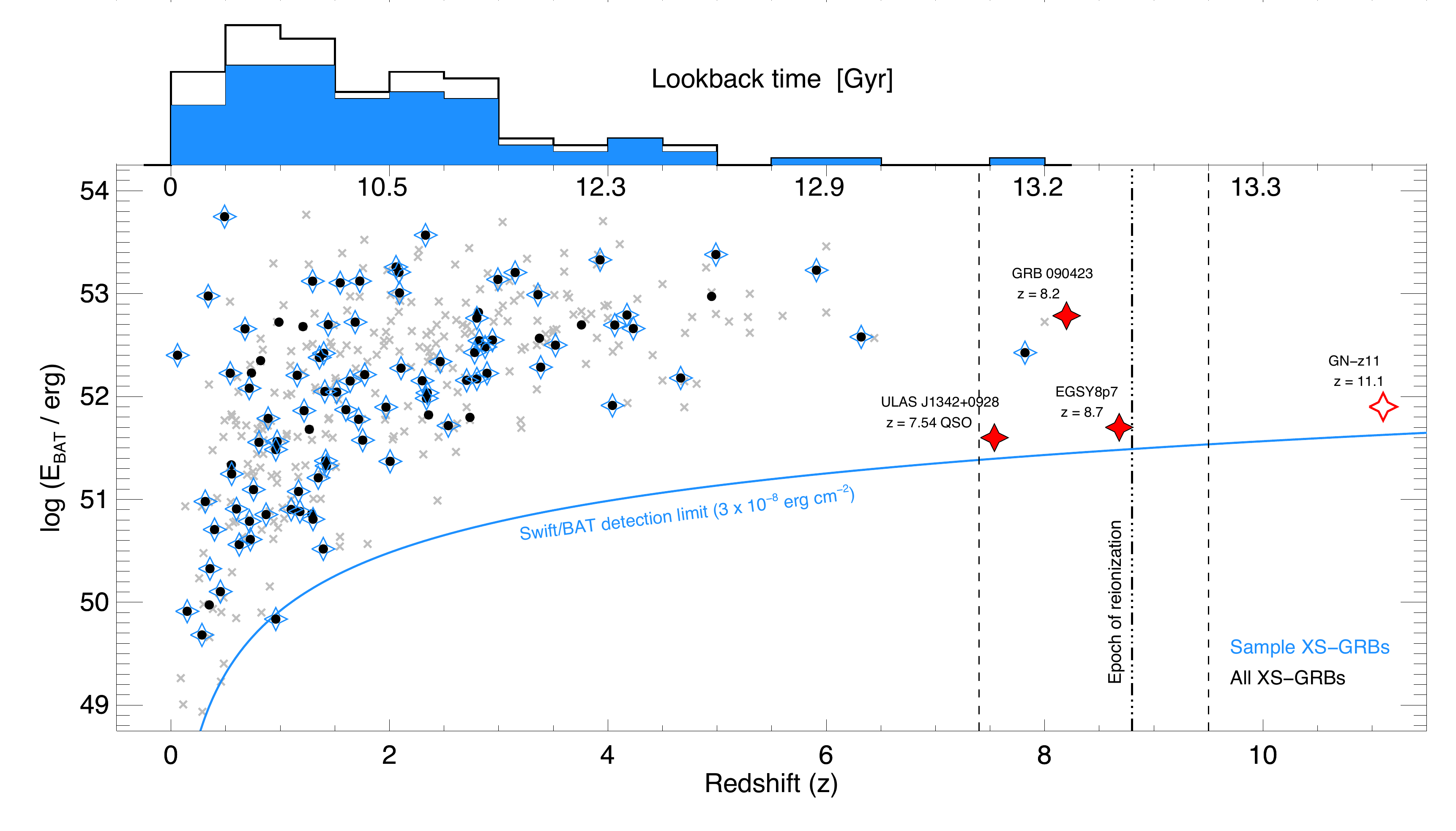}
\caption{Redshift distribution as a function of intrinsic BAT $\gamma$-ray
	energy, $E_{\mathrm{BAT}}$. Bursts that are a part of the statistical sample
	are marked by blue stars whereas black dots show all GRBs observed with
	X-shooter. All \swift~GRBs with measured redshifts are shown in gray. For
	comparison, we overplot with red stars the GRB \citep{Tanvir2009b,
		Salvaterra2009a}, quasar \citep{Banados2017}, and galaxies \citep{Zitrin2015,
		Oesch2016} with the highest spectroscopically confirmed redshifts, the latter
	three shown at arbitrary $E_{\mathrm{BAT}}$. The blue solid line represents the
	so-called 1 s BAT sensitivity limit described in the text. The estimated epoch
	of reionization is shown by the black dot-dashed line, with the uncertainty
	shown as the black dashed lines, from the most recent measurement by
	\citet{Planck2015}. On the top of the plot is shown the marginalization of the
	redshift. Again the blue histogram represents the bursts than enter our sample
	and the white histogram the full GRB population. } \label{fig:z}
\end{figure*}

In Fig.~\ref{fig:z} we show the redshift distribution of all  the observed GRBs.
In the top panel we show a histogram for the full sample and the statistical
sample and in the main panel, we show the redshifts of the individual bursts as
a function of GRB energy in the observed 15 -- 150 keV band. To calculate the
energy, $E_{\mathrm{BAT}}$, we follow a similar procedure as \citet{Lien2016}
and define $E_{\mathrm{BAT}} = F_{\gamma}\,4 \pi\,d_L^2\,(1+z)^{-1}$, where
$F_{\gamma}$ is the observed BAT fluence in the 15-150 keV band and $d_L$ is the
luminosity distance to the burst at the given redshift. Note that this measure
of luminosity does not include any $k$-correction. As an indication of the
effect of the \textit{Swift} sensitivity limit on the redshift distribution,
also in the figure, we have shown the so called $\sim 1$ s flux BAT sensitivity
limit ($\sim 3 \times 10^{-8}~\mathrm{erg}~\mathrm{s}^{-1}~\mathrm{cm}^{-2}$;
\citealt{Baumgartner2013, Lien2016}). Due to the complex triggering mechanism of
\swift, this sensitivity limit should be interpreted with some caution as the
effective limit depends on the light curve of the prompt emission signal. Due to
the dilution of light with distance, the \swift~GRB luminosity detection limit
is almost an order of magnitude brighter at $z=2$ than at $z=1$. At $z\geq4$ we
are only able to observe GRBs that are $\sim$ hundred times brighter than the
faintest bursts at $z=1$ and below. The effect of GRB redshift on the
\swift~triggering criteria have previously been studied in detail
\citep{Littlejohns2013a}.

We compare the point estimates for the redshift distributions of previous
complete samples of GRBs in Tab. \ref{tab:redshift_comparison}. We see that when
we compare to other complete samples, the XS-GRB presented here has the lowest
average redshift. However, the other samples also exhibit a large spread in the
redshift distributions. A 2-sided KS test reveals that the XS-GRB sample is
consistent with being drawn from the same parent sample with the following
$p$-values: SHOALS ($p$-value = 0.13), BAT6 ($p$-value = 0.95), Fynbo09
($p$-value = 0.08) and TOUGH ($p$-value = 0.09). As a small note, the redshift
distribution of BAT6 are not expected to be identical to the other complete
samples due to the additional cut on the GRB peak flux in the BAT band.

Because the redshift completeness of our statistical sample is 88 per cent,
making an inference of the \textit{true} redshift distribution of GRBs based on
this sample is impossible. 
For instance, only the brightest GRBs are seen above redshift $z \gtrsim 1$ as
shown in Fig. \ref{fig:z}. As described in detail in \citet{Hjorth2012} and
\citet{Perley2016b}, bursts for which the redshift is measured from the
afterglow are systematically found in host galaxies with a lower luminosity
than bursts for which the redshift is measured from the host galaxy. Only a few
GRBs hosted in galaxies, with stellar masses more than
$10^{10}~\mathrm{M}_\odot$ have the redshift measured based on the afterglow
continuum. This is likely related to the presence of higher contents of dust in
more massive galaxies, leading to a larger fraction of extincted afterglows.

\subsection{Sample darkness} \label{darkness}

A fraction of all GRBs exhibit no detectable or very faint optical afterglows
\citep{Groot1998, Djorgovski2001, Fynbo2001}. The degree of optical extinction
relative to the X-ray brightness has been parametrized in terms of their optical
darkness, using the measurement of, or limit on, the optical to X-ray spectral
index $\beta_{OX}$ \citep{Jakobsson2004, Rol2005, VanderHorst2009}. The X-ray
properties of such bursts have previously been investigated
\citep{DePasquale2003, Fynbo2009, Melandri2012} and there are some indications
that dark bursts have somewhat higher X-ray luminosity and $N_{\mathrm{H, X}}$
compared to the optically bright bursts \citep{Campana2012, Watson2012}. The
X-ray column density has been shown to be roughly correlated with the gas column
density, which, for a given metallicity, also correlates with the dust
extinction \citep{Watson2013, Covino2013a}, though the range in metallicity
introduces a large additional scatter in the correlation between the extinction
and the $N_{\mathrm{H, X}}$. This indicates along with investigations of host
galaxy properties \citep{Greiner2011, Kruhler2011, Hjorth2012, Perley2016b},
that the extinction of the optical afterglows is primarily driven by the
presence of dust in the host galaxies and not solely by unfortunate placement of
the synchrotron spectral break frequencies. \citet{Hjorth2012} find that systems
with no optical afterglow have higher $N_{\mathrm{H, X}}$, irrespective of the
nature of the host -- which however also turn out to be redder. Additionally,
the ISM absorption lines in dark sight-lines are found to be stronger compared
to optically brighter bursts \citep{Christensen2011a}, which is consistent with
the dark bursts being found in more metal-rich and dustier galaxies.

For all bursts with follow-up within 100 hours we calculate the
"darkness"-parameter, $\beta_{OX}$ \citep{Jakobsson2004}. This requires the
simultaneous measurement of the X-ray flux density and the optical flux density
which is in practice possible, but in reality extremely rarely available. As
a proxy, we use the measured acquisition camera magnitude reported in Table
\ref{tab:sample_overview} to get the optical flux density at the beginning of
the spectroscopic integration. Because we know the delay between the follow-up
and the \textit{Swift} trigger, we can use the measured XRT lightcurve
\citep{Evans2007,
	Evans2009}\footnote{http://www.swift.ac.uk/xrt\_curves/trigger\_numer/flux.qdp}
to infer the corresponding X-ray flux density at the time of the optical
observation. This is done by either linearly interpolating between temporally
neighboring XRT measurements or by extrapolating the last few X-ray data points
to the time of the spectroscopic observation. When the afterglow is not detected
in the acquisition camera, an upper limit of $> 24$ mags is used, which
propagates into an upper limit on $\beta_{OX}$.

\begin{figure}[!ht]
	\centerline{\includegraphics[width=\columnwidth]{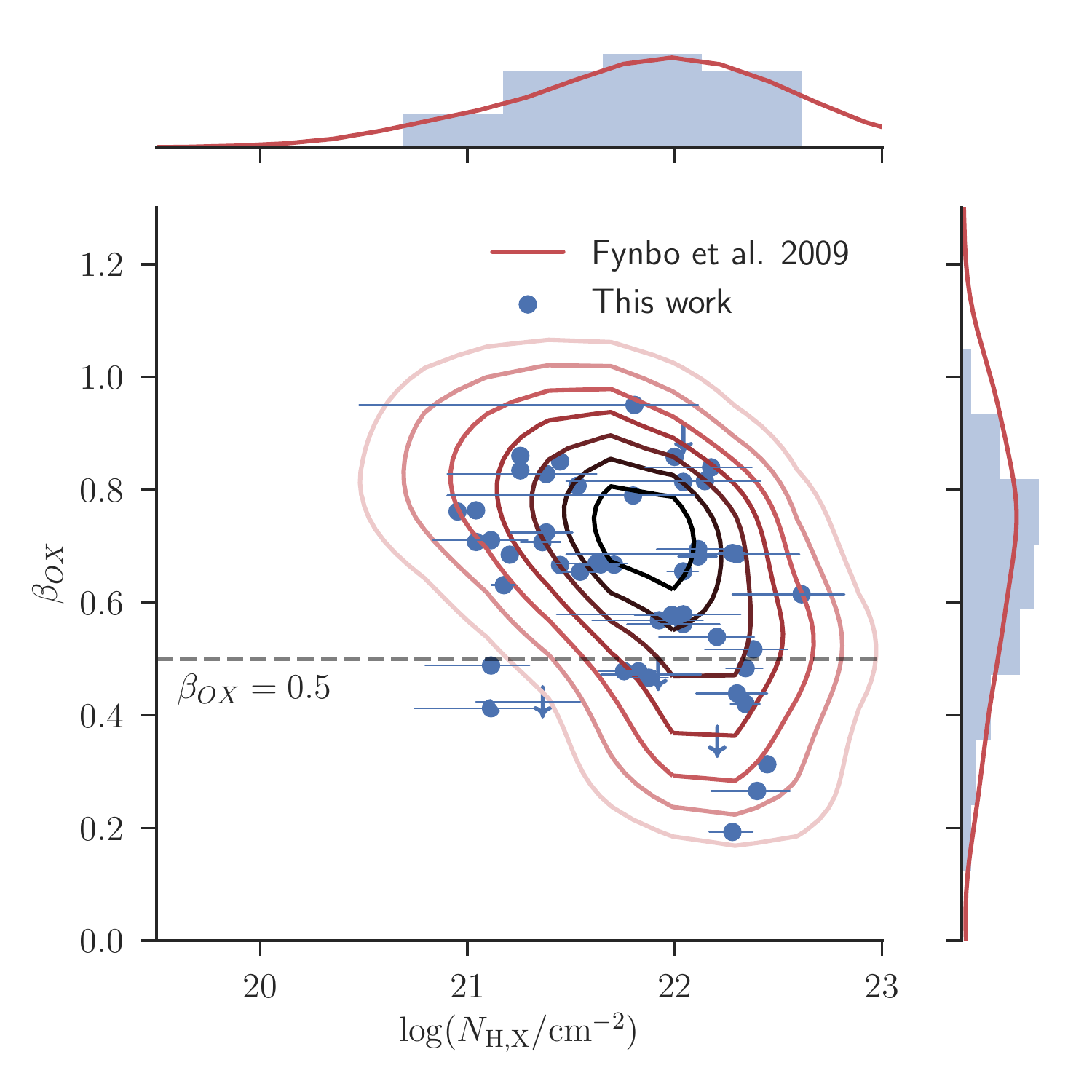}}
\caption{$\beta_{OX}$ against X-ray derived hydrogen column density. In red is
	shown the sample presented in \citet{Fynbo2009} where the lines indicate the
	kernel density estimate of the distribution. For the kernel density estimate,
	the limits have been replaced with values. We note that contrary to what is
	presented in \citet{Fynbo2009}, we exclude all bursts for which we do not have
	a redshift. Darker colors represent a higher density of points. In blue are the
	points for the bursts presented here along with the marginal histograms. Limits
	on $\beta_{OX}$ are shown by downwards facing arrows. The corresponding
	marginal distributions are shown along the edges of the plot. A Kendall's
	$\tau$ test, indicates a statistically significant, low degree of negative
	correlation ($\Gamma = -0.21$ at a $p$-value = 0.01)} \label{fig:betaOX}
\end{figure}

In Fig. \ref{fig:betaOX} we compare the $\beta_{OX}$ - $N_{\mathrm{H, X}}$
distribution with the one presented in \citet{Fynbo2009}. We take the
$N_{\mathrm{HI,X}}$ values from the
\href{http://www.swift.ac.uk/xrt_spectra}{XRT spectral fits} \citep{Evans2009}.
The values from \citet{Fynbo2009} have been treated as detections, meaning that
we artificially bias the distribution towards higher $\beta_{OX}$-values. The
two distributions exhibit a large degree of overlap. We confirm the
result by \citet{Fynbo2009}, that dark bursts, $\beta_{OX} < 0.5$, have higher
\nhx. Specifically, for bursts with measured redshift either from the afterglow
or the host galaxy, we find the following: For bursts with $\beta_{OX} \geq 0.5$
we find \nhx~$ = 21.4_{-1.0}^{+0.7}$, whereas for $\beta_{OX} < 0.5$ we find
\nhx~$21.8_{-0.9}^{+0.5}$ where 68 per cent of the probability mass is contained
within the error intervals. A 2-sided KS test fails to reject the null
hypothesis that they are drawn from the same distribution with $p = 0.11$,
meaning that there is no strong evidence for a discrepancy. A Kendall's $\tau$
test, however, suggest a statistically significant, low degree of negative
correlation ($\Gamma = -0.21$ at a $p$-value = 0.01).

Using the table maintained by
J. Greiner\footnote{http://www.mpe.mpg.de/~jcg/grbgen.html}, we can see how the
presence of an optical afterglow affects the follow-up statistics. 50.5 per cent
of all \textit{Swift}-triggered bursts in this list do not have a detected
optical afterglow. This number also includes bursts where no optical
observations were available, so the real number is likely to be lower. For the
bursts that enter our statistical sample, the percentage of bursts without a
detected optical afterglow is 28 per cent, close to the upper limit on the
fraction of dark bursts found in a complete sample with a very high degree of
redshift completeness \citep{Melandri2012}. Of the bursts for which follow-up
has actually been attempted, this number is 23 per cent, suggesting a slight
bias against bursts without a detected optical afterglow in the spectroscopic
sample.

However, the fraction of dark bursts for which we have measured redshifts is
lower than the ones with a detected optical afterglow. For the afterglows we
have observed as part of the statistical sample that do not have a detected
optical afterglow, the redshift completeness is 53 per cent. For comparison, for
the afterglows in the statistical sample we have observed with an optical
afterglow detected, the redshift completeness is 92 per cent. This also shows
that the lack of redshift completeness in the sample presented here is in part
due to the increased difficulty of measuring a redshift for bursts without an
afterglow. To measure a redshift, we either need a detected afterglow to obtain
a spectrum or to locate the host galaxy and determine the redshift from there.
It is more difficult to correctly associate a galaxy with a burst when there is
no detected optical afterglow and hence a correct redshift measurement is more
difficult to make, see \citet{Jakobsson2005, Levesque2010} and
\citet{Perley2017}.

Regardless the fraction of dark bursts being lower in the observed sample,
compared to the statistical sample, the X-ray properties do not differ
significantly, as shown in Sect. \ref{completeness}. This is despite
spectroscopic follow-up only being carried out in cases where either a
detectable optical afterglow or a clear counterpart are seen, which naively
should be biased against dark bursts occurring in more obscured galaxies, which
is shown to exhibit different galactic properties \citep{Perley2009,
	Kruhler2011, Rossi2012, Perley2013b, Perley2015b}. That the decreased fraction
of dark bursts in observed sample does not alter the observed prompt X-ray
brightness distribution, potentially reflects the independence of the X-ray
brightness on the density of the circumburst medium \citep{Freedman2001,
	Berger2003, Nysewander2009}, if the measured $N_{\mathrm{H, X}}$ is primarily
driven by the gas column in the neighborhood of the burst. Because we only use
values for $N_{\mathrm{H, X}}$ in the comparison for which the GRB has a
measured redshift, this measure is likely biased toward optically brighter
bursts \citep[e.g.][]{Watson2012}.

 \subsection{Hydrogen column densities}

The locations of long GRBs are associated with intensely star-forming regions
\citep{Hogg1999, Bloom2002, Fruchter2006, Lyman2017}. Because a significant fraction
of the hydrogen along the line of sight in these regions has not been ionized,
the optical depth at the wavelength of \lya~is very high, saturating not only
the line center, but also the damping wings. This causes a strong absorption
system from the \lya-transition to appear in the afterglow continuum. For bursts
with $z \gtrsim 1.7$, the position of \lya~moves into the spectroscopic coverage
of X-shooter, meaning that we are able to detect this absorption trough due to
\lya. The exceptionally good UV response of X-shooter, allows us to robustly
measure H{\sc i}-columns - also at these relatively low redshifts.

\begin{figure*}[!h]
	\centering
	\includegraphics[page=1, width=\linewidth]{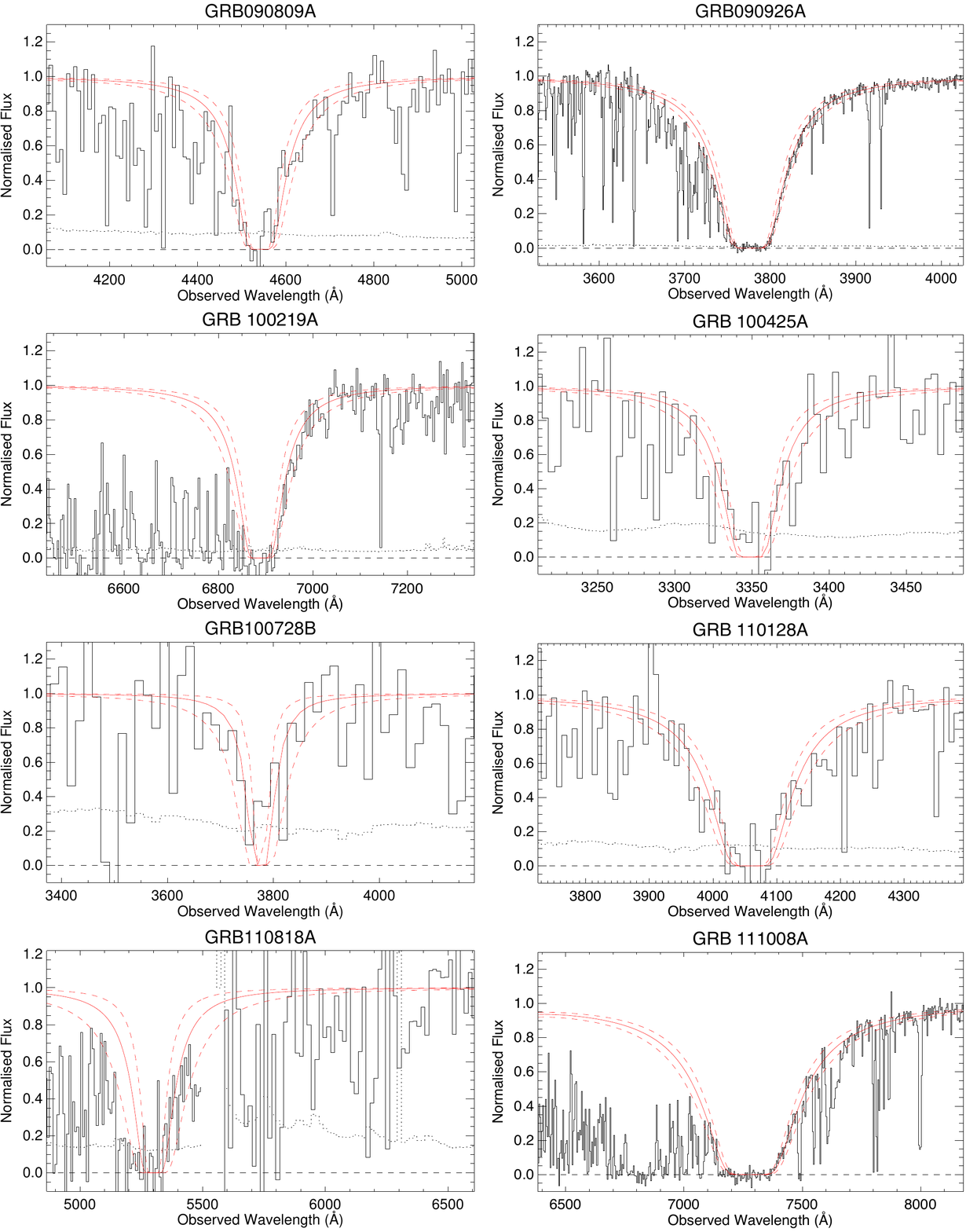}
	\caption{Measurements of the hydrogen column densities for all bursts with a
		clear Lyman alpha absorption system. In solid black is shown the spectrum with
		black dotted giving the corresponding 1-$\sigma$ error. The black dashed line shows zero
		flux density. The solid red line is the absorption of column density equal to
		the value presented in Tab. \ref{tab:HI} with the 1-$\sigma$ interval shown
		with dashed lines.}
	\label{fig:HI1}
\end{figure*}

Due to the stochastic nature of the \lya-forest, the blue wing of the
Lyman-$\alpha$ absorption line is randomly superposed with Lyman-$\alpha$ forest
systems, along with strong absorption from \mnii~and \SIiii, making it
notoriously difficult to model. Additionally, the red wing has ISM signatures
imprinted on it, especially strong absorption due to \SIii, \sii~and \nv, which
can exhibit significant velocity structure. Along with instrumental effects, the
generative model for the data that we would use in a likelihood-based analysis
would be very complicated. We have therefore decided not to make formal $\chi^2$
fitting of the hydrogen column densities, but instead use a more subjective
visual measurement to the absorption profile. Using an analytic approximation to
the absorption profile from \citet{TepperGarcia2006}, we overplot a synthetic
absorption line with a specified column density on our observed, normalised
spectrum. By tuning the value of the hydrogen column density until the synthetic
absorption line matches the spectrum, we can thereby infer the actual column
density of the GRB sight line in a manual way. Similarly, the uncertainty on the
hydrogen column can be estimated by adjusting the error, until the confidence
bounds contain the continuum variation. We show the results of this procedure
for all bursts where possible in Fig. \ref{fig:HI1} and the inferred hydrogen
column densities in Table~\ref{tab:HI}.

\begin{figure}[!t]
	\centering \includegraphics[width=\columnwidth]{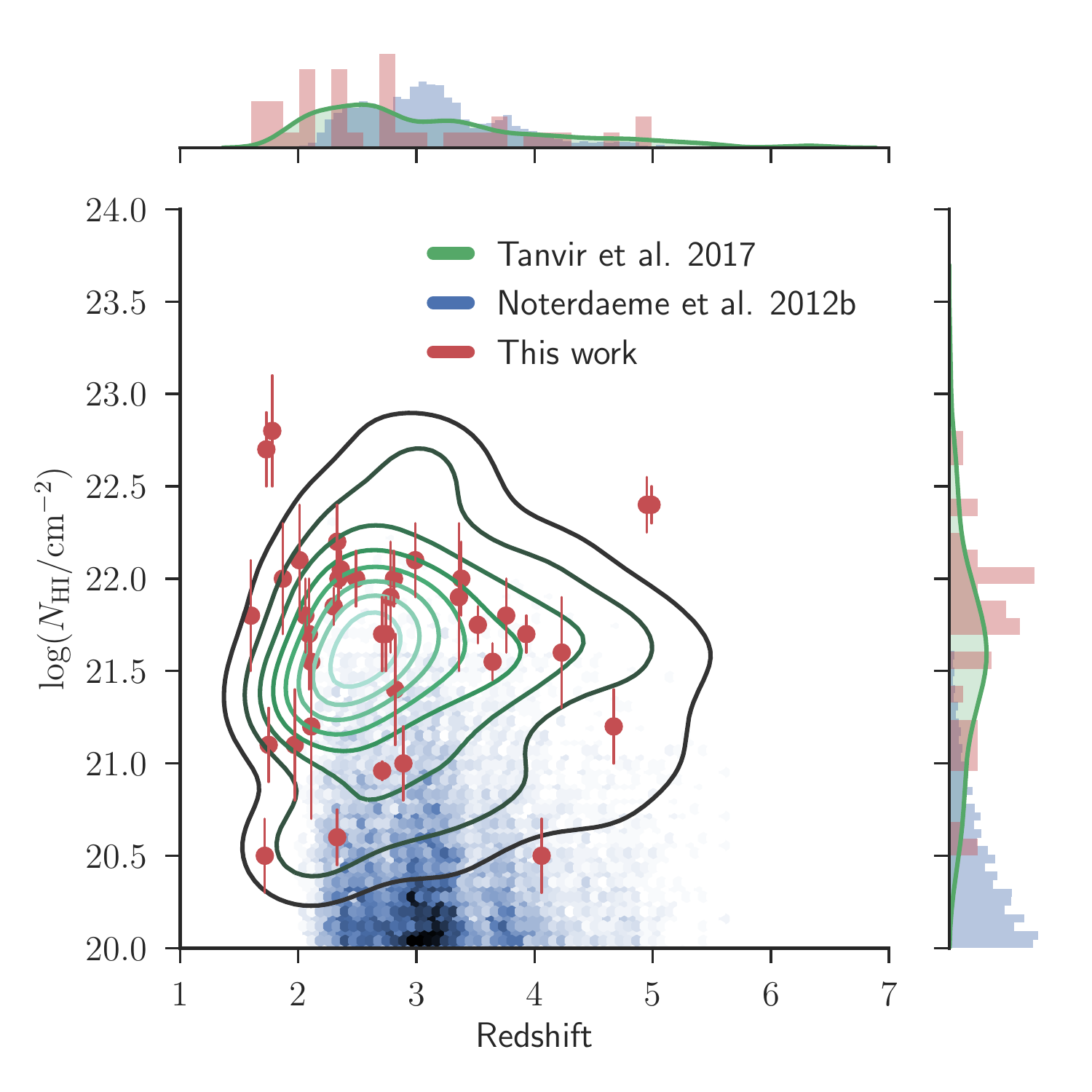}
\caption{Distributions of hydrogen column densities for absorbers found in
	quasar absorption lines, from \citet{Noterdaeme2012b} in blue. Overplotted in
	green is the kernel density estimate of absorbers in GRB sightlines. Values are
	taken from the compilation in Tanvir et al. (submitted), along with the new
	values presented in this sample. We also show in the red the values derived in
	this work. The marginal distributions for the three samples are also shown
	along the left side and on the top, where the different environments probed are
	clearly visible in the hydrogen column densities, as previously also noted in
	\citet{Fynbo2009}.} \label{fig:NH_dist}
\end{figure}

12 of the $N_{\mathrm{HI}}$ measurements for these spectra have previously been
presented in \citet{Cucchiara2015} (See Table \ref{tab:HI}). We provide new
measurements here for completeness. In the compilation of $N_{\mathrm{HI}}$
measurements towards GRBs in Tanvir et al. (submitted) there are 93 published
$N_{\mathrm{HI}}$ values, excluding the measurements provided here. We here
provide 30 new neutral hydrogen column density measurements -- an increase of
the number of optically derived hydrogen column densities of $\sim$ 33 per cent.
We show the two distributions in  Fig. \ref{fig:NH_dist}. We compare the median,
the 16th, and 84th percentiles of the two distributions. The sample presented
has a \nh~$= 21.8_{-0.8}^{+0.3}$ and the rest of the literature values has
\nh~$= 21.5_{-1.5}^{+0.4}$. We see that the two distributions have a large
degree of overlap due to the large width of the distributions, but we find a
slightly higher median value for the new sample presented here. A 2-sided KS
test gives a $p$-value of p = 0.006, meaning relatively strong evidence against the
null hypothesis that the two samples are drawn from the same underlying distribution.
Because the bursts that have measurements of the hydrogen column density are
selected solely based on our ability to infer a column, it is difficult to make
any strong conclusions about the population statistics in terms of gas content.

In Fig. \ref{fig:NH_dist}, we also show the column density distribution
for the 12081 quasar absorbers with \nh~$> 20$ from \citet{Noterdaeme2012b}. The
fact that GRBs are systematically located behind the highest \nh, previously
noted \citep[e.g.,][]{Prochaska2007, Fynbo2009}, is very clear in this
figure. The reason for this is that quasar sample sight-lines through galaxies
that are cross-section selected, whereas GRB sight-lines probe the dense,
star-forming regions in their hosts.

\section{Discussion and conclusions}\label{conclusions}

In this paper we have presented the results of a dedicated effort over the years
2009 -- 2017 to use the X-shooter spectrograph on the ESO-VLT to secure
spectroscopic observations of afterglows and host galaxies of GRBs detected by
\swift. This work was initiated by the consortium that built X-shooter and
included this project as a key part of the GTO program, but over the years the
project continued in open time.

The sample presented here includes spectroscopic observations of 93 systems
fulfilling our sample criteria, including 18 spectra that are late-time
observations of the underlying host galaxies. All spectra have been made
publicly available in the reduced form used in this paper.

Our sample serves the purpose to further characterize the environments of GRBs
that was also much advanced by the previous surveys based primarily on
lower-resolution spectroscopy. GRB afterglow sight-lines are unique in the
sense that only after observing more than 12000 damped Lyman-$\alpha$ absorbers
(DLAs) towards about 10$^5$ quasars, a handful systems with \nh~$ > 22$ have been
identified \citep[e.g., five in][]{Noterdaeme2012b}. Long GRB afterglow spectra,
by contrast, reveal such systems routinely \citep[][and this
work]{Jakobsson2006b, Fynbo2009, Cucchiara2015}. With afterglow spectroscopy
(throughout the electromagnetic spectrum from X-rays to the sub-mm) we are able
to characterize the properties of star-forming galaxies over cosmic history in
terms of redshifts, metallicities, molecular contents, ISM temperatures, UV-flux
densities, extinction curves, etc.  A number of independent papers have been
published or submitted for publication focusing on many of these specific issues
of our sample such as extinction curves (\citealt{Japelj2015}, Zafar et al.
submitted, see also \citealt{Fynbo2014, Heintz2017a}), emission lines from the
underlying host galaxies \citep{Kruhler2015}, the frequency of intervening
\mgii~ absorbers \citep{Christensen2017}, \citet{Arabsalmani2018} on the
metallicity-scaling relations, and escape of ionizing radiation (Tanvir et al.,
submitted). A number of additional companion papers are also planned,
investigating the detailed properties of the sample presented here, including
equivalent width distributions (de Ugarte Postigo et al., in preparation),
metallicities and kinematics (Th{\"o}ne et al., in preparation), high ionization
lines (Heintz et al., in submitted), molecular lines (Bolmer et al., in
preparation), fine-structure lines (Vreeswijk et al., in preparation), and
composite GRB afterglow spectrum (Selsing et al., in preparation).

The potential of using GRB sightlines as probes is far from fully harvested. The
sample of sightlines probed by our spectra are not representative for all GRB
sightlines as we have shown and consistent with earlier findings from samples
based on low-resolution spectroscopy \citep[e.g.,][]{Fynbo2009} and from studies
of complete samples of GRB host galaxies \citep{Hjorth2012, Covino2013,
	Perley2016a}. \cite{Kruhler2013} argue, that very rich sightlines like that
probed by the remarkable GRB~080607 \citep{Prochaska2009, Sheffer2009, Perley2011} are
probably significantly more frequent than in the sightlines sampled by our
spectra. However, with current instrumentation, these sightlines are out of
reach except under very fortunate circumstances as in the case of GRB~080607
when the afterglow could be observed only a few minutes after the burst with a
10-m class telescope. Observations of such sightlines with X-shooter-like
spectrographs on the next generation of 20--40-m telescopes is likely to be very
rewarding, given that a suitable GRB detector will be available.

\input{tables/HI_columns.tex}

\begin{acknowledgements}
Much of the work done here would not have been done without Neil Gehrels, who has now passed away. We owe him a large debt for his invaluable work with the \swift~satellite. We also want to remember Javier Gorosabel, who was taken from us too early. 
JPUF, BMJ and DX acknowledge support from the ERC-StG grant EGGS-278202.
The Dark Cosmology Centre was funded by the Danish National Research
Foundation. 
This work was supported by a VILLUM FONDEN Investigator grant to JH
(project number 16599). 
TK acknowledges support by the European Commission
under the Marie Curie Intra-European Fellowship Programme in FP7.  
LK and JJ acknowledges
support from NOVA and NWO-FAPESP grant for advanced instrumentation in
astronomy. 
KEH and PJ acknowledge support by a Project Grant (162948--051) from
The Icelandic Research Fund. 
AG acknowledges the financial support from the
Slovenian Research Agency (research core funding No. P1-0031 and project grant
No. J1-8136). 
CT acknowledges support from a Spanish National Research Grant of Excellence
under project AYA 2014-58381-P and funding associated to a Ramón y Cajál
fellowship under grant number RyC-2012-09984.
AdUP acknowledges support from a Ramón y Cajal fellowship, a BBVA Foundation
Grant for Researchers and Cultural Creators, and the Spanish Ministry of Economy
and Competitiveness through project AYA2014-58381-P.
ZC acknowledges support from the Spanish research project AYA 2014-58381-P and
support from Juan de la Cierva Incorporaci\'on fellowships IJCI-2014-21669.
DAK acknowledges support from the Spanish research project AYA 2014-58381-P and
support from Juan de la Cierva Incorporaci\'on fellowships IJCI-2015- 26153.
RSR acknowledges AdUP's BBVA Foundation Grant for Researchers and Cultural
Creators and support from ASI (Italian Space Agency) through the Contract n. 2015-046-R.0 and from European Union Horizon 2020 Programme under the AHEAD project (grant agreement n. 654215).
GL is supported by a research grant (19054) from VILLUM FONDEN.
SDV acknowledges the support of the French National Research Agency (ANR) under contract ANR-16-CE31-0003 BEaPro
DM acknowledges support from the Instrument Center for Danish Astrophysics (IDA).
This work made use of data supplied by the UK {\it Swift} Science Data Centre at
the University of Leicester. Finally, it is with pleasure that we acknowledge
expert support from the ESO staff at the Paranal and La Silla observatories in
obtaining these target of opportunity data.
This research made use of Astropy, a community-developed core Python package for Astronomy \citep{TheAstropyCollaboration2013}. The analysis and plotting was achieved using the Python-based packages Matplotlib \citep{Hunter2007}, Numpy, and Scipy \citep{scipy, VanderWalt2011}, along with other community-developed packages.

\end{acknowledgements}

\bibliographystyle{aa_arxiv}
\bibliography{XSGRB_sample,GCN}

\clearpage

\clearpage
\begin{figure*}[!h]
	\centering
	\includegraphics[page=2, width=\linewidth]{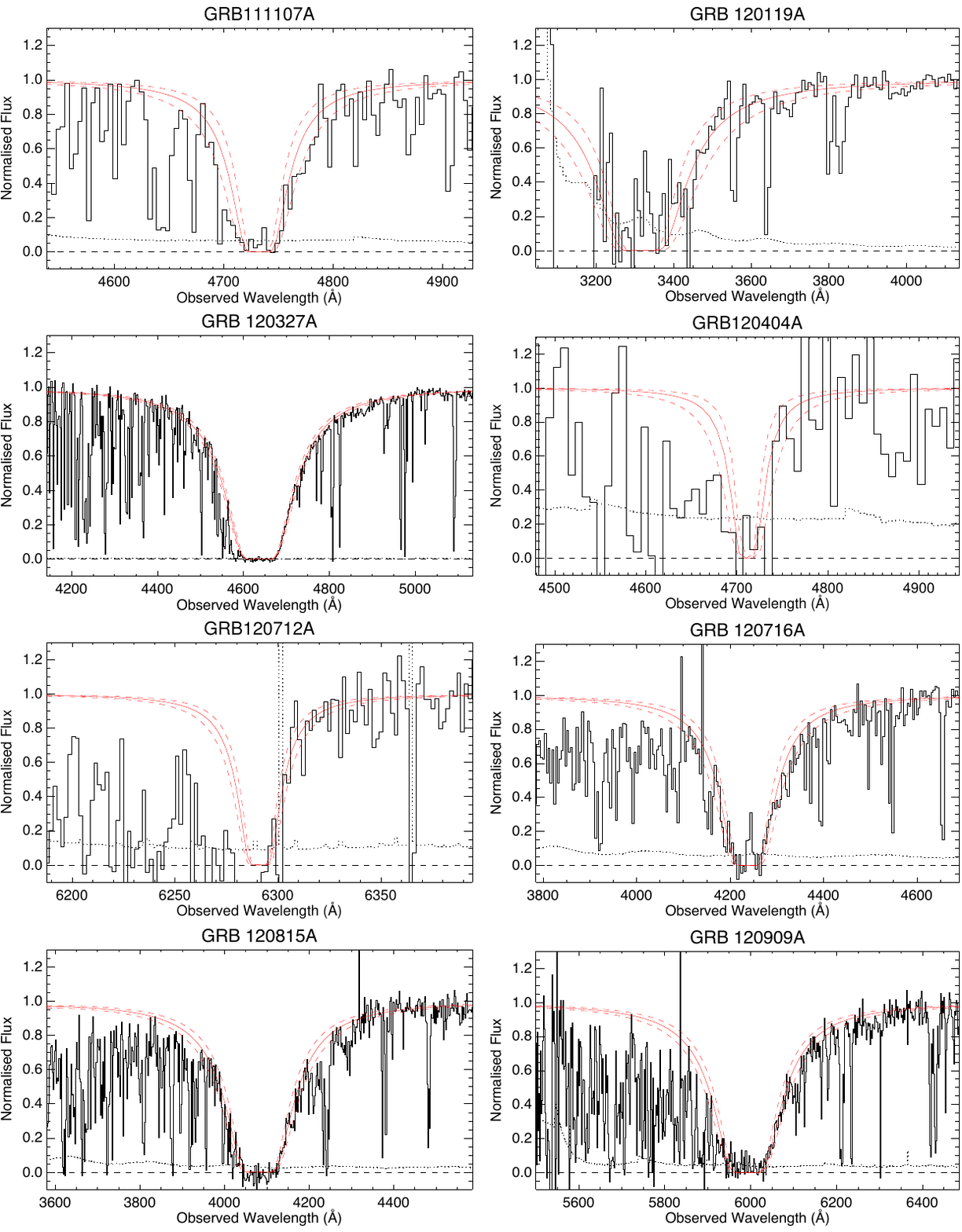}
	\caption*{Fig. \ref{fig:HI1}. continued.}
	\label{fig:HI2}
\end{figure*}
\clearpage
\begin{figure*}[!h]
	\centering
	\includegraphics[page=3, width=\linewidth]{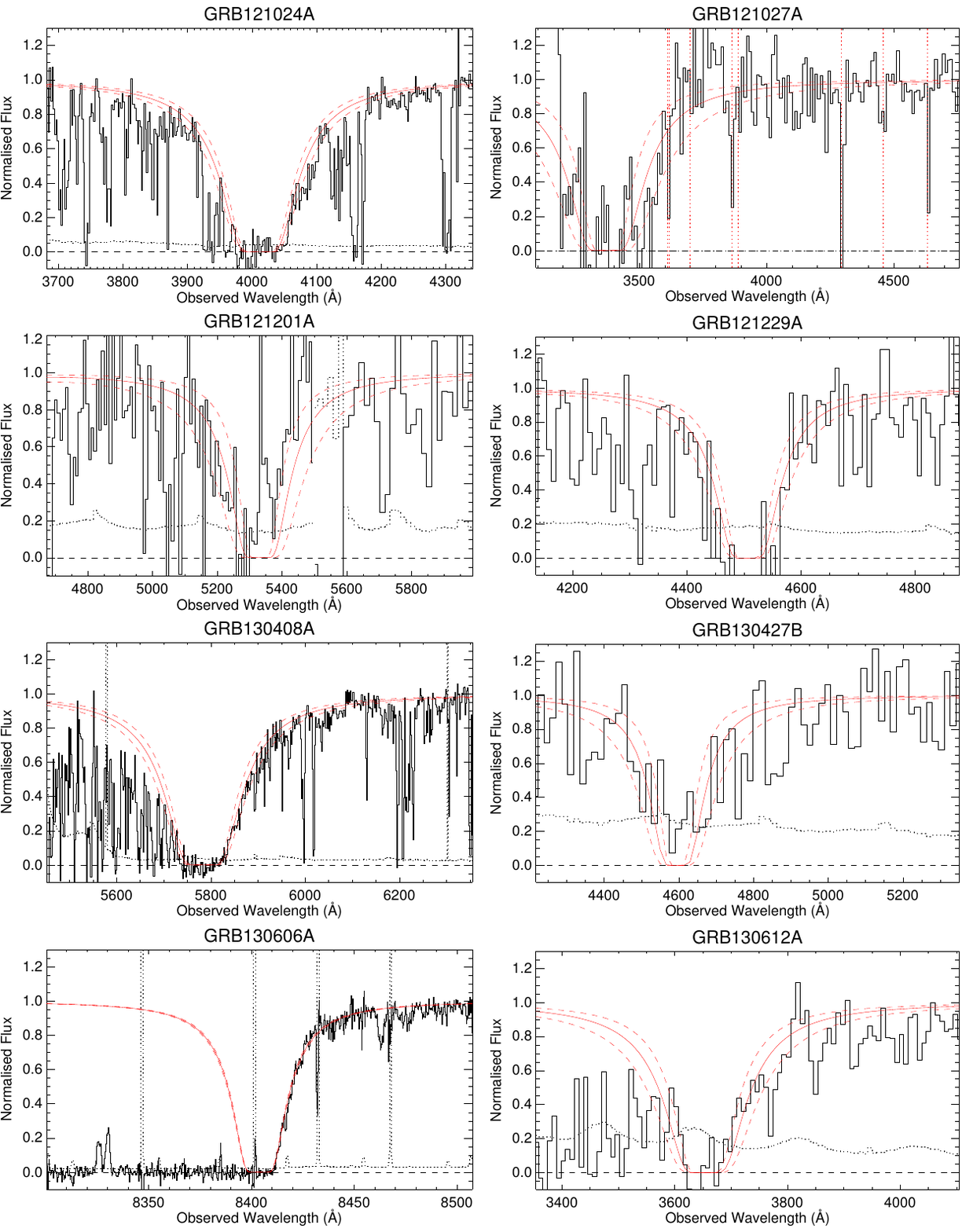}
	\caption*{Fig. \ref{fig:HI1}. continued.}
	\label{fig:HI3}
\end{figure*}
\clearpage
\begin{figure*}[!h]
	\centering
	\includegraphics[page=4, width=\linewidth]{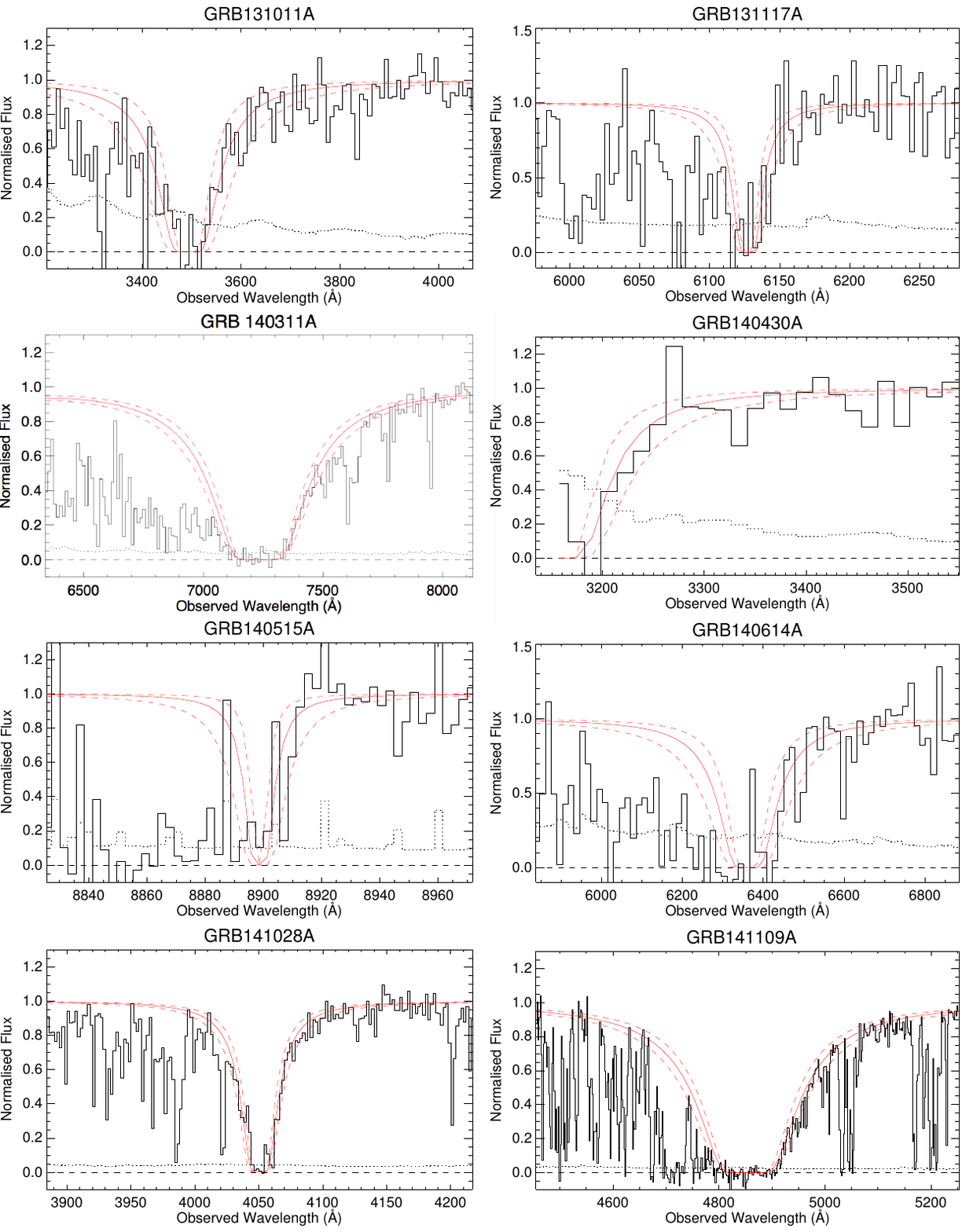}
	\caption*{Fig. \ref{fig:HI1}. continued.}
	\label{fig:HI4}
\end{figure*}
\clearpage
\begin{figure*}[!h]
	\centering
	\includegraphics[page=5, width=\linewidth]{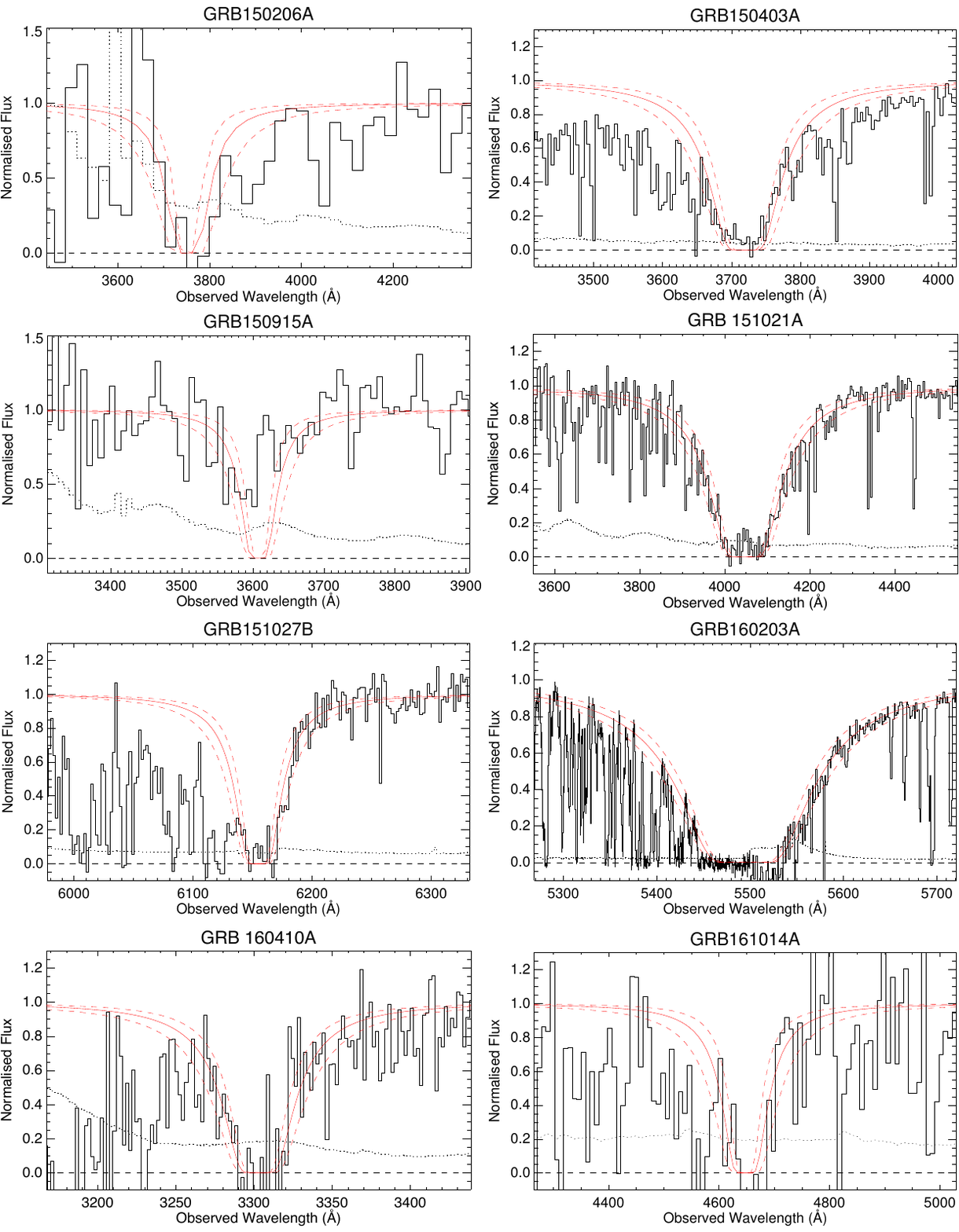}
	\caption*{Fig. \ref{fig:HI1}. continued.}
	\label{fig:HI5}
\end{figure*}
\clearpage
\begin{figure*}[!h]
	\centering
	\includegraphics[page=6, width=\linewidth]{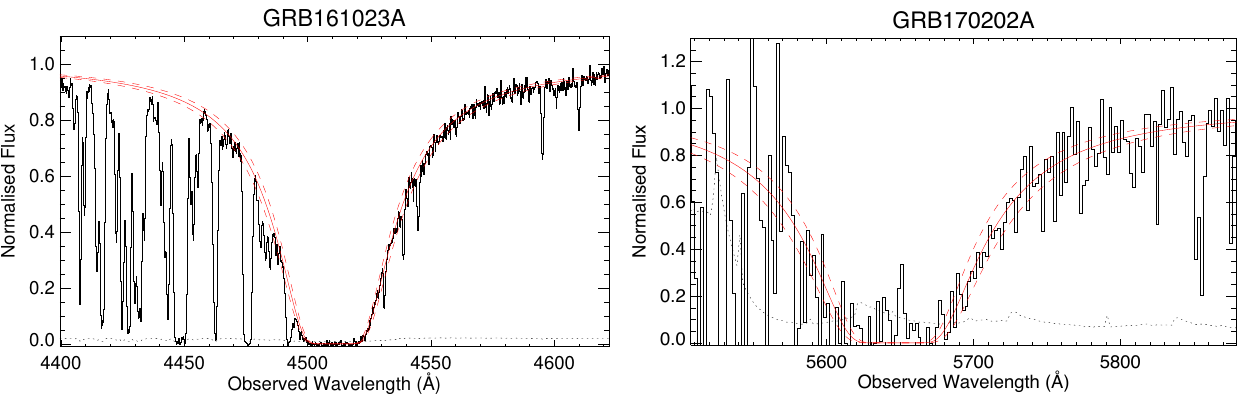}
	\caption*{Fig. \ref{fig:HI1}. continued.}
	\label{fig:HI6}
\end{figure*}
\clearpage

\appendix


\section{The complex error function and the Voigt profile} \label{voigt}
When modeling the spectral PSF, we need to evaluate the Voigt-profile. The Voigt profile, which is the convolution of the Gaussian and Lorentzian profiles, can, centered at zero, be written as \citep{pagnini2010} 
\begin{equation} 
\begin{split}
V(\lambda,\sigma, \gamma)  
& = G(\lambda, \sigma)  \otimes L(\lambda, \gamma) \\
& = \int_{-\infty}^{\infty} G(\xi, \sigma) L(\lambda - \xi, \gamma) d\xi \\
& = \int_{-\infty}^{\infty} \frac{1}{\sqrt{2 \pi} \sigma} e^{- \left( \frac{\xi}{\sqrt{2} \sigma}  \right)^2 } \frac{1}{\gamma \pi} \frac{\gamma^2}{(\lambda - \xi)^2 + \gamma^2} d\xi \\
& = \frac{\gamma}{\sqrt{2} \sigma} \frac{1}{ \pi^{3/2}}   \int_{-\infty}^{\infty} \frac{e^{- \left( \frac{\xi}{\sqrt{2} \sigma}  \right)^2 }}{(\lambda - \xi)^2 + \gamma^2} d\xi.
\end{split}
\end{equation}
We can by making the following substitution, $\xi = \sqrt{2} \sigma$ and $d\xi = \sqrt{2} \sigma dt$, write it as
\begin{equation} 
\begin{split}
V(\lambda,\sigma, \gamma)  
& =  \frac{\sqrt{2} \sigma}{ \sqrt{{\pi}}} \frac{\frac{\gamma}{\sqrt{2} \sigma}}{\pi}  \int_{-\infty}^{\infty} \frac{e^{- t^2 }}{(\lambda - \sqrt{2} \sigma t)^2 + \gamma^2} dt \\
& = \frac{1}{\sqrt{2 \pi} \sigma}  \frac{\frac{\gamma}{\sqrt{2} \sigma}}{\pi}  \int_{-\infty}^{\infty} \frac{e^{- t^2 }}{\left(\frac{\lambda}{\sqrt{2} \sigma} -  t\right)^2 + \left(\frac{\gamma}{\sqrt{2} \sigma}\right)^2} dt.	
\end{split}
\end{equation}
This form of the convolution is closely related to the complex probability function \citep{letchworth2007, abrarov2015a},
\begin{equation} 
\begin{split}
W(z)  
& = \frac{i}{\pi} \int_{-\infty}^{\infty} \frac{e^{-t^2}}{z - t}  dt
\end{split}
\end{equation}
for any complex argument, $z = x + iy$. The complex probability function can be expressed as a sum of a real and an imaginary part \citep{benner1995, abrarov2015b},
\begin{equation} 
\begin{split}
W(x, y)  
& = K(x, y) + i L(x, y) \\
& = \frac{y}{\pi}  \int_{-\infty}^{\infty} \frac{e^{- t^2 }}{(x -  t)^2 +y^2} dt  + \frac{i}{\pi}  \int_{-\infty}^{\infty} \frac{(x - t)e^{- t^2 }}{(x -  t)^2 +y^2} dt,
\end{split}
\end{equation}
where the real part, $\mathtt{Re}[W(x, y)] =  \sqrt{2 \pi} \sigma
V(\lambda,\sigma, \gamma)$ if $x = \frac{\lambda}{\sqrt{2} \sigma}$ and $y =
\frac{\gamma}{\sqrt{2} \sigma}$, which can be obtained by using the complex
argument, $z = \frac{\lambda + i\gamma}{\sqrt{2} \sigma}$, in the complex
probability function. If $\mathtt{Im}[z] \geq 0$, which is always guaranteed for
the width of a spectral profile, the complex probability function equals the
complex error function. The complex error function has numerous, fast, numerical
approximations where in this work we use the \texttt{scipy.special.wofz}
\citep{scipy} implementation.


\section{Notes on Individual objects} \label{notes}

\subsection{GRB~090313 (z = 3.373)} \label{090313}

The first GRB ever observed with X-shooter, during the commissioning of the
instrument, these data formed the basis of GCN 9015 \citep{GCN.9015} and are
published in \citet{DeUgartePostigo2010}. Due to the lingering brightness of
GRB~090313, 6.9 ks spectroscopic integration starting 45 hours after the BAT
trigger reveals a wealth of absorption features superposed on the afterglow
continuum at a common redshift of $z = 3.373$. Two intervening systems at $z =
1.959$ and $z = 1.800$ are identified based on strong \mgii-absorption. Because
this burst is observed before the instrument is science-verified, it does not
enter into the statistical sample.

\subsection{GRB~090530 (z = 1.266)}\label{090530}

Observed during paranalization of the instrument, these data forms the basis of
GCN 15571 \citep{GCN15571}, but are not published elsewhere. Observations began
20.6 hours after the BAT trigger and 4.8 ks spectroscopic integration in all
three arms reveals the absorption signature for a host at $z = 1.266$ from the
detection of \mgii, \mgi, \SIii, \feii, \aliii. Because this burst was observed
before the instrument was science-verified, it does not enter into the
statistical sample.

\subsection{GRB~090809 (z = 2.737)} \label{090809}

Observed during the first science verification period and was reported in GCN
9761 \citep{GCN.9761} and is additionally used as the basis for the master
thesis by \'Asa Sk\'ulad\'ottir (2010). 7.2 ks integration starting 10.2 hours
after the GRB trigger yields a clear afterglow continuum in all arms. The
simultaneous detection of absorption lines identified as \lya, \SIii, \oi,
\SIi$^*$, \SIiv, \civ, \feii, \alii, \aliii~and \mgii~at $z = 2.737$ sets it as
the redshift of the GRB. Because this burst is observed before the instrument
was science-verified, it does not enter into the statistical sample.

\subsection{GRB~090926A (z = 2.106)} \label{090926}

Obtained during the second science
verification period, this dataset forms the basis of GCN 9942 \citep{GCN.9942} and is
additionally published in \citet{DElia2010}. Spectroscopic integration started
22 hours after the BAT trigger and from the acquisition camera the optical
afterglow is measured to R = 17.9 mag at the beginning of the observations which
causes a strong continuum to be seen in all arms. An absorption trough due to
\lya~is clearly visible along with numerous metal resonance lines \civ, \SIii,
\SIii$^*$ \feii, \mgii, all at $z = 2.106$. Because this burst was observed
before the instrument was science-verified, it does not enter into the
statistical sample.

\subsection{GRB~091018 (z = 0.971)}\label{091018}

The first burst observed during normal operation after science verification was
completed and the first burst that enters the statistical sample. These data are
the basis for GCN 10042 \citep{GCN10042} and is published in
\citet{Wiersema2012}. With a bright afterglow and a rapid follow-up, this
spectrum is of pristine quality. The afterglow continuum is bright throughout
all spectroscopic arms which allows the ready detection of \alii, \aliii, \feii,
\mnii, \mgii, \mgi, and \caii~- all located at $z = 0.971$, setting it as the
redshift of the GRB.

\subsection{GRB~091127 (z = 0.490)} \label{091127}

Obtained 4 days after the burst trigger, these data forms the basis for GCN 10233
\citep{GCN10233} and are published in \citet{Vergani2011}. Due to the late
follow-up and a nearby moon, the signal-to-noise of the afterglow continuum is
low, especially in the UVB arm. This is why no clear absorption lines are detected
against the afterglow continuum, although see \citet{Vergani2011} which report a
tentative detection of \mgii. Emission lines from the underlying host are clearly
visible with lines from \oii, \hb, \oiii, and \ha~all at $z = 0.490$. This
bursts is additionally associated with SN2009nz \citep{Cobb2010, Berger2011, Olivares2015}.

\subsection{GRB~100205A  (z = na)} \label{100205}

Observed 3 days after the \textit{Swift} trigger. No afterglow or host detected
is in 10.8 ks. The GRB is likely located at high redshift \citep[GCN
GCN10399;][]{GCN10399}. The spectrum has not otherwise been published
previously.

\subsection{GRB~100219A (z = 4.667)} \label{100219}

These data presented here have also formed the basis of GCN 10441 \citep{GCN10441} and
are published in \citet{Thone2013}. Observations started 12.5 hours after the
\textit{Swift} trigger and has a total exposure time of 4.8 ks. Absorption
features, including those of \lya~and from a multitude of ions are detected
against the afterglow continuum at $z = 4.667$. Additionally, absorption from an
intervening system is found at $z = 2.181$.

\subsection{GRB~100316B (z = 1.180)} \label{100316B}

The data presented here also formed the basis of GCN 10495 \citep{GCN10495}. The
spectrum has not otherwise been published. Observations started 44 minutes after
the \textit{Swift} trigger and have a total exposure time of 2.4 ks. Absorption
features from \feii, \alii, \aliii,	\mgii~and \mgi~are well detected against the
afterglow continuum at $z = 1.180$. Additionally, strong absorption lines from
\feii~and \mgii~from an intervening system are found at $z = 1.063$.

\subsection{GRB~100316D (z = 0.059)} \label{100316D}

The data presented here also formed the basis of GCN 10512 \citep{GCN10512}, GCN
10513 \citep{GCN10513}, GCN 10543 \citep{GCN10543} and are published in
\citet{Bufano2012} and \citet{Starling2011}. This GRB is very close by and has
an associated SN, SN2010bh, and has therefore undergone intense follow-up
\citep{Olivares2011, Cano2011a, Izzo2017}. The spectra presented here consists
of a subset of the entire VLT/X-shooter campaign, covering the four first
observing days while the afterglow still contributes significantly to the total
emission. The first observations started 10 hours after the burst, before the SN
was discovered, and targeted the star-forming 'A'-region \citep{Starling2011},
not the GRB. A very rich spectrum containing a multitude of emission lines puts
the host at $z = 0.059$. For three consecutive nights, 58, 79 and 101 hours
after the \textit{Swift} trigger, the afterglow was observed as it transitioned
into the spectrum of a high-velocity Ic-BL SN. The observations taken 79 and 101
hours after the burst are taken under programme 084.D-0265(A) (PI: Benetti), but
with an identical setup to the first two observations.

\subsection{GRB~100418A (z=0.624)} \label{100418}

The data presented here also formed the basis of GCN 10620 \citep{GCN10620} and
GCN 10631 \citep{GCN10631} and are published in \citet{DeUgartePostigo2011}. The
burst has been followed up in three epochs of observations, 0.4, 1.4, and 2.4
days after the burst, each lasting 4.8 ks. The unambiguous redshift of the host,
$z=0.624$, is found from the simultaneous detection of emission features
belonging to nebular lines, including \hi, \oii, \oiii, \neiii, \nii, [\sii],
[\siii], and \hei~as well as absorption features due to the presence of \znii,
\crii, \feii, \mnii, \mgii, \mgi, \tiii, and \caii, all at a consistent
redshift. Temporal evolution of the fine structure lines belonging to \feii$^*$
is found between the epochs.

\subsection{GRB~100424A (z=2.465)} \label{100424}

The data presented here also formed the basis of GCN 
14291 \citep{GCN14291}. The spectrum has
not otherwise been published. Observations were carried out long after the burst
afterglow had faded. Emission lines from the host are detected at $z=2.465$.

\subsection{GRB~100425A (z=1.1755)} \label{100425}

The spectra presented here also formed the basis of GCN 10684 \citep{GCN10684}
and are used in Skuladottir (2010), but not published elsewhere. Observations
started 4 hours after the \textit{Swift} trigger, totaling 2.4 ks. Absorption
features from \mgii~and \feii~in the afterglow continuum are detected at
$z=1.1755$.

\subsection{GRB~100615A (z=1.398)} \label{100615}

The data presented here also formed the basis of GCN 
14264 \citep{GCN14264}, but are not
published elsewhere. Host observation of a dark burst \citep{DElia2011} taken
long after the afterglow had faded. Emission lines from the host belonging to
\oii, \neiii, \oiii~and \ha~are detected at a common redshift of $z=1.398$.

\subsection{GRB~100621A (z=0.542)} \label{100621}

The data presented here also formed the basis of GCN 
10876 \citep{GCN10876}, but are not
published elsewhere. Beginning 7.1 hours after the GRB, 2.4 ks observations
reveal emission lines from \oii, \hb~and \oiii~at a common redshift of $z=0.542$
and a very weak afterglow continuum.

\subsection{GRB~100625A (z=0.452)} \label{100625}

The data presented here is of the candidate host galaxy, taken long after the
burst had faded and have not previously been published. 4.8 ks of exposure
reveals a weak continuum present in all arms, but an absence of emission lines.
This could indicate that the host primarily contains an older stellar
population. The redshift, $z=0.452$, is taken from \citet{Fong2013}.

\subsection{GRB~100724A (z = 1.288} \label{100724}

The data presented here also formed the basis of GCN 10971 \citep{GCN10971}. The
spectrum has not otherwise been published previously. The observations were
carried out in RRM starting 11 min after the GRB trigger. See section \ref{RRM},
for a description of the RRM scheme. Absorption lines from several ionic species
are detected in the afterglow continuum at a common redshift of $z = 1.288$.

\subsection{GRB~100728B (z=2.106)} \label{100728}

The data presented here also formed the basis of GCN 11317 \citep{GCN11317}. The
spectrum has not otherwise been published previously. Starting 22 hours after
the burst trigger, 7.2 ks of observations reveals a faint afterglow continuum
with \lya- and \mgii-absorption at $z=2.106$. Due to a malfunctioning ADC, the
sensitivity of X-shooter is depressed with respect to normal operations,
resulting in a lower throughout. Additionally, the position of the trace on the
slit moves due to atmospheric differential refraction. The presence of the DLA
is confirmed in the 2D image and despite the observational challenges that
affects this spectrum, we measure \nh~$=21.2 \pm 0.15$.

\subsection{GRB~100814A (z=1.439)} \label{100814}

The spectra presented here has not been published previously. The observations
consist of three visits, the first beginning only 0.9 hours after the
\textit{Swift} trigger, the other two visits were 2.13 and 98.40 hours after the
trigger, respectively. A bright afterglow continuum is present in all visits,
allowing identification of absorption features belonging to a wide range of ions
at $z=1.439$. A complex velocity structure in the absorption features belonging
to \mgii, shows several components, separated by as much as 500 km/s, pointing to
a likely merger scenario in the host or starburst driven outflows.

\subsection{GRB~100816A (z=0.805)} \label{100816}

The data presented here also formed the basis of GCN 11123 \citep{GCN11123}. The
spectrum has not otherwise been published previously. This short GRB was
observed 28.4 hours after the GRB trigger. 4 x 1200 s of exposure reveals two
distinct sets of emission lines, spatially offset $\lesssim 1 \arcsec $, very
close in redshift space, $z=0.8034$ and $z=0.8049$, indicating either an
interacting host or some complex velocity structure of the host. Faint
underlying continua are present under both sets of emission lines.

\subsection{GRB~100901A (z=1.408)} \label{100901}

The data presented here have been published in \citet{Hartoog2013}. Because of
the unusual lingering brightness of this GRB, 2.4 ks of observations taken 65.98
hours after the GRB trigger still reveals an afterglow continuum visible across
the entire spectral coverage of X-shooter. Absorption lines from a wide range of
ions sets the redshift at $z=1.408$, with intervening absorption systems at $z =
1.3147$ and $z = 1.3179$.

\subsection{GRB~101219A (z=0.718)} \label{101219A}

These data have not been published before. Starting 3.7 hours after the GRB
trigger, 7.2 ks of exposure time reveal a very faint continuum in the visual and
near-infrared, only visible when heavily binning the images. No redshift
estimate is available from these observations.  Late-time Gemini-North
observations reveal emission lines from the host at $z=0.718$ \citep{GCN11518}.

\subsection{GRB~101219B (z=0.552)} \label{101219B}

The data presented here also formed the basis of GCN 11579 \citep{GCN11579} and
are published in \citet{Sparre2011}. The first observation, taken 11.6 hours
after the burst trigger and lasting 4.8 ks, reveals absorption from \mgii~and
\mgi~in the host located at $z = 0.552$ on a featureless continuum visible
across the entire coverage of X-shooter.  Subsequent observations taken 16 and
37 days after the trigger shows the fading spectral signature of a SN, SN2010ma.

\subsection{GRB~110128A (z=2.339)} \label{110128}

These observations form the basis of GCN 11607 \citep{GCN11607}, but have not
been published before. Spectroscopic integration started 6.55 hours after the
\swift~trigger and lasted for a total of 7.2 ks. The afterglow continuum is
detected across the entire spectral coverage at moderate signal-to-noise.
Absorption lines in the continuum are detected from \lya, \oi, \cii, \SIiv, \civ,
\SIii~and \feii, all at a common redshift of $z=2.339$. From the broad
\lya~trough, a hydrogen column density $\log (N_{\mathrm{HI}}/\mathrm{cm}^{-2})
= 22.6 \pm 0.2$ is derived. An intervening system at $z=2.20$ is tentatively
identified from an absorption feature, likely due to \civ.

\subsection{GRB~110407A (z=na)} \label{110407}

These observations have not been published before. Starting 12.36 hours after
the BAT trigger, 4.8 ks spectroscopic integration yields a very faint trace down
to $\sim$430 nm, only visible after binning heavily. This could indicate a
redshift, $z ~\sim 2.5$, but no emission lines or absorption lines are
immediately visible to support this.

\subsection{GRB~110709B (z=2.109)} \label{110709}

This is a late-time observation (> 1 year) and has previously been used in
\citet{Perley2016a}. In this reduction of the 7.2 ks spectroscopic integration,
the tentative detection of \oiii~reported in \citet{Perley2016a} is confirmed
along with low-significance detection of \ha~at the end of the spectral
coverage, both at a consistent redshift, $z=2.109$, securing it as the redshift
of the GRB host.

\subsection{GRB~110715A (z=0.823)} \label{110715}

These observations, starting 12.3 hours after the trigger, have been published
in \citet{Sanchez-Ramirez2017} and additionally formed the basis of GCN 12164
\citep{GCN12164}. Only a single exposure of 600 s was obtained, before strong
winds interrupted the observations. A red continuum is detected across all arms
and a multitude of absorption lines are superposed on the afterglow continuum.
We identify lines belonging to \alii, \aliii, \znii, \crii, \feii, \mgii, \mgi,
\caii, and \caii, all at  $z=0.823$, marking it as the redshift of the GRB.

\subsection{GRB~110721A (z=0.382)} \label{110721}

This is a Fermi burst with a LAT detection and thus outside the statistical
sample, but nonetheless followed up due to the extremely high peak energy
\citep{Axelsson2012}. Starting 28.7 hours after the burst trigger, 2.4 ks
spectroscopic observation reveals after heavy binning, a wide, faint trace down
to $\sim$ 580 nm, offset by 2.5\arcsec~relative to the centering of the slit.
No good redshift measurement can be inferred from this. We have adopted the
redshift from GCN 12193 \citep{GCN12193}.

\subsection{GRB~110808A (z=1.348)} \label{110808}

This spectrum has already formed the basis of GCN 12258 \citep{GCN12258}, but is
not published otherwise. Starting 3 hours after the \swift~trigger, a rich
spectrum is obtained in 2.4 ks spectroscopic integration. The GRB afterglow
continuum is visible across all three spectrscopic arms of VLT/X-shooter with
emission lines identified as \oii, \oiii, \ha~all at $z = 1.348$. At the same
redshift, we identify absorption lines superposed on the afterglow continuum
from \mgii~and \feii.

\subsection{GRB~110818A (z=3.36)} \label{110818}

Starting 6.15 hours after the BAT trigger, spectroscopic integration for 4.8 ks
reveals a moderate signal-to-noise GRB afterglow continuum, down to $\sim$ 500 nm. The simultaneous detection of absorption features identified as \lya,
\SIii, \civ, \alii, \cah, \cak, and \mgii, and emission from the \oiii-doublet,
securely sets $z = 3.36$ as the redshift of the GRB. These data form the basis
of GCN 12284 \citep{GCN12284}, but is not published elsewhere.

\subsection{GRB~111005A (z=0.013)} \label{111005}

The data presented here have previously been published in \citet{Michaowski2016}.
2.4 ks spectroscopic integration of the host galaxy, obtained long after the
burst had faded, contains bright emission lines filling the entire slit on top
of a broad, underlying stellar continuum. We identify emission lines from \oii,
\hd, \hg, \hb, \oiii, \nii, \hb, [\sii], \ariii, and [\siii], all at $z=0.013$.
Significant velocity structure of the lines across the spatial direction of the
slit indicates a large degree of coherent motion relative to the line-of-sight.

\subsection{GRB~111008A  (z=4.989)} \label{111008}

These data formed the basis of GCN 12431 \citep{GCN12431} and are additionally
published in \citet{Sparre2014}. Observations of this GRB afterglow were
initiated 7.71 hours after the BAT trigger and had a duration of 8.4 ks. A second
observational epoch started 20.1 hours after the GRB trigger and lasted for
6.6 ks. The GRB afterglow continuum is well detected down to $\sim$760 nm, with
several strong absorption features imprinted, all at a common $z = 4.990$.
\lya~is clearly detected and we additionally detect lines identified as \SIii,
\feii, \civ, \mgii, \SIii*, \sii*, \oi*. An intervening DLA system is
additionally detected at  $z = 4.61$ as seen from \lya~and \mgii~absorption.

\subsection{GRB~111107A (z=2.893)} \label{111107}

GCN 12542 \citep{GCN12542} is based on this spectrum, but it is not published
elsewhere. Spectroscopic integration started 5.26 hours after the \swift~trigger
and consists of $4 \times 1200$ s integration in the UVB and VIS and $16 \times
300$ s in NIR, the observations ending in twilight. The GRB afterglow continuum
is well detected across the arms with absorption lines from \lya, \civ, \feii,
and \mgii, all at a consistent redshift of $z = 2.893$. Additionally an
intervening \mgii~system is detected at $z = 1.998$. From the \lya~absorption
trough, we additionally infer $\log (N_{\mathrm{HI}}/\mathrm{cm}^{-2}) = 21.0
\pm 0.2$.

\subsection{GRB~111117A (z=2.211)} \label{111117}

These data have previously been used to form some of the basis of
\citet{Selsing2017}. Starting 37.3 hours after the BAT trigger, 4.8 ks of
spectroscopic integration yields faint emission lines identified at \oii, \hb,
\oiii~and \ha, all at a common $z = 2.211$, marking it as the redshift of the
GRB host. No afterglow continuum is detected. This is the highest redshift short
GRB detected to date.

\subsection{GRB~111123A  (z=3.151)} \label{111123}

These data formed the basis of GCN 14273 \citep{GCN14273}, but are not published
elsewhere. Observed twice, the first time shortly after the GRB and the second
long after the burst had faded, securely sets the redshift of the host at $z =
3.151$ based on the detection of emission lines identified as \oii~and \oiii.

\subsection{GRB~111129A (z=1.080)} \label{111129}

Starting 8.26 hours after the GRB trigger and lasting 3.6 ks, these observations have previously
been published in \citet{Kruhler2015}. A very faint continuum is visible after
severe binning and a redshift is suggested in \citet{Kruhler2015}, based on the
detection of \oii. At this redshift, \hb~and \oii~are located in the gap between
the VIS and NIR arm and \ha~is located in the middle of the $JH$-bandgap and is
therefore not detected.

\subsection{GRB~111209A (z=0.677)} \label{111209}

These spectra have previously been used in \citet{Levan2013, Greiner2015,
	Kruhler2015, Kann2017} and additionally formed the basis for GCN 12648
\citep{GCN12648}. The first epoch of spectroscopic observations was initiated
17.7 hours after the BAT trigger and lasted for 4.8 ks. A very bright
afterglow continuum is detected across the entire spectral coverage of
X-shooter, with several absorption features imprinted. The absorption features
are identified as \feii, \mgii, \mgi, \cah, and \cak~- all at a common redshift
of $z = 0.677$. The second epoch, taken 20 days later, still contains a faint
continuum detected across all arms. The detection of nebular emission lines
identified as \oii, \oii~and \ha~at the same redshift, securely marks it at the
redshift of this ultra-long GRB with accompanying GRB-SN.

\subsection{GRB~111211A (z=0.478)} \label{111211}

These data formed the basis for GCN 12677 \citep{GCN12677} and are also
published in \cite{Kruhler2015}. Observations began 31 hours after the AGILE
trigger and consisted of $4 \times 600$ s. A bright GRB afterglow continuum is
detected across the entire spectral coverage of X-shooter with absorption and
emission features visible. We identify absorption features due to \feii, \mgii,
and \caii~and emission lines from \oiii~and \ha, all at a common $z = 0.478$,
which we suggest is the redshift of the GRB. Additionally detected in the GRB
afterglow continuum are broad undulation, suggesting an accompanying SN.

\subsection{GRB~111228A (z=0.716)} \label{111228}

These data formed the basis of GCN 12770 \citep{GCN12770} and are also published
in \cite{Kruhler2015}. Observations began 15.9 hours after the BAT trigger and
consist of $4 \times 600$ s. The GRB afterglow continuum is clearly detected in
all the spectroscopic arms and superposed on the continuum are absorption
features identified as due to \feii, \mnii, \mgii, \mgi, \cah, and \cak, all at
$z = 0.716$. Supporting this redshift measurement as the redshift of the GRB is
the detection of nebular emission from \oiii.

\subsection{GRB~120118B (z = 2.943)} \label{120118}

The data presented here also formed the basis of GCN 14225 \citep{GCN14225}, but
are not published otherwise. This late-time observation of the host of
GRB~120118B consists of 3.6 ks exposures and contains emission lines belonging
to \oii~and \oiii~at $z = 2.943$, suggested to be redshift of the host.

\subsection{GRB~120119A (z = 1.728)} \label{120119}

The data presented here have been examined by \citet{Japelj2015} and Zafar et al.
(submitted), who both find a significant amount of extinction, $A_V\approx 1$
mag, also supported by previous measurements \citep{Morgan2014a}. Three epochs
of observations have been obtained, the first two immediately after the burst,
and the last one long after the afterglow had faded. Starting 1.4 hours after
the \textit{Swift} trigger, the first epoch contains bright afterglow continuum.
Rich in absorption features belonging to a multitude of ions at $z = 1.728$ is
estimated for the host with intervening systems at $z = 1.476$, $z = 1.214$, $z
= 0.662$ and $z = 0.632$. The second epoch, obtained 4.5 hours after the burst
contains the fading afterglow. A third epoch is obtained $>1$ year after the GRB
in which emission lines from \hb~and \ha~are found at the redshift of the host,
confirming the association of the absorption line system and the host. We also
detect C\,\textsc{i} in absorption which indicates the presence of cold gas.

\subsection{GRB~120211A (z = 2.346)} \label{120211}

The data presented here have been published in \citet{Kruhler2015}. Two
observations of the host of GRB~120211A has been obtained, starting 2013.02.17,
$> 1$ year after the burst has faded. A redshift for this object has been
reported by \citet{Kruhler2015} and the features seen by those authors are
reproduced in these reductions, confirming $z =	2.346$.

\subsection{GRB~120224A (z = 1.10)} \label{120224}

The data presented here have formed the basis of GCN 12991 \citep{GCN12991}, and
have also been published in \citet{Kruhler2015}. Starting 19.8 hours after the
GRB trigger, a total exposure time of 2.4 ks reveals a faint continuum, starting
at $\sim$ 700 nm~and extending all the way through 2500 nm. In the 2D-spectrum
we detect a $\sim 2 \sigma$ emission line which, if interpreted as \ha, gives $z
= 1.10$, supporting the photometric redshift ($0.9 < z_\mathrm{phot} < 1.3$)
derived by \citet{Kruhler2015}.

\subsection{GRB~120311 (z = 0.350)} \label{120311}

The data presented here have formed the basis of GCN 12991 \citep{GCN12991}, but
are not published otherwise. Starting just before twilight, 3.65 hours after the
burst, a faint afterglow continuum is detected at all wavelengths. Due to the
faintness of the afterglow, no absorption features are discernible superposed on
the continuum. Displaced from the afterglow continuum by 1\farc4, emission lines
belonging to \hb, \oiii~and \ha~are detected at $z = 0.350$. The line belonging
to \ha~shows some extended emission toward the afterglow continuum. The angular
distance between the two sources correspond to a projected distance in the host
plane of 6 kpc, posing a potential problem for the host redshift, unless the GRB
occurred in a merging system. The extended emission in \ha, supports this
interpretation. This burst is not a part of the statistical sample.

\subsection{GRB~120327A (z = 2.813)} \label{120327}

The data presented here also formed the basis of GCN 13134 \citep{GCN13134} and
are published in \citet{DElia2014}. The observation consists of two visits, 2.13
hrs and 29.98 hrs after the burst, with an afterglow continuum visible in all
arms for both visits. We detect absorption features from Ly-limit, \lya,
\cii/\cii$^*$, \SIii/\SIii$^*$, \ali, \feii ~and \mgii~at a consistent redshift,
$z = 2.813$.

\subsection{GRB~120404A (z = 2.876)} \label{120404}

The data presented here have formed the basis of GCN 13227 \citep{GCN13227}, but
are not published otherwise. 9.6 ks integration, starting 15.7 hours after the
\textit{Swift}-trigger reveals a low-intensity afterglow continuum on which
absorption from \lya~is detected in two distinct regions at redshifts $z=2.876$
and $z=2.55$. These absorption systems are confirmed by ionic absorption
features at both of these redshifts.

\subsection{GRB~120422A (z = 0.283)} \label{120422}

The data presented here also formed the basis of GCN 13257 \citep{GCN13257} and
are published in \citet{Schulze2014}. Being a GRB-SN, this burst has been followed up
multiple times. The data presented here only contain the first epoch in which
the afterglow is still visible and before the rise of SN2012bz. Starting 16.5
hours after the burst, 4.8 ks integration time captures both the host and the
burst in emission. A blue afterglow continuum is detected at all wavelengths
covered by X-shooter, on which \mgii~absorption at $z = 0.283$ is found. Offset
by 1\farc75, the host is clearly detected at a consistent redshift with a rich
emission line spectrum, the lines extending towards to burst.

\subsection{GRB~120712A (z = 4.175)} \label{120712}

The data presented here also formed the basis of GCN 13460 \citep{GCN13460}, but
are not published elsewhere. 4.8 ks integration time, starting 10.5 hours after
the BAT trigger, shows a bright afterglow continuum starting at $\sim$ 472 nm,
signifying the onset of the Lyman alpha forest, for a GRB located at $z =
4.175$. Absorption features from \lya, \feii, \mgii~and \SIii~are readily
detected at a consistent redshift.

\subsection{GRB~120714B (z = 0.398)} \label{120714}

The data presented here also formed the basis of GCN 13477 \citep{GCN13477} and
are discussed in Klose et al. (submitted), but are not published anywhere.
Observations of this burst started 7.8 hours after the GRB trigger, lasting 4.8
ks. A continuum is visible across the entire spectral coverage of X-shooter,
with both emission lines from  \oii, \hb, \oiii~and \ha, as well as absorption
from \mgii~detected at $z = 0.398$, securely setting it as the redshift of the
GRB.

\subsection{GRB~120716A (z = 2.486)} \label{120716}

The data presented here also formed the basis of GCN 13494 \citep{GCN13494}, but
are not published elsewhere. Despite observations starting 62 hours after the
\textit{Swift} trigger and lasting 3.6 ks, a bright afterglow is clearly seen,
along with a plethora of absorption features. Absorption of \lya-photons in the
host leaves a broad trough, beyond which the Lyman alpha forest is visible
bluewards, all the way down to the Lyman limit. Metal absorption lines from
\cii, \SIii, \oi, \feii, \civ, \SIiv, including fine structure transitions
identified as \cii$^*$, \SIii$^*$, \feii$^*$ and metastable \NIii~lines are all
detected at $z = 2.486$

\subsection{GRB~120722A (z = 0.959)} \label{120722}

The data presented here also formed the basis of GCN 13507 \citep{GCN13507}, but
are not published elsewhere. In 4.8 ks integration time, starting 10 hours after
the burst trigger, the simultaneous detection of absorption features belonging
to \mgii~and \feii~superposed on a blue continuum, and emission lines from \oii,
\hg, \hb, \oiii~and \ha, all at $z = 0.959$, confidently sets it as the redshift
of the GRB.

\subsection{GRB~120805A (z = 3.9)} \label{120805}

A separate reduction of this burst has been published in \citet{Kruhler2015},
but is not used otherwise. Starting 9 days after the burst trigger, this is a host
observation and does not contain any afterglow continuum. In 3.6 ks integration
time, we detect a faint continuum visible from 450 nm and all the way through
2100 nm, in contrast to what is found previously. The continuum from 450 - 600
nm is detected at very low significance. If the drop at 450 nm is the Lyman
limit, this fits with Lyman alpha at $\sim$ 600 nm, giving $z \sim 3.9$. The
absence of nebular lines is due to \oii~falling in a telluric absorption band
and the rest of the strong nebular lines being shifted out of the wavelength coverage.

\subsection{GRB~120815A (z = 2.358)} \label{120815}

Not a part of the statistical sample, this burst also formed the basis of GCN
13649 \citep{GCN13649} and is published in \citet{Kruhler2013}. Observations
started 1.69 hours after the BAT trigger and consist of 2.4 ks integration. A
bright afterglow continuum is detected across the entire spectral coverage of
X-shooter, with a multitude of absorption lines superposed. Absorption features
from the host at $z = 2.358$ include a DLA as well as metal absorption lines
from \nv, \sii, \SIii, \oi, \civ, \SIiv, \feii, \alii, \aliii, \mnii, \mgii, and
\mgi. Also fine structure lines from \NIii~and \feii~are excited in the local
environment of the GRB. Additionally, this spectrum is one of the rare cases in
which we detect lines from molecular hydrogen, \h2. From the \lya-line we
measure \nh=$22.1\pm0.10$. Intervening systems are found at $z = 1.539$, $z =
1.693$, and $z = 2.00$.

\subsection{GRB~120909A (z = 3.929)} \label{120909}

The data presented here have formed the basis of GCN 13730 \citep{GCN13730}, but
are not published otherwise. Follow-up, started only 1.7 hours after the BAT
trigger. This 1.2 ks observation captures a very bright afterglow continuum,
starting at 450 nm, signifying the onset of the Lyman limit for a system at $z =
3.929$. Absorption from high-column density hydrogen leaves very prominent
absorption features in the form of \lya, \lyb, and \lyg, visible in the Lyman
alpha forest. Metal absorption lines arising from \feii, \NIii, \SIii, \sii,
\alii, \aliii, \cii, \oi, \civ, and \znii~are all detected along with the
corresponding fine structure lines (\feii$^*$, \SIii$^*$, \oi$^*$, \oi$^**$,
\cii$^*$), securely anchoring the redshift of the host.

\subsection{GRB~120923A (z = 7.84)} \label{120923}

This spectrum has previously been published in \citet{Tanvir2017a}. Starting
18.5 hours after the BAT trigger, the final spectrum is based on $2 \times 4
\times 1200$ s spectroscopic integration for a total of 160 minutes. Nothing is
immediately visible in the 2D-spectra, however after severe binning in the
dispersion direction, a faint trace shows up in the NIR arm, suggesting a very
high redshift. We here adopt the redshift ($z = 7.84$) suggested by
\citet{Tanvir2017a}.

\subsection{GRB~121024A (z = 2.300)} \label{121024}

The data presented here also formed the basis of GCN 13890 \citep{GCN13890} and
are published in \citet{Friis2015}. Starting 1.8 hours after the
\textit{Swift} trigger, a bright afterglow continuum is visible across all arms.
A broad absorption feature from Lyman alpha, along with narrow lines from \civ,
\SIii, \SIiv, \feii, \sii, and \alii, as well as fine structure lines associated
with \SIii$^*$ are all detected at $z = 2.300$, securely setting it as the
redshift of the GRB.

\subsection{GRB~121027A (z = 1.773)} \label{121027}

The data presented here have formed the basis of GCN 13930 \citep{GCN13930}, but
are not published otherwise. Starting 69.6 hours after the GRB trigger, we detect
the afterglow continuum at high significance in all arms with 8.4 ks
integration, testifying to the brightness of this burst. The concurrent
identification of emission lines from \oiii~and absorption from \civ, \alii,
\aliii, \mgi, \mgii, and \feii, tightly constrains the redshift of the burst to
be $z = 1.773$.

\subsection{GRB~121201A (z = 3.385)} \label{121201}

These data formed the basis for GCN 14035 \citep{GCN14035} and are additionally
published in \citet{Kruhler2015}. These observations started 12.9 hours after
the \swift~trigger and consist of 4.8 ks spectroscopic integration under good
atmospheric conditions. The GRB afterglow continuum is well detected at all
arms. A broad absorption trough due to \lya~is visible at $z = 3.385$, which
along with the detection of absorption features identified as \SIiv, \civ,
\alii, and \aliii, marks it as the redshift of the GRB. In the middle of the
\lya~trough, we additionally detect \lya~emission. By modelling the
\lya~absorption, we infer $\log (N_{\mathrm{HI}}/\mathrm{cm}^{-2}) = 22.0 \pm
0.3$.

%

\subsection{GRB~121229A (z = 2.707)} \label{121229}

These data formed the basis for GCN 14120 \citep{GCN14120}, but are not
published elsewhere. Taken under poor seeing conditions, a total of 4.8 ks
spectroscopic integration starting 2 hours after the \swift~trigger yields a low
signal-to-noise GRB afterglow continuum is all arms. Binning the spectrum
reveals broad absorption troughs, which we identify as \lyb~and \lya~at $z =
2.707$. Additionally, an intervening system at $z = 1.658$ is detected from
absorption features of \mgii. From the absorption trough due to \lya, we infer
$\log (N_{\mathrm{HI}}/\mathrm{cm}^{-2}) = 21.7 \pm 0.2$. Due to strong
contamination in the slit, the background is slightly over subtracted, causing
the background to be negative in the center of the \lya~trough.

\subsection{GRB~130131B (z = 2.539)} \label{130131}

These data formed the basis for GCN 14286 \citep{GCN14286} and are additionally
published in \citet{Kruhler2015}. This is a late-time observation, taken long
after the GRB afterglow had faded. In 7.2 ks spectroscopic integration, emission
lines identified as \oii~and \oiii~are detected at a common $z = 2.539$, which
we suggest is the redshift of the GRB.

\subsection{GRB~130408A (z = 3.758)} \label{130408}

The data presented here also formed the basis of GCN 14365 \citep{GCN14365}. The
spectrum has not otherwise been published previously. The observations consists
of two 600 s spectra taken 1.9 hours after the burst. We detect absorption
features from a wide range of ions. We also detect intervening absorption at
$z=1.255$ and $z=3.248$.

\subsection{GRB~130418A (z = 1.222)} \label{130418}

GCN 14390 \citep{GCN14390} is based on this spectrum, but it is not published
elsewhere. Starting only 4.57 hours after the \swift~trigger, 1.2 ks observations
contain a bright GRB afterglow continuum, visible across the entire spectral
coverage of X-shooter. Superposed on the afterglow continuum are absorption
features which we identify as \civ, \feii, and \mgii, caused by an absorber at
$z = 1.217$, and additional absorption from \civ~ at $z = 1.222$. The two
systems are offset by $\sim 1500~\mathrm{km}~\mathrm{s}^{-1}$ and the proximity
of the two absorption systems in velocity space, suggests a possible association
of the two systems with peculiar velocity affecting the measured redshift. We
adopt $z = 1.222$ as the redshift of the GRB. Note, that this value is slightly
different from the one reported in GCN 14390.

\subsection{GRB~130427A (z = 0.340)} \label{130427A}

This spectrum is also published in \citet{Xu2013b} and \citet{Kruhler2015} and
additionally has formed the basis for GCN 14491 \citep{GCN14491}. Starting 16.5
hours after the BAT trigger, these observations lasting $2 \times 600$ s contain
a very bright GRB afterglow continuum across the total spectral coverage of
X-shooter. In absorption we identify features from the following metal resonance
lines: \feii, \mnii, \mgii, \mgi, \TIii~and additional line absorption from
\caii~and \nai. Simultaneously, we find emission lines from \ha, \hb, \oiii,
\oii - all at common redshift of $z = 0.340$, which is the redshift of the GRB.
This is one of the most energetic GRBs observed, and its proximity along with
the associated broad-lined Type Ic SN, 2013cq has caused it to be one of the
more well-studied GRBs \citep{Maselli2014, Perley2014, Ackermann2014}.

\subsection{GRB~130427B (z = 2.780)} \label{130427B}

This spectrum formed the basis of GCN 14493 \citep{GCN14493}, but are not
published otherwise. A short, 2x600 s spectroscopic integration obtained before
twilight, 20.6 hours after the BAT trigger, captures a faint afterglow continuum
visible across the entire spectral coverage at low signal-to-noise. Due to the
low signal-to-noise, the metal lines are weak, but the broad absorption trough
due to \lya~is detected. From the \lya~line we measure the redshift to be $z =
2.780$ and provide a measure of the neutral hydrogen column density, \nh~$= 21.9
\pm 0.3$. The redshift is confirmed by the presence of \feii~absorption at a
consistent redshift.

\subsection{GRB~130603B (z = 0.356)}\label{130603}

This burst is the first short GRB observed with a potential associated kilonova
\citep{Tanvir2013a, Berger2013a}.  GCN 14757 \citep{GCN14757} was based on this
spectrum and it is additionally published in \citet{DeUgartePostigo2014e}.
Starting 8.2 hours after the \swift~trigger, a total of 2.4 ks spectroscopic
integration was obtained. Spectral continuum is clearly detected across all arms
from both host and afterglow and superposed are both absorption (\cahk~and
\mgii) and emission lines (\oii, \hb, \oiii, \ha, and [\sii]), all at a
consistent redshift of $z = 0.356$, which is the redshift of the GRB.

\subsection{GRB~130606A (z = 5.913)}\label{130606}

The data presented here also formed the basis of GCN 14816 \citep{GCN14816} and
are published in \citet{Hartoog2015}. The observations consist of three
$2\times600$ s visits starting 7.1 hours after the burst at fairly high airmass.
We detect absorption features from a wide range of ions at $z=5.913$ as well as
intervening absorption at $z=2.3103, 2.5207, 3.4515, 4.4660, 4.5309, 4.5427,
4.6497 $ and $ 4.7244$.

\subsection{GRB~130612A (z = 2.006)}\label{130612}

The spectral features of this spectrum have previously been reported in GCN 14882
\citep{GCN14882}, but are not published elsewhere. Starting only 1.1 hours after
the \swift~trigger, $2\times600$ s spectroscopic integration captures a moderate
signal-to-noise afterglow continuum across the total spectral coverage of
X-shooter. At a consistent redshift of $z = 2.006$, absorption from the metal
resonance lines \feii, \mnii, \mgii, \mgi~are identified. Additionally, \lya~is
visible as a broad absorption trough, from which we can infer \nh~= $22.2 \pm 0.2$,
which is in the upper end of the hydrogen column density distribution. The blue
part of the GRB continuum exhibits a downturn in the continuum level which could
indicate the presence of a significant amount of dust along the line-of-sight.

\subsection{GRB~130615A (z = 2.9)} \label{130615}

This spectrum has not previously been published. Starting only 45 minutes after
the BAT trigger, $2\times600$ s spectroscopic integration carried on into the
beginning twilight. Observed at very high airmass with a quickly varying
background, a faint afterglow trace is visible across all arms of X-shooter,
down to ~ 480 nm, which if interpreted as the break due to \lya~suggests $z =
2.9$. This supports the approximate redshift suggested in GCN 14898
\citep{GCN14898}.

\subsection{GRB~130701A (z = 1.155)}\label{130701}

These data formed the basis for GCN 14956 \citep{GCN14956} and are additionally
published in \citet{Kruhler2015}. Starting 5.5 hours after the GRB trigger,
$2\times600$ s reveals a bright continuum visible across the entire spectral
coverage of X-shooter. Superposed are absorption features which we identify as
due to \feii, \mgii, \mgi, and \caii~- all at a consistent redshift of $z =
1.155$, which we take to be the redshift of the GRB.

\subsection{GRB~130925A (z = 0.347)}\label{130925}

This spectrum has already been used in GCN 15250 \citep{GCN15250} and is
additionally published in \citet{Schady2015} and \citet{Kruhler2015}.
Observations of this burst began with X-shooter 3.5 hours after the
\swift~trigger. 6 ks spectroscopic integration captures a heavily dust obscured
afterglow \citep[A$_\mathrm{V} = 5.9 \pm 0.7$;][]{Greiner2014}, with the
spectrum primarily dominated by host emission lines. All the nebular lines
(\oii, \hg, \hb, \oii, \ha, \nii, [\sii]) are well detected at $z = 0.347$, as
well as those from \pad, \pag, and \pab. We take this as the redshift of the
GRB. This spectrum is taken under ESO programme ID: 091.A-0877(A) (PI: Schady).

\subsection{GRB~131011A (z = 1.874)}\label{131011}

These data formed the basis for GCN 15330 \citep{GCN15330}, but are not published
elsewhere. Starting $\sim$1.5 days after the Fermi-GBM trigger, 4.5 ks
spectroscopic integration captures a modest signal-to-noise GRB afterglow
continuum all the way down to $\sim$320 nm. Imprinted on the continuum are
absorption features, which we identify as due to \lya, \feii,  \mgii, \mgi~at
the same redshift, which we measure to be $z = 1.874$. From the broad absorption
trough due to \lya, we infer \nh~$= 22.0 \pm 0.3$. This spectrum is taken under
ESO programme ID: 092.D-0056(A) (PI: Rau).

\subsection{GRB~131030A (z = 1.296)}	
\label{131030}

These data has not been published before. Starting 3.4 hours after the
\swift~trigger, $6\times600$ s exposure were taken under good conditions,
containing a bright GRB afterglow continuum across the entire spectral coverage
of X-shooter. A myriad of absorption features are superposed on the afterglow
continuum which we identify as being caused by \SIiv, \SIii, \civ, \alii,
\aliii, \znii, \crii, \NIii, \feii, \NIii$^*$, and \feii$^*$ at $z = 1.296$. A
very strong \mgii-absorber is also detected, intervening the line-of-sight at $z
= 1.164$ with lines from \SIii, \civ, \aliii, \aliii, \feii, \mnii, and many
more.

\subsection{GRB~131103A (z = 0.599)}\label{131103}

This spectrum has already been used to form the basis for GCN 15451
\citep{GCN15451} and is additionally published in \citet{Kruhler2015}. Starting
5.8 hours after the BAT trigger, $4\times600$ s exposure captures a modest
signal-to-noise continuum across all arms. Imprinted on the continuum are
absorption features identifed as due to \feii~and \mgii~as well as emission
lines from \oii, \hd, \hg, \hb, \oiii, \ha, and \niil. All the lines are measured
at a consistent redshift of $z = 0.599$, which we take as the redshift of the
GRB.

\subsection{GRB~131105A (z = 1.686)}\label{131105}

This spectrum has already been used in GCN 15450 \citep{GCN15450} and is
additionally published in \citet{Kruhler2015}. Starting only 1.3 hours after the
\swift~trigger, a total of 4.8 ks spectroscopic integration contains a low
signal-to-noise GRB afterglow continuum across the entire spectral coverage of
X-shooter. There are deviations from the continuum at both emission and
absorption. We identify lines from \hb, \oiii, and \ha~in emission and \feii,
and \mgii~in absorption. All lines are at a consistent $z = 1.686$, which is
probably the redshift of the GRB. Absorption lines at shorter wavelengths are
also detected, but at low significance due to an apparent downturn in the
continuum caused by the presence of dust local to the burst.

\subsection{GRB~131117A (z = 4.042)}\label{131117}

This spectrum has previously been used in GCN 15494 \citep{GCN15494}, but is not
published in a refereed paper. Starting only 68 minutes after the BAT trigger,
4.8 ks spectroscopic integration secures afterglow continuum for this burst,
which is measured to be at $z = 4.042$. A moderate signal-to-noise GRB afterglow
continuum is detected down to $\sim$ 610 nm, signifying the onset of the
\lya~forest, with part of the forest also visible. Metal absorption lines from
\SIii~and \SIiv~are detected at a consistent redshift.

\subsection{GRB~131231A (z = 0.642)}\label{131231}

This spectrum has previously been published in \citet{Kruhler2015} and
addtionally forms the basis for GCN 15645 \citep{GCN15645}. This spectrum,
observed the following year (20.2 hours after the \swift~trigger), consists of
$4\times600$ s exposures. A high signal-to-noise GRB afterglow continuum is
detected all the way through the X-shooter arms. We identify absorption features
imprinted in the continuum as caused by \feii, \mgii, and \cahk~at a consistent
$z = 0.642$. By subtracting off the bright afterglow continuum, we readily
detect emission lines arising from \oii, \hg, \hb, \oiii, and \ha~in the GRB
host galaxy. This GRB is the subject of a forthcoming work Kann et al. (in
prep).

\subsection{GRB~140114A (z = 3.0)}	\label{140114}

This spectrum has previously been published in \citet{Kruhler2015}. This is a
late time observation, taken long after the GRB had faded. Despite a long
integration time of 5.4 ks, no clear features stand out to clearly secure a
redshift measurement. By heavily binning the spectrum, a faint trace is visible
down to 485 nm, which if interpreted at the onset of \lya~signifies $z \sim
3$. We adopt the redshift inferred in \citet{Kruhler2015}.

\subsection{GRB~140213A (z = 1.208)}\label{140213}

The data presented here also formed the basis of GCN 15831 \citep{GCN15831} and
are additionally published in \citet{Kruhler2015}. Starting 5.8 after the
\swift~trigger alert, $2\times600$ s spectroscopic integration contains a high
signal-to-noise GRB afterglow continuum across the entirety of X-shooter.
Imprinted on the afterglow continuum are absorption features, which we identify
as metal resonance lines from \civ, \alii, \aliii, \feii, \mgii, and \mgi. These
lines are likely formed by metals in the GRB host, which we measure to be at $z
= 1.208$.

\subsection{GRB~140301A (z = 1.416)}\label{140301}

These data formed the basis for GCN 15900 \citep{GCN15900} and are additionally
published in \citet{Kruhler2015}. Spectroscopic follow-up began 9 hours after
the BAT trigger and lasted for $12\times600$ s. A low signal-to-noise, spatially
extended continuum is visible across the entire spectral coverage of X-shooter.
The GRB afterglow continuum is visible at moderate signal-to-noise on top of the
underlying host continuum. The host exhibits usual nebular emission from \oii,
\hb, \oiii, \ha, \nii, and [\sii] which puts it at $z = 1.416$. Supporting this as
the redshift of the GRB are absorption features from the \mgii-doublet in the
GRB afterglow continuum.

\subsection{GRB~140311A (z = 4.954)}	
\label{140311}

This spectrum has not been published previously. Starting 32.5 hours after the
GRB trigger onboard \swift, this observation lasted $14\times600$ s for a total
of 8.4 ks. Some loss occurred during the observations, which reduced integration
time in the UVB and VIS arm slightly. The GRB afterglow continuum is clearly
visible in the VIS and NIR arm of X-shooter. The continuum is very rich in
absorption features, with at least the following lines identified: \lyg, \lyb,
\lya, \SIii, \SIiv, \civ, \alii, \aliii, \feii, \mgii, and \mgi. All of these
lines are at $z = 4.954$, which we take as the redshift of the GRB. This spectrum
is taken under ESO programme ID: 092.D-0633(E) (PI: Greiner).

\subsection{GRB~140430A (z = 1.601)}\label{140430}

A separate reduction of this burst has been published in \citet{Kruhler2015},
and additionally the spectrum has also been used in GCN 16194 \citep{GCN16194}.
Observations for this burst began in twilight, 2.5 hours after the BAT trigger
and lasted for $2\times600$ s. The spectrum contains a moderate signal-to-noise
GRB afterglow continuum all the way through the spectroscopic arms of X-shooter.
We identify absorption features in the afterglow continuum from \SIii, \civ,
\alii, \feii, and \mgii, and emission lines from \oii~and \oiii~- all at $z =
1.601$ which is likely the redshift of the GRB.

\subsection{GRB~140506A  (z = 0.889)} \label{140506}

The data presented here has formed the basis of GCN 16217 \citep{GCN16217} and
is published in \citet{Fynbo2014}, \citet{Kruhler2015}, and \citet{Heintz2017a}.
The observations consists of $4\times 600$ s at 8.8 and 33 hours after the
burst. We detect absorption features from a wide range of ions, together with
molecular absorption from CH+, all at $z=0.889$. The optical/near-infrared
afterglow reveals an unusual steep extinction curve which is found to be caused
by dust very close to the burst.

\subsection{GRB~140515A (z = 6.327)}\label{140515}

This spectrum has previously been used in \citet{Melandri2015}. Starting 15.5
hours after the \swift~trigger, $8\times600$ s spectroscopic integration
captures this very high redshift GRB afterglow. The Gunn-Peterson trough is
visible against the GRB afterglow continuum, starting at 890 nm, which along
with absorption lines from the \mgii~doublet securely sets the redshift of this
GRB at $z = 6.327$. From the red wing of the \lya-profile we measure \nh~$=19.0
\pm 0.5$, which is very low compared to the measured distribution of
$N_{\mathrm{HI}}$.

\subsection{GRB~140614A (z = 4.233)}\label{140614}

This spectrum forms the basis for GCN 16401 \citep{GCN16401}, but is not
published elsewhere. $4\times600$ s spectroscopic integration, starting 3.8
hours after the BAT trigger catches this GRB afterglow, which turns out to be at
very high redshift. The continuum is detected at moderate signal-to-noise in
both the VIS and NIR arms of X-shooter, heavily affected by absorption features.
We identify the the lines belonging to \lya, \SIii, \cii, \cii$^*$, \alii,
\aliii, \feii, and \mgii~at a consistent redshift of $z = 4.233$. From the shape
of the \lya~absorption trough we measure \nh~$=21.3 \pm 0.3$.

\subsection{GRB~140622A (z = 0.959)}\label{140622}

The characteristics of this short \citep[T90 = $0.13 \pm 0.04$
s;][]{Lien2016} GRB spectrum has been published in GCN 16437 \citep{GCN16437},
but does not appear in the refereed literature. Spectroscopic observations began
with X-shooter only 34 minutes after the BAT trigger and lasted for $2\times600$
s. In the spectrum, we detect continuum across all three arms of X-shooter. In
the UVB arm, the continuum is only visible after heavily binning the spectrum in
the dispersion direction. It is unclear how much of the continuum is from the
host galaxy and how much is from the potential GRB afterglow. Superposed on the
continuum, emission lines from \oii, \hb, \oiii, \ha~are all detected at $z =
0.959$. We use this as the likely redshift of the GRB.

\subsection{GRB~141028A (z = 2.332)}\label{141028}

This spectrum has already been used in GCN 16983 \citep{GCN16983}, but is
not published elsewhere. Starting 15.4 hours after the \swift~trigger and
lasting for a total of 2.4 ks, these observations captures a bright GRB
afterglow. Very rich in absorption, the continuum is detected across the entire
spectral coverage of X-shooter, except the blue half of the UVB arm, where the
\lya-forest absorbs the continuum. The redshift of the host of the GRB is
measured based in the detection of features from \lya, \SIii, \civ, \cii, \feii,
and \mgii - all at a consistent $z = 2.332$. From the \lya~trough we infer
\nh~$=20.6 \pm 0.15$. Two intervening system at $z = 1.823$ and $z = 2.09$ are
also found in the spectrum based on the detection of \civ.

\subsection{GRB~141031A  (z = na)}	\label{141031}

This spectrum is an attempt at a late-time host redshift measurement for
GRB~141031A. The spectrum is taken long after the burst has faded and consists
of $4\times600$ s spectroscopic integration. In the 2D spectrum are two sources,
which are both offset from the targeted host position and are thus likely foreground
objects. The spectrum does not contain anything that can be used to measure a
redshift from.

\subsection{GRB~141109A (z = 2.993)}\label{141109}

The gross content of the spectrum has been issued in GCN 17040 \citep{GCN17040},
but it is not used otherwise. Starting only 1.9 hours after the BAT trigger,
these $4\times600$ s spectra contain a high signal-to-noise GRB afterglow. The
afterglow continuum is readily detected all the way down to 370 nm, although
the bluest part is affected by the \lya-forest. A broad absorption trough from
neutral hydrogen is clearly visible, with additional, narrower absorption
features identified as due to \SIii, \SIii$^*$, \cii, \cii$^*$, \SIiv, \civ,
\feii, \feii$^*$, \oi$^*$, and \nii$^*$. These are all detected at a consistent
$z = 2.993$, which due to the detection of the locally excited fine-structure
lines securely sets this as the redshift of the GRB. Additionally, two
\mgii~absorption systems are detected at $z = 1.67$ and $z=2.5$. From the broad
\lya-absorption, we measure a neutral hydrogen column density of \nh~$=22.1 \pm
0.1$.

\subsection{GRB~150206A (z = 2.087)}\label{150206}

This spectrum has previously been used to form the basis for GCN 17420
\citep{GCN17420}, but it is not published in the refereed literature. Due to the
short visibility of this burst, it does not enter into the statistical sample.
Beginning in twilight, $4\times600$ s integration time contains a weak/moderate
signal-to-noise GRB afterglow continuum throughout the X-shooter spectral
coverage, down to $\sim$ 375 nm. In the continuum are absorption features,
which we identify as due to metal lines from \znii, \feii, and \mgii.
Additionally, \lya-absorption is seen at the end of the trace. The spectral
position of the lines means that this GRB is at $z = 2.087$. From the
\lya~trough, we infer \nh~$=21.7 \pm 0.4$. The afterglow continuum appears
depressed in the blue end of the spectrum, suggestion dust extinction in the
host.

\subsection{GRB~150301B (z = 1.517)}\label{150301}

This spectrum has already been used in GCN 17523 \citep{GCN17523}, but it
is not used elsewhere. Starting 5.1 hours after the BAT trigger, this spectrum
is based on $6 \times 600$ s spectroscopic integration. The GRB afterglow
continuum is well detected across the entire spectral coverage of X-shooter at
moderate signal-to-noise. Imprinted on the continuum are absorption features
which we identify as being caused by \SIii, \civ, \alii, \feii, \mgii, and \mgi
- all at a similar redshift of $z = 1.517$, which is likely the redshift of the
GRB.

\subsection{GRB~150403A (z = 2.057)}\label{150403}

This spectrum has previously been used to form the basis for GCN 17672
\citep{GCN17672}. $4 \times 600$ s spectroscopic integration, starting 10 hours
after the GRB trigger, captures a bright GRB afterglow. A broad absorption
trough centered at $\sim$ 370 nm due to \lya~signals the redshift of the GRB,
which is refined to $z = 2.057$ based on the additional detection of metal
absorption lines. We readily identify features associated with \sii, \SIiv, \oi,
\SIii, \SIii$^*$, \cii, \cii$^*$, \civ, \alii, \feii, \feii$^*$, \mni, \mgii,
and \mgi~in the host of the GRB. An intervening \civ~absorber is additionally
detected at $z=1.76$. From the absorption trough due to \lya, we infer the
amount of neutral hydrogen in the host along the line of sight to be \nh~$=21.8
\pm 0.2$.

\subsection{GRB~150423A (z = 1.394)}\label{150423}

The gross content of these observations have previously been presented in GCN
17755 \citep{GCN17755}, but is not published as part of any refereed paper. This
bona-fide short GRB \citep[T90 = is $ 0.22 \pm 0.03$ s;][]{Lien2016} was
observed in RRM mode and spectroscopic integration started after only 22
minutes. A series of stare mode observations, increasing in exposure time, and
ending with a nodding sequence, totaling $\sim$ 5000 s, are combined to form
this spectrum. A faint, almost featureless continuum is detected at low
signal-to-noise all the way to the bluest part of the spectrum. An absorption
doublet is detected in the VIS arm against the GRB afterglow continuum, which we
identify as \mgii~at $z = 1.394$.

\subsection{GRB~150428A (z = na)}	\label{150428}

This spectrum is empty, but is included here for completeness. It is not
published anywhere. $4\times600$ s spectroscopic observation, starting 3.7
hours after the trigger does not reveal anything conclusive. The host
association is additionally ambiguous for this, likely extincted \citep[GCN
17767;][]{GCN17767}, GRB.

\subsection{GRB~150514A (z = 0.807)}\label{150514}

This spectrum has already been used in GCN 17822 \citep{GCN17822}. Spectroscopic
observations began 28.4 hours post trigger and consist of a $4 \times 600$ s
nodding sequence. The GRB afterglow continuum is detected at moderate
signal-to-noise across the entire spectral coverage of X-shooter. Narrow
absorption features are imprinted in the continuum where we identify features
from multiple \feii~transitions as well as the \mgii-doublet and the
\mgi$\lambda2852.96$ resonance line. These lines are all found at a position
matching $z = 0.807$, suggesting it is the redshift of the GRB.

\subsection{GRB~150518A (z = 0.256)}\label{150518}

This gross content of this data has previously been issued in GCN 17832
\citep{GCN17832}, but is not published. Starting $> 1$ day after the GRB
trigger, $4 \times 600$ s spectroscopic integration securely allows us to
measure the redshift of the host. A continuum is detected all the way through
the spectral coverage with multiple emission lines superposed. It is not clear
to what degree the GRB afterglow contributes to the continuum. We identify the
emission lines as \oii, \hb, \oiii, \ha, \niil, \ha, and [\sii]~- all at $z =
0.256$. Due to the spatial proximity of this burst, it is a candidate for SN
follow-up and indeed there are indications of re-brightening at the burst
position \citep[GCN 17903;][]{GCN17903}.

\subsection{GRB~150616A (z = 1.188)}\label{150616}

These host observations are taken long after the burst had faded  and have not
been published before. They are included here for completeness. This bursts is
excluded from the sample because an observing constraint delayed the
\swift~slew, causing the XRT observations to begin 16 minutes post trigger. No
continuum is detected, but emission lines from \oii, \oiii, and \ha~are all
detected at $z = 1.188$, setting it as the redshift of the GRB host.

\subsection{GRB~150727A (z = 0.313)}\label{150727}

The overall content of these observations have previously been reported in GCN
18080 \citep{GCN18080}, but is not published. Starting 5 hours after the
\swift~GRB trigger, these observations consist of a combined $2 \times 1200$ s
and $2 \times 600$ s integration. A blue continuum is detected across the entire spectral
coverage of X-shooter, suggesting a significant contribution from the GRB
afterglow. Superposed on the continuum are emission lines which we identify as
\hb, \oiii, and \ha~with a measured redshift of z = 0.313. Supporting this
redshift is the tentative detection of the \mgii~absorption doublet in the
afterglow continuum. The relative strength of the lines suggest that the
line-forming region is dust obscured, contrary to the story told by the blue
afterglow continuum.

\subsection{GRB~150821A (z = 0.755)}\label{150821}

The gross content of this spectacular spectrum has already been issued in GCN
18187 \citep{GCN18187}, but the spectrum has not been published. Starting just
12.4 minutes after the onboard trigger of \swift, a total of $4 \times 600$ s
spectroscopic integration was obtained. The observations began just before dawn
and the last two exposures are heavily affected by the brightening sky. A
bright, high signal-to-noise GRB afterglow continuum is detected all across the
the spectral window and imprinted on this are a myriad of absorption features.
We identify individual lines from transitions in \aliii, \crii, \znii,
\NIii$^*$, \feii, \feii$^*$, \scii, \mnii, \mgii, \mgi, \TIii, and  \caii~ - all
at a consistent redshift of $z = 0.755$. The detection of fine-structure lines,
excited local to the burst, clearly marks this as the redshift of the GRB.

\subsection{GRB~150910A (z = 1.359)}\label{150910}

This spectrum has not previously been published. Starting $\sim$ 20 hours after
the BAT trigger, these observations were stopped after $2 \times 600$s, when it
became apparent that only a modest signal-to-noise was obtainable, and the
redshift had already been published in GCN 18273 \citep{GCN18273} and GCN 18274
\citep{GCN18274}. The continuum is detected at low signal-to-noise all across
X-shooter and \mgii~is detected at the suggested redshift, which we take as the
redshift of the GRB.

\subsection{GRB~150915A (z = 1.968)}\label{150915}

The gross content of these spectra have already been issued in GCN 18318
\citep{GCN18318}. Observations began 3.3 hours after the \swift~trigger and
lasted for $4 \times 1200$ s. A moderate signal-to-noise afterglow continuum is
detected across the entire spectral coverage of X-shooter, where the
bluest part is affected by \lya-forest absorption. Imprinted on the afterglow
continuum are a wealth of both emission and absorption features, all caused in a
system for which we measure a redshift of $z = 1.968$. We identify in emission
\oii, \hb, \oiii, and \ha~and absorption due to \lya, \civ, \alii, \SIii, \feii,
and \mgii. We additionally identify fine-structure absorption lines from
\SIii$^*$ and \feii$^*$, at a similar redshift, which unequivocally sets the
suggested redshift as the redshift of the GRB. The \lya-line is affected by an
atmospheric transmission drop and \lya~emission in the trough, making the
measurement of the neutral hydrogen column density difficult. We measure
\nh~$21.2 \pm 0.3$.

\subsection{GRB~151021A (z = 2.330)}\label{151021}

The data presented here also formed the basis of GCN 18426 \citep{GCN18982} and
are not published elsewhere. The observation was carried out in RRM starting 44
minutes after the GRB trigger. We detect absorption features from a wide range
of ions at $z=2.330$ as well as intervening absorption at $z=1.49$.

\subsection{GRB~151027B (z = 4.063)}\label{151027}

The content of these spectra has previously been issued in GCN 18506
\citep{GCN18506}, but is not published. Beginning 5 hours after the BAT trigger,
4 spectroscopic integrations obtained in a nodding sequence, each lasting for
600s, securely allows us to measure a redshift for this GRB. Beginning at $\sim$
470 nm, the \lya-forest is clearly detected at high signal-to-noise leading up
to the broad \lya~absorption trough and the onset of a bright GRB afterglow
continuum. Imprinted in the afterglow continuum are a range of absorption lines
which we identify as due to \SIii, \SIii$^*$, \oi, \cii, \cii$^*$, \civ, \alii,
\alii, \feii, and \feii$^*$. These lines are all detected at $z = 4.063$ and the
presence of fine-structure lines at a consistent redshift securely sets this as
the redshift of the GRB. GRB~151027B was the 1000th GRB detected by \swift. The
afterglow of this burst is the subject of a forthcoming publication
\citep{Greiner2018}.

\subsection{GRB~151029A (z = 1.423)}\label{151029}

The gross content of these observations have previously been described in GCN
18524 \citep{GCN18524}. Starting only 1 hour after the BAT trigger, $2 \times
600$ s observations, extending into the morning twilight, were obtained of this
GRB. The brightening sky affects the signal-to-noise of these spectra -
especially in the blue end. The GRB afterglow is well detected over the entire
spectral coverage of X-shooter, with spectral features from \feii~and
\mgii~imprinted by an absorber at $z = 1.423$. The absorber is likely the host
of the GRB and we therefore consider $z = 1.423$ the redshift of the GRB.

\subsection{GRB~151031A (z = 1.167)}\label{151031}

The spectral features in this spectrum has previously been reported in GCN 18540
\citep{GCN18540}. Observed in RRM mode, these observations were initiated only
19 minutes after the BAT trigger. As per the standard RRM scheme, a series of
stare observation, increasing in en exposure time and ending with a regular $4
\times 600$ s nodding sequence, were acquired for this burst. A moderate
signal-to-noise continuum is detected all the way through the X-shooter spectral
coverage. In the GRB afterglow continuum, we identify absorption features as
\feii, \mgii, \mgi, and \caii~at a consistent redshift of $z = 1.167$.
Supporting this as the redshift of the GRB host galaxy are the simultaneous
detection of nebular emission lines from \oii, \hb, and \oiii.

\subsection{GRB~160117B (z = 0.870)}	\label{160117}

A preliminary analysis of these spectra have previously been presented in GCN
18886 \citep{GCN18886}. Starting 13.5 hours after the \swift~trigger, these
spectra are formed on the basis of $4 \times 1200$ s spectroscopic integration.
The GRB afterglow is clearly detected at high signal-to-noise across the entire
spectral coverage of X-shooter. Superposed on the continuum are absorption and
emission features, which are identified as \feii, \mgii, \mgi~in absorption, and
\oii, \hb, and \oiii. The spectral positions of these lines all correspond to a
redshift of $z = 0.870$, which most likely is the redshift of the GRB.

\subsection{GRB~160203A (z = 3.518)}\label{160203}

The data presented here also formed the basis of GCN 18982 \citep{GCN18982} and
are not published elsewhere. The observation was carried out in RRM starting 18
minutes after the GRB trigger. We detect absorption features from a wide range
of ions at $z=3.518$ as well as intervening absorption at $z=2.203$. From the
very high signal-to-noise spectrum we are able to measure \nh~$=21.75 \pm 0.10$,
based on the shape of the \lya~profile. This burst is the subject of a
forthcoming publication (Pugliese et al. in prep.).

\subsection{GRB~160228A (z = 1.640)} \label{160228}

The gross content of these spectra have previously been presented in GCN 19186
\citep{GCN19186}. Taken long after the burst had faded, the target is the likely
host galaxy. In $4\times1200$ s spectroscopic integration we detected emission
lines from the host. We find \oiii~and \ha~at a consistent $z = 1.640$.  We
therefore suggest that this is the redshift of the GRB.

\subsection{GRB~160303A (z = na)} \label{160303}

This is potentially a short burst with extended emission \citep[GCN
19148;][]{GCN19148}. The content of the spectra have previously been reported in
GCN 19154 \citep{GCN19154}. Observations began 19.1 hours after the BAT trigger
and consist of $4 \times 1200$ s spectroscopic integration. A very faint
continuum is detected in the UVB and VIS arm after binning in the dispersion
direction, but no clear redshift can be inferred from the continuum shape. No
emission lines are readily apparent, suggesting little or no star formation in
the host.

\subsection{GRB~160314A (z = 0.726)} \label{160314}

The gross content of these spectra have previously been issued in GCN 19192
\citep{GCN19192}. Spectroscopic integration began 13 hours post GRB trigger and
lasted for $4 \times 1200$ s. In the data we see a moderate signal-to-noise
continuum across the entire spectral coverage of X-shooter. Clearly visible on
top of the continuum are emission lines from \oii, \hd, \hg, \hb, \oiii, \ha,
\niil, and [\sii]~at a consistent redshift of $z = 0.726$. It is unclear how much
of the continuum is due to the host and how much is from the GRB afterglow. At
the same redshift, we see the tentative detection of \feii-absorption,
supporting the redshift measurement and suggesting a significant afterglow
contribution to the continuum.

\subsection{GRB~160410A (z = 1.717)}\label{160410}

This short GRB with extended emission \citep[GCN 19276;][]{GCN19276} was
observed in RRM mode, with observations starting after only 7.7 minutes. The
gross content has already been presented in GCN 19274 \citep{GCN19274}. These
observations have the shortest delay between trigger time and start of
observation. Spectroscopic integration began after 8.4 minutes and lasted for $3
\times 600$ s. The last image in the nodding sequence was aborted because the
object was setting and the telescope reached the hardware limit. A blue,
moderate signal-to-noise continuum is detected across the entire spectral
coverage of X-shooter with multiple absorption features imprinted on it. At $z =
1.717$ we detected the broad absorption trough due to \lya~and we additionally
find \alii~and \feii~at a consistent redshift, setting it as the redshift of the
GRB. Additionally, two intervening \civ~absorption systems are detected at  $z =
1.444$ and $z = 1.581$. From the \lya~absorption trough, we measure a neutral
hydrogen column density of \nh$= 21.2 \pm 0.2$. This is the only short GRB for
which we have a measurement of $N_{\mathrm{HI}}$ and is the subject of an
forthcoming paper (Selsing et al. in prep).

\subsection{GRB~160425A (z = 0.555)}\label{160425}

These spectra were already used to form the basis for GCN 19350
\citep{GCN19350}, but are not published anywhere else. Spectroscopic integration
started after 7.2 hours and lasted for $4 \times 1200$ s. A low signal-to-noise
continuum is detected across the spectral coverage of X-shooter, but no clear
absorption lines are discernible. It is therefore unclear how much of the
continuum is due to the host and how much is due to the GRB. Superposed on the
continuum are two sets of emission lines, offset spatially, but at very similar
redshifts. We identify lines from \oii, \hb, \oiii, \ha, \niil, and [\sii]~in both
systems -- all at $z = 0.555$. As per GCN 19350 \citep{GCN19350}, the GRB
probably occurred in an interacting pair of galaxies.

\subsection{GRB~160625B (z = 1.406)}\label{160625}

GRB~160625B is the first GRB for which a signicant linear polarization was
measured \citep{Troja2017}. Because this burst was detected by \textit{Fermi}, it
is therefore not a part of the statistical sample. The gross content of these
spectra has previously been presented in GCN 19600 \citep{GCN19600}. Starting 30
hours after the trigger, these spectra are obtained in $4 \times 600$ s. The GRB
afterglow continuum is detected at high signal-to-noise throughout the spectral
coverage of X-shooter. Imprinted on the afterglow continuum are a multitude of
absorption features, from which we identify the responsible elements as \SIii,
\oi, \SIiv, \civ, \alii, \aliii, \feii, \znii, \mgii, and \mgi~at a consistent
redshift of $z = 1.406$. An additional intervening \mgii~absorber is detected at
$z = 1.319$.

\subsection{GRB~160804A (z = 0.736)}\label{160804}

The data presented here also formed the basis of GCN 19773 \citep{GCN19773}, and
are published in \citet{Heintz2017b}. Observations started 22.37 hours after the
BAT trigger and lasted for 2.4ks. The afterglow continuum is detected across the
entire spectral coverage of X-shooter and absorption lines from \mgi, \mgii,
\feii~and \alii~are found at $z = 0.736$. At the same redshift, emission lines
from \oii, \oiii, \ha, \hb, \hg, \niil, [\sii], and [\siii]~are found. A second epoch,
lasting 3.6ks, is obtained after the afterglow had faded, confirming the
emission line detections. The host galaxy is found to have a roughly solar
metallicity and is among the most luminous GRB hosts at $z < 1$.

\subsection{GRB~161001A (z = 0.891)}	\label{161001}

The content and scope of these spectra has previously been issued in GCN 19971
\citep{GCN19971}. $4 \times 600$ s spectroscopic integration was obtained of
this potentially short GRB (GCN 19974; \citealt{GCN19974}), starting 6.1 hr
after the GRB trigger. Close to the
\href{http://www.swift.ac.uk/xrt_positions/00020702/}{XRT error circle}
\citep{GCN19969} there is a bright, point-like source, also visible in archival
images. Our spectrum reveals that this is an M-type star, based on the presence
of TiO bands, and is thus unrelated to the GRB. Two more objects lie within or
in the near proximity of the X-ray error circle, as visible in our images from
the acquisition camera. One is blended with the above mentioned star, and is
serendipitously covered by the X-shooter slit, revealing emission lines from
\oii, \hb, and \ha~at a consistent redshift of $z = 0.891$. The second source,
first noted by \citet{GCN19975}, is extended and lies just outside the XRT error
circle. Based on GROND imaging, \citet{GCN19975} report moderate variability,
and suggest this object as an afterglow candidate. Following our original
imaging, we thus secured further late-time photometry with the X-shooter
acquisition camera, starting on 2016 Oct 5.26 UT (4.22 days after the GRB),
securing $3 \times 120$~s in each of the $g$ and $r$ filters. No variability is
apparent in either bands. In offline analysis of the GROND data (T. Kr\"uhler,
priv. comm.), the evidence for fading is less significant than originally
reported. The situation is thus left ambiguous, but due to the consistency of
the $z = 0.891$ object with the XRT localization, we consider this as the most
likely redshift of GRB\,161001A.

\subsection{GRB~161007A (z = na)} \label{161007}

These data have not been published elsewhere. Observations for GRB~161007A
started 323 hr after the burst trigger and were centered on the potential host
galaxy (see GCN 20014 \citep{GCN20014} and GCN 20020 \citep{GCN20020}). $4
\times 600$ s of observations were obtained revealing a faint continuum rising
abruptly above the noise at $\sim 685$ nm and extending up to 2100 nm. A very
low significance continuum is detected at shorter wavelengths, down to $\sim
600$ nm. No significant emission features could be identified. Two possibilities
can explain the observed continuum break, but neither are fully satisfactory.
The first option is the Lyman alpha cutoff at $z \sim 4.6$; at this redshift, the
usually detected strong nebular lines would be shifted out of the covered
wavelength range. The host would be however exceptionally bright, with an
absolute magnitude at $\approx 150$~nm{} $M = -23.8$~AB, that is about 2.8~mag
brighter than $M^*$ at that redshift (e.g. \citealt{Bouwens2015}). The inferred
X-ray hydrogen-equivalent column density would also be a record-high $4 \times
10^{23}$~cm$^{-2}$ (see e.g. \citealt{Campana2012}). Alternatively, the
continuum discontinuity could be the 400~nm{} break at $z = 0.71$; such
feature is however prominent only in early-type galaxies, which are unlikely
hosts of long-duration GRBs (but see \citealt{Rossi2014}). Given the lack of
optical variability, it cannot be excluded that the object within the XRT error
circle is a chance association, as the probability of a random alignment is
small but non negligible (of the order of 1\%). In conclusion, we cannot provide
a secure redshift for GRB 161007A.

\subsection{GRB~161014A (z =2.823)} \label{161014}

The data presented here also formed the basis of GCN  20061 \citep{GCN20061},
but are not published elsewhere. Starting 11.6 hours after the GRB trigger, 4.8
ks of integration time captures the afterglow continuum across all three
spectroscopic arms. A broad absorption trough due to Lyman alpha is visible,
along with metal absorption features from \mgii, \SIii, \cii, \civ, \alii,
\aliii, and	\feii, all at $z =2.823$. Similar to GRB~140506 \citep{Fynbo2014,
	Heintz2017b}, a break in the continuum shape is tentatively detected bluewards
of 600 nm, possible signifying some anomalous form of extinction.

\subsection{GRB~161023A (z = 2.710)}\label{161023}

The gross content of these spectra have previously been reported in GCN 20104
\citep{GCN20104}. This extremely bright \textit{INTEGRAL} burst is not a part of
the statistical sample. Starting 3 hours after the trigger, the spectra are
based on $2 \times 600$ s. The GRB afterglow continuum is detected with the
highest signal-to-noise of all the spectra presented here. A myriad of
absorption features are visible against the continuum, many of which would not be
visible against lower signal-to-noise spectra. We identify at least 15
intervening absorbers between $z = 1.243$ and $z = 2.710$. In the host system we
clearly see absorption from \lyb, \lya, \sii, \SIii, \oi, \SIiv, \civ, \alii,
\aliii, \feii, \mgii, and \mgi. Several fine-structure transitions are
additionally seen, clearly associating the $z = 2.710$ system with the GRB. From
the \lya-line we infer \nh$=20.96 \pm 0.05$. This spectrum will be further analyzed in de Ugarte Postigo et al. (in prep).

\subsection{GRB~161117A (z = 1.549)}\label{161117}

The majority of the spectral features of these spectra have previously been
reported in GCN 20180 \citep{GCN20180}. Starting only $\sim$ 40 minutes after
the BAT trigger, these observations consist of $4 \times 600$ s, taken under
relatively poor observing conditions. The GRB afterglow continuum is detected
throughout most of the X-shooter spectral coverage at low-to-moderate
signal-to-noise, however the continuum level drops below the noise in the bluest
part of the UVB arm at $\sim$ 380 nm. Imprinted in the continuum are
absorption features, which we identify as due to \feii, \mgii, and \mgi~in an
absorption system at $z = 1.549$. We consider this the likely redshift of the
GRB.

\subsection{GRB~161219B (z = 0.148)}\label{161219}

The core signatures in these spectra have already been issued in GCN 20321
\citep{GCN20321}, and are additionally used in both \citet{Ashall2017} and
\citet{Cano2017}. This GRB is very close-by and is associated with the SN
Ic-BL SN2016jca \citep[GCN 20342;][]{GCN20342}. The observations consist of $4
\times 600$ s spectroscopic integration, starting 1.5 days after the GRB
trigger. The GRB afterglow continuum is detected at high signal-to-noise across
the entire spectral coverage of X-shooter. Superposed on the continuum are
absorption features, which we identify as being formed by the \mgii, \mgi, and
the \cahk-lines at $z = 0.148$. Underneath the glaring afterglow continuum we
additionally detect nebular emission from \oii, \hb, \oiii, \ha, and [\sii] at a
consistent redshift, which is the likely redshift of the GRB.

\subsection{GRB~170113A (z=1.968)}\label{170113}

These gross content of these data have previously been issued in GCN 20458
\citep{GCN20458}. Starting 15.2 hours after the BAT trigger, these observations
consist of $4 \times 1200$ s spectroscopic integration. The GRB afterglow
continuum is detected at moderate signal-to-noise across across the entire
spectral coverage of X-shooter. In the bluest end of the spectral coverage, the
background noise increases, thus lowering the continuum below detection.
Superposed on the afterglow continuum are both absorption lines and emission
lines. In absorption we identify lines from \SIii, \feii, and \mgii, and in
emission we detect \oii, \hb, and \oiii. All the lines are detected at a
consistent redshift of $z=1.968$, which we take as the redshift of the GRB.

\subsection{GRB~170202A (z=3.645)}\label{170202}

GRB~170202A is the last GRB afterglow observed in this work. It concludes the
follow-up effort that has lasted for 8 years. The majority of the spectral
features in this spectrum have previously been reported in GCN 20589
\citep{GCN20589}. Spectroscopy began 9.7 hours after the BAT trigger, and lasted
for $4 \times 600$ s. The GRB afterglow is well detected at moderate-to-high
signal-to-noise throughout the spectral coverage of X-shooter, down to 430 nm.
The lowest wavelength continuum from 430 nm to 560 nm is absorbed by the
\lya~forest, with the broad \lya~absorption trough centered at 565 nm. In the
continuum, we additionally identify metal absorption features from \SIii, \civ,
\SIiv, \feii, and \mgii, with the additional detection of
\SIii$^*$~fine-structure lines. With an additional detection of \oiii~and all
lines detected at $z=3.645$, this unequivocally marks this as the redshift of
the GRB. There is an additional \civ~absorber at $z = 3.077$. From the
\lya~absorption we measure \nh$= 21.55 \pm 0.10$.

%
%

\end{document}

%% file: definitions.tex
\newcommand{\farc}{\hbox{$.\!\!^{\prime\prime}$}}

\newcommand{\lya}{Ly$\alpha$} 
\newcommand{\lyb}{Ly$\beta$} 
\newcommand{\lyg}{Ly$\gamma$} 
\newcommand{\hb}{H$\beta$} 
\newcommand{\ha}{H$\alpha$} 
\newcommand{\hg}{H$\gamma$} 
\newcommand{\hd}{H$\delta$} 
\newcommand{\pab}{Pa$\beta$} 
 
\newcommand{\pag}{Pa$\gamma$} 
\newcommand{\pad}{Pa$\delta$}

\newcommand{\cah}{Ca H} 
\newcommand{\cak}{Ca K} 
\newcommand{\cahk}{Ca H \& K} 
\newcommand{\nai}{Na ID}

\newcommand{\hi}{\ion{H}{i}} 
\newcommand{\hei}{\ion{He}{i}} 
\newcommand{\oi}{\ion{O}{i}} 
\newcommand{\sii}{\ion{S}{ii}} 
\newcommand{\siii}{\ion{S}{iii}} 
\newcommand{\oii}{[\ion{O}{ii}]} 
\newcommand{\oiii}{[\ion{O}{iii}]}
\newcommand{\nii}{\ion{N}{ii}} 
\newcommand{\niil}{[\ion{N}{ii}]} 
\newcommand{\nv}{\ion{N}{v}} 
\newcommand{\neiii}{[\ion{Ne}{iii}]} 
\newcommand{\NIii}{\ion{Ni}{ii}} 
\newcommand{\feii}{\ion{Fe}{ii}} 
\newcommand{\civ}{\ion{C}{iv}} 
\newcommand{\cii}{\ion{C}{ii}}
\newcommand{\mgi}{\ion{Mg}{i}} 
\newcommand{\mgii}{\ion{Mg}{ii}} 
\newcommand{\ali}{\ion{Al}{i}} 
\newcommand{\alii}{\ion{Al}{ii}} 
\newcommand{\aliii}{\ion{Al}{iii}} 
\newcommand{\SIi}{\ion{Si}{i}} 
\newcommand{\SIii}{\ion{Si}{ii}} 
\newcommand{\SIiii}{\ion{Si}{iii}} 
\newcommand{\SIiv}{\ion{Si}{iv}} 
\newcommand{\znii}{\ion{Zn}{ii}} 
\newcommand{\crii}{\ion{Cr}{ii}} 
\newcommand{\mnii}{\ion{Mn}{ii}} 
\newcommand{\mni}{\ion{Mn}{i}} 
\newcommand{\tiii}{\ion{Ti}{ii}} 
\newcommand{\caii}{\ion{Ca}{ii}} 
\newcommand{\ariii}{\ion{Ar}{iii}} 
\newcommand{\TIii}{\ion{Ti}{ii}} 
\newcommand{\scii}{\ion{Sc}{ii}} 
\def\h2{\mbox{{\sc H}$_2$}\xspace}

\newcommand{\startdate}{$14/03/2009$} 

\newcommand{\termdate}{$31/03/2017$} 

\newcommand{\swift}{\textit{Swift}} 
\newcommand{\nhx}{$\log (N_{\mathrm{H, X}}/\mathrm{cm}^{-2})$} 
\newcommand{\nh}{$\log (N_{\mathrm{HI}}/\mathrm{cm}^{-2})$}

%% file: authorlist.tex
\author{
J.~Selsing \inst{1}$^,$\thanks{\email{jselsing@dark-cosmology.dk}}\addtocounter{footnote}{2}
\and D.~Malesani \inst{1, 2, 3}\fnmsep\thanks{On-call observer}
\and P.~Goldoni \inst{4}\fnmsep$^\dagger$
\and J.~P.~U. Fynbo \inst{1, 2}\fnmsep$^\dagger$
\and T.~Kr\"{u}hler \inst{5}\fnmsep$^\dagger$
\and L.~A.~Antonelli \inst{6}\fnmsep$^\dagger$
\and M.~Arabsalmani \inst{7, 8}
\and J.~Bolmer \inst{5, 9}\fnmsep$^\dagger$
\and Z.~Cano \inst{10}\fnmsep$^\dagger$
\and L.~Christensen \inst{1}
\and S.~Covino \inst{11}\fnmsep$^\dagger$
\and P.~D'Avanzo \inst{11}\fnmsep$^\dagger$
\and V.~D'Elia \inst{12}\fnmsep$^\dagger$
\and A.~De~Cia \inst{13}
\and A.~de~Ugarte~Postigo \inst{1, 10}\fnmsep$^\dagger$
\and H.~Flores \inst{14}\fnmsep$^\dagger$
\and M.~Friis \inst{15, 16}
\and A.~Gomboc \inst{17}
\and J.~Greiner \inst{5}
\and P.~Groot \inst{18}
\and F.~Hammer \inst{14}
\and O.E.~Hartoog \inst{19}\fnmsep$^\dagger$
\and K.~E.~Heintz \inst{1, 2, 20}\fnmsep$^\dagger$
\and J.~Hjorth \inst{1}\fnmsep$^\dagger$
\and P.~Jakobsson \inst{20}\fnmsep$^\dagger$
\and J.~Japelj \inst{19}\fnmsep$^\dagger$
\and D.~A.~Kann \inst{10}\fnmsep$^\dagger$
\and L.~Kaper \inst{19}
\and C.~Ledoux \inst{9}
\and G.~Leloudas \inst{1}
\and A.J.~Levan \inst{21}\fnmsep$^\dagger$
\and E.~Maiorano \inst{22}
\and A.~Melandri \inst{11}\fnmsep$^\dagger$
\and B.~Milvang-Jensen \inst{1, 2}
\and E.~Palazzi \inst{22}
\and J.~T.~Palmerio \inst{23}\fnmsep$^\dagger$
\and D.~A.~Perley \inst{24}\fnmsep$^\dagger$
\and E.~Pian \inst{22}
\and S. ~Piranomonte \inst{6}\fnmsep$^\dagger$
\and G.~Pugliese \inst{19}\fnmsep$^\dagger$
\and R.~S\'{a}nchez-Ram\'{\i}rez \inst{25}\fnmsep$^\dagger$
\and S.~Savaglio \inst{26}
\and P.~Schady \inst{5}
\and S.~Schulze \inst{27}\fnmsep$^\dagger$
\and J.~Sollerman \inst{28}
\and M.~Sparre \inst{29}\fnmsep$^\dagger$
\and G.~Tagliaferri \inst{11}
\and N.~R.~Tanvir \inst{30}\fnmsep$^\dagger$
\and C.~C.~Th\"{o}ne \inst{10}
\and S.D.~Vergani \inst{14}\fnmsep$^\dagger$
\and P.~Vreeswijk \inst{18, 26}\fnmsep$^\dagger$
\and D.~Watson \inst{1, 2}\fnmsep$^\dagger$
\and K.~Wiersema \inst{21, 30}\fnmsep$^\dagger$
\and R.~Wijers \inst{19}
\and D.~Xu \inst{31}\fnmsep$^\dagger$
\and T.~Zafar \inst{32}
}
\institute{Dark Cosmology Centre, Niels Bohr Institute, University of Copenhagen, Juliane Maries Vej 30, 2100 K\o benhavn \O, Denmark 
\and The Cosmic Dawn Center, Niels Bohr Institute, University of Copenhagen, Juliane Maries Vej 30, 2100 K\o benhavn \O, Denmark
\and DTU Space, National Space Institute, Technical University of Denmark, Elektrovej 327, 2800 Lyngby, Denmark
\and APC, Astroparticule et Cosmologie, Universit\'e Paris Diderot, CNRS/IN2P3, CEA/Irfu, Observatoire de Paris, Sorbonne Paris Cit\'e, 10, Rue Alice Domon et L\'eonie Duquet, 75205, Paris Cedex 13, France
\and Max-Planck-Institut f\"{u}r extraterrestrische Physik, Giessenbachstra\ss e, 85748 Garching, Germany
\and INAF - Osservatorio Astronomico di Roma, Via Frascati 33, 00078, Monte Porzio Catone (Roma), Italy
\and IRFU, CEA, Universit\'e Paris-Saclay, F-91191 Gif-sur-Yvette, France
\and Universit\'e Paris Diderot, AIM, Sorbonne Paris Cit\'e, CEA, CNRS, F-91191 Gif-sur-Yvette, France
\and European Southern Observatory, Alonso de C\'{o}rdova 3107, Vitacura, Casilla 19001, Santiago 19, Chile
\and Instituto de Astrof\'isica de Andaluc\'ia (IAA-CSIC), Glorieta de la Astronom\'ia s/n, E-18008, Granada, Spain. 
\and INAF - Osservatorio Astronomico di Brera, via Bianchi 46, 23807, Merate (LC), Italy
\and Space Science Data Center - Agenzia Spaziale Italiana, via del Politecnico, s.n.c., I-00133, Roma, Italy
\and European Southern Observatory, Karl-Schwarzschild Str. 2, 85748 Garching bei M\"unchen, Germany
\and GEPI, Observatoire de Paris, PSL University, CNRS, 5 place Jules Janssen 92195, Meudon, France
\and KTH Royal Institute of Technology, Department of Physics, 106 91 Stockholm, Sweden
\and The Oskar Klein Centre for Cosmoparticle Physics, AlbaNova University Centre, 106 91 Stockholm, Sweden
\and Center for Astrophysics and Cosmology, University of Nova Gorica, Vipavska 11c, 5270 Ajdov\v s\v cina, Slovenia
\and Department of Astrophysics, IMAPP, Radboud University Nijmegen, PO Box 9010, 6500 GL Nijmegen, the Netherlands
\and Anton Pannekoek Institute for Astronomy, University of Amsterdam, Science Park 904, 1098 XH Amsterdam, the Netherlands
\and Centre for Astrophysics and Cosmology, Science Institute, University of Iceland, Dunhagi 5, 107, Reykjavik, Iceland
\and Department of Physics, University of Warwick, Coventry, CV4 7AL, UK 
\and IASF/INAF Bologna, via Piero Gobetti 101, 40129 Bologna, Italy
\and Sorbonne Universit\'e, CNRS, UMR7095, Institut d'Astrophysique de Paris, F-75014, Paris, France
\and Astrophysics Research Institute, Liverpool John Moores University, IC2, Liverpool Science Park, 146 Brownlow Hill, Liverpool L3 5RF, UK
\and INAF, Istituto Astrofisica e Planetologia Spaziali, Via Fosso del Cavaliere 100, I-00133 Roma, Italy
\and Physics Department, University of Calabria, I-87036 Arcavacata di Rende, Italy
\and Department of Particle Physics and Astrophysics, Weizmann Institute of Science, 234 Herzl Street, Rehovot, 761000, Israel
\and Department of Astronomy, Stockholm University, AlbaNova, 10691 Stockholm, Sweden 
\and Institut f\"ur Physik und Astronomie, Universit\"at Potsdam, Karl-Liebknecht-Str.\,24/25, 14476 Golm, Germany
\and Department of Physics and Astronomy, University of Leicester, University Road, Leicester, LE1 7RH, United Kingdom
\and CAS Key Laboratory of Space Astronomy and Technology, National Astronomical Observatories, Chinese Academy of Sciences, Beijing, 100012, P.R. China 
\and Australian Astronomical Observatory, PO Box 915, North Ryde, NSW 1670, Australia.
}

%% file: tables/burst_overview.tex
\begin{longtab}
\begin{longtable}{lccccccccc} 
\caption{The full sample of afterglows and hosts observed in the program.
	We here list the burst names and details of the spectroscopic observations. The
	exposure times and slit widths are given in the order UVB/VIS/NIR. The column
	$\Delta t$ shows the time after trigger when the spectroscopic observation was
	started. Mag$_\mathrm{acq}$ gives the approximate magnitude (typically in the
	$R$-band) of the afterglow or the host in the acquisition image. \label{tab:sample_overview}}  \\
\hline\hline
{GRB} &  Obs Date & Exptime & Slit width & Airmass & Seeing & $\Delta t$ & Mag$_\mathrm{acq}$ & Redshift & Notes \\[1.5pt]
\hline
{} & {} &  (ks)   & (arcsec) & {}  &(arcsec) & (hr)   & {} & {} &  \\ [1.5pt]
\hline
\endfirsthead
\caption{The full sample of afterglows or hosts observed in the program (continued).}\\
\hline\hline
{GRB} &  Obs Date & Exptime & Slit width & Airmass & Seeing & $\Delta t$ & Mag$_\mathrm{acq}$ & Redshift & Notes \\[1.5pt]
\hline
{} & {} &  (ks)   & (arcsec) & {}  &(arcsec) & (hr)   & {} & {} &  \\ [1.5pt]
\hline
\endhead
\input{tables/sample_list.tex}

\hline\noalign{\smallskip}

\end{longtable}
\centering
\begin{minipage}{5.3in}
\tablefoot{
\tablefoottext{a}{Not part of the statistical sample}
\tablefoottext{b}{Spectrum dominated by light from the host galaxy}
\tablefoottext{c}{Spectrum of the host galaxy taken long after the burst}
\tablefoottext{d}{RRM observation}
\tablefoottext{e}{ADC malfunction during observation}
\tablefoottext{f}{Short burst}
}

\end{minipage}
\end{longtab}

%% file: tables/sample_list.tex
GRB090313\tablefootmark{a} 		                &        2009-03-15         &   6.9/6.9/6.9  	& 1.0/0.9/0.9		& 1.2--1.4  	& 1.5   	& 45      	&  21.6    	& 3.374 		& \ref{090313} \\
GRB090530\tablefootmark{a}  	                &        2009-05-30         &   4.8/4.8/4.8  	& 1.0/1.2/1.2		& 1.6--2.2  	& 1.7   	& 20.6      &  22    	& 1.266 		& \ref{090530} \\
GRB090809\tablefootmark{a} 		                &        2009-08-10         &   7.2/7.2/7.2  	& 1.0/0.9/0.9		& 1.2--1.1  	& 1.1   	& 10.2      &  21    	& 2.737  		& \ref{090809} \\
GRB090926A\tablefootmark{a}  	                &        2009-09-27         &   7.2/7.2/7.2  	& 1.0/0.9/0.9		& 1.4--1.5  	& 0.7   	& 22      	&  17.9    	& 2.106 		& \ref{090926} \\
GRB091018     		                            &        2009-10-18         &   2.4/2.4/2.4  	& 1.0/0.9/0.9		& 2.1--1.8  	& 1.0   	& 3.5      	&  19.1    	& 0.971 		& \ref{091018} \\
GRB091127     		                            &        2009-12-02         &   6.0/6.0/6.0  	& 1.0/0.9/0.9		& 1.1--1.2  	& 1.0   	& 101      	&  21.2    	& 0.490  		& \ref{091127} \\
GRB100205A     		                            &        2010-02-08         &   10.8/10.8/10.8 	& 1.0/0.9/0.9		& 1.9--1.8  	& 0.9   	& 71      	&   >24    	&  --    		& \ref{100205} \\
GRB100219A     		                            &        2010-02-20         &    4.8/4.8/4.8	& 1.0/0.9/0.9		& 1.3--1.1  	& 0.8   	& 12.5      &  23    	& 4.667  		& \ref{100219} \\
GRB100316B     		                            &        2010-03-16         &    2.4/2.4/2.4	& 1.0/0.9/0.9		& 2.0--2.4  	& 0.6   	& 0.7      	&  18.2    	& 1.180   		& \ref{100316B} \\
GRB100316D-1\tablefootmark{b}	                &        2010-03-17         &    3.6/3.6/3.6	& 1.0/0.9/0.9		& 1.2--1.3  	& 0.8   	& 10      	&  21.5     & 0.059  		& \ref{100316D} \\
GRB100316D-2   		                            &        2010-03-19         &    2.4/2.4/2.4	& 1.0/0.9/0.9		& 1.1--1.2  	& 0.9   	& 58      	&  20.2     & 0.059  		& \ref{100316D} \\
GRB100316D-3   		                            &        2010-03-20         &    2.6/2.6/3.2	& 1.0/0.9/0.9		& 1.1--1.2  	& 1.1   	& 79      	&  19.9     & 0.059  		& \ref{100316D} \\
GRB100316D-4   		                            &        2010-03-21         &    2.6/2.6/3.2	& 1.0/0.9/0.9		& 1.1--1.2  	& 1.5   	& 101      	&  19.9     & 0.059  		& \ref{100316D} \\
GRB100418A-1   		                            &        2010-04-19         &    4.8/4.8/4.8	& 1.0/0.9/0.9		& 1.6--1.3  	& 0.7   	& 8.4      	&  18.1    	& 0.624 		& \ref{100418} \\
GRB100418A-2   		                            &        2010-04-20         &    4.8/4.8/4.8	& 1.0/0.9/0.9		& 1.2--1.3  	& 0.6   	& 34      	&  19.2     & 0.624 		& \ref{100418} \\
GRB100418A-3   		                            &        2010-04-21         &    4.8/4.8/4.8	& 1.0/0.9/0.9		& 1.2--1.4  	& 0.7   	& 58      	&   >24    	& 0.624 		& \ref{100418} \\
GRB100424A\tablefootmark{c} 	                &        2013-03-11         &    4.8/4.8/4.8	& 1.0/0.9/0.9		& 1.1--1.2  	& 0.9   	& 25239     &   >24    	& 2.465  		& \ref{100424} \\
GRB100425A     		                            &        2010-04-25         &    2.4/2.4/2.4	& 1.0/0.9/0.9		& 1.5--1.3  	& 0.7   	& 4      	&  20.6    	& 1.755  		& \ref{100425} \\
GRB100615A\tablefootmark{c}		                &        2013-03-05         &    4.8/4.8/4.8	& 1.0/0.9/0.9		& 1.0--1.1  	& 0.9   	& 23859     &   >24   	& 1.398  		& \ref{100615} \\
GRB100621A     		                            &        2010-06-21         &    2.4/2.4/2.4	& 1.0/0.9/0.9		& 1.3--1.4  	& 1.0   	& 7.1      	&   22 	    & 0.542  		& \ref{100621} \\
GRB100625A\tablefootmark{c}\tablefootmark{f}    &        2010-07-07         &    4.8/4.8/4.8	& 1.0/0.9/0.9		& 1.1--1.0  	& 0.8   	& 278.7    	&  >24	    & 0.452  		& \ref{100625} \\
GRB100724A\tablefootmark{a}\tablefootmark{d} 	&        2010-07-24         &    4.2/4.2/4.2	& 1.0/0.9/0.9		& 1.5--2.3  	& 0.7   	& 0.2      	&  19.52    & 1.288  		& \ref{100724} \\
GRB100728B\tablefootmark{e} 	                &        2010-07-29         &    7.2/7.2/7.2	& 1.0/0.9/0.9		& 1.5--1.1  	& 0.6   	& 22      	&  23    	& 2.106  		& \ref{100728} \\
GRB100814A-1\tablefootmark{d} 	                &        2010-08-14         &    0.9/0.9/0.9	& 1.0/0.9/0.9		& 1.9--1.7  	& 0.5   	& 0.9      	&  19    	& 1.439   		& \ref{100814} \\
GRB100814A-2   		                            &        2010-08-14         &    4.8/4.8/4.8	& 1.0/0.9/0.9		& 1.5--1.2  	& 0.7   	& 2.1      	&  19    	& 1.439   		& \ref{100814} \\
GRB100814A-3   		                            &        2010-08-18         &    4.8/4.8/4.8	& 1.0/0.9/0.9		& 1.2--1.0  	& 0.6   	& 98      	&  20    	& 1.439   		& \ref{100814} \\
GRB100816A\tablefootmark{f}		                &        2010-08-17         &    4.8/4.8/4.8	& 1.0/0.9/0.9		& 1.8--1.6  	& 0.8   	& 28.4      &  21.6     & 0.805  		& \ref{100816} \\
GRB100901A     		                            &        2010-09-04         &    2.4/2.4/2.4	& 1.0/0.9/0.9		& 1.5--1.5  	& 1.9   	& 66      	&  >24    	& 1.408  		& \ref{100901} \\
GRB101219A     		                            &        2010-12-19         &    7.2/7.2/7.2	& 1.0/0.9/0.9		& 1.1--1.7  	& 1.8   	& 3.7      	&  >24   	& 0.718  		& \ref{101219A} \\
GRB101219B-1\tablefootmark{a}                   &        2010-12-20         &    4.8/4.8/4.8	& 1.0/0.9/0.9		& 1.6--2.6  	& 1.4   	& 11.6      &  20    	& 0.552 		& \ref{101219B} \\
GRB101219B-2\tablefootmark{a}                   &        2011-01-05         &    7.2/7.2/7.2	& 1.0/0.9/0.9		& 1.2--2.0  	& 1.0   	& 394      	&  22.7    	& 0.552 		& \ref{101219B} \\
GRB101219B-3\tablefootmark{a}                   &        2011-01-25         &    7.2/7.2/7.2	& 1.0/0.9/0.9		& 1.4--2.1  	& 0.7   	& 886      	&   >24    	& 0.552 		& \ref{101219B} \\
GRB110128A     		                            &        2011-01-28         &    7.2/7.2/7.2	& 1.0/0.9/0.9		& 2.0--1.6  	& 0.6   	& 5.5      	&  22.5    	& 2.339  		& \ref{110128} \\
GRB110407A     		                            &        2011-04-08         &    9.6/9.6/9.6	& 1.0/0.9/0.9		& 1.4--1.3  	& 2.1   	& 12.4      &  23    	&  --    		& \ref{110407} \\
GRB110709B\tablefootmark{c}  	                &        2013-03-19         &    7.2/7.2/7.2	& 1.0/0.9/0.9		& 1.6--1.1  	& 0.9   	& 14835     &   >24    	& 2.109 		& \ref{110709} \\
GRB110715A\tablefootmark{a}     		        &        2011-07-16         &    0.6/0.6/0.6	& 1.0/0.9/0.9		& 1.1--1.1  	& 1.6   	& 12.3      &  18.5    	& 0.823  		& \ref{110715} \\
GRB110721A\tablefootmark{a}     		        &        2011-07-22         &    2.4/2.4/2.4	& 1.0/0.9/0.9		& 1.2--1.4  	& 2.3   	& 28.7      &   >24    	& 0.382  		& \ref{110721} \\
GRB110808A     		                            &        2011-08-08         &   2.4/2.4/2.4 	& 1.0/0.9/0.9		& 1.2--1.1  	& 1.0   	& 3.0      	&  21.2    	& 1.349 		& \ref{110808} \\
GRB110818A     		                            &        2011-08-19         &   4.8/4.8/4.8 	& 1.0/0.9/0.9		& 1.3--1.3  	& 0.9   	& 6.2      	&  22.3    	& 3.36   		& \ref{110818} \\
GRB111005A\tablefootmark{a}\tablefootmark{c}    &        2013-04-01         &   1.2/1.2/1.2 	& 1.0/0.9/0.9		& 1.3--1.3  	& 0.7   	& 13052     &   >24    	& 0.013 		& \ref{111005} \\
GRB111008A-1   		                            &        2011-10-09         &   8.8/8.8/8.4 	& 1.0/0.9/0.9		& 1.1--1.0  	& 1.3   	& 8.5      	&  21    	& 4.990 		& \ref{111008} \\
GRB111008A-2   		                            &        2011-10-10         &   8.0/8.0/7.2 	& 1.0/0.9/0.9		& 1.3--1.0  	& 0.9   	& 20.1      &  22    	& 4.990 		& \ref{111008} \\
GRB111107A     		                            &        2011-11-07         &   4.8/4.8/4.8 	& 1.0/0.9/0.9		& 1.8--1.5  	& 0.8   	& 5.3      	&  21.5    	& 2.893  		& \ref{111107} \\
GRB111117A\tablefootmark{f}		                &        2011-11-19         &   4.8/4.8/4.8 	& 1.0/0.9/0.9		& 1.5--1.4  	& 0.7   	& 38      	&   >24    	& 2.211   		& \ref{111117} \\
GRB111123A-1   		                            &        2011-11-24         &   6.2/6.6/6.6 	& 1.0/0.9/0.9		& 1.6--1.1  	& 0.8   	& 12.2      &   >24    	& 3.152 		& \ref{111123} \\
GRB111123A-2\tablefootmark{c} 	                &        2013-03-07         &    2.4/2.4/2.4	& 1.0/0.9/0.9		& 1.0--1.0  	& 0.5   	& 11266     &   >24    	& 3.152 		& \ref{111123} \\
GRB111129A     		                            &        2011-11-30         &   3.6/3.6/3.6 	& 1.0/0.9/0.9		& 1.6--2.1  	& 1.9   	& 8.7      	&  >24 	    & 1.080    		& \ref{111129} \\
GRB111209A-1   		                            &        2011-12-10         &   4.8/4.8/4.8 	& 1.0/0.9/0.9		& 1.1--1.2  	& 0.8   	& 17.7      &  20.1    	& 0.677  		& \ref{111209} \\
GRB111209A-2   		                            &        2011-12-29         &   9.6/9.6/9.6 	& 1.0/0.9/0.9		& 1.2--2.0  	& 1.0   	& 497      	&  23    	& 0.677  		& \ref{111209} \\
GRB111211A\tablefootmark{a}  		            &        2011-12-13         &   2.4/2.4/2.4 	& 1.0/0.9/0.9		& 1.4--1.6  	& 0.6   	& 31      	&  19.5    	& 0.478  		& \ref{111211} \\
GRB111228A     		                            &        2011-12-29         &   2.4/2.4/2.4 	& 1.0/0.9/0.9		& 1.4--1.4  	& 0.7   	& 15.9      &  20.1    	& 0.716  		& \ref{111228} \\
GRB120118B\tablefootmark{c} 		            &        2013-02-13         &   3.6/3.6/3.6 	& 1.0/0.9/0.9		& 1.1--1.0  	& 0.7   	& 9393      &   >24    	& 2.943  		& \ref{120118} \\
GRB120119A-1   		                            &        2012-01-19         &   2.4/2.4/2.4 	& 1.0/0.9/0.9		& 1.1--1.1  	& 0.6   	& 1.4      	&  17    	& 1.728  		& \ref{120119} \\
GRB120119A-2   		                            &        2012-01-19         &   1.2/1.2/1.2 	& 1.0/0.9/0.9		& 1.8--1.9  	& 0.5   	& 4.5      	&  20    	& 1.728  		& \ref{120119} \\
GRB120119A-3\tablefootmark{c} 	                &        2013-02-26         &    4.8/4.8/4.8	& 1.0/0.9/0.6JH 	& 1.0--1.1  	& 1.8  	    & 9694      &   >24    	& 1.728  		& \ref{120119} \\
GRB120211A-1\tablefootmark{c}   		        &        2013-02-17         &   4.8/4.8/4.8   	& 1.0/0.9/0.9       & 1.1--1.4      & 1.3   	& 8919      &   >24     & 2.346 		& \ref{120211} \\
GRB120211A-2\tablefootmark{c}   		        &        2013-03-20         &   3.6/3.6/3.6   	& 1.0/0.9/0.9       & 1.1--1.2      & 1.2   	& 9660      &   >24     & 2.346 		& \ref{120211} \\
GRB120224A     		                            &        2012-02-25         &   2.4/2.4/2.4 	& 1.0/0.9/0.9		& 1.7--2.1  	& 1.3   	& 19.8      &  22.3    	& 1.10    		& \ref{120224} \\
GRB120311A\tablefootmark{a}      	            &        2012-03-11         &   2.4/2.4/2.4 	& 1.0/0.9/0.9		& 1.6--1.4  	& 0.7   	& 3.7      	&  21.6    	& 0.350    		& \ref{120311} \\
GRB120327A-1\tablefootmark{a}   		        &        2012-03-27         &   2.4/2.4/2.4 	& 1.0/0.9/0.9		& 1.6--1.4  	& 0.6   	& 2.1      	&  18.8    	& 2.815  		& \ref{120327} \\
GRB120327A-2\tablefootmark{a}   		        &        2012-03-28         &   4.2/4.2/4.2 	& 1.0/0.9/0.9		& 1.0--1.1  	& 0.6   	& 29      	&  22.5    	& 2.815  		& \ref{120327} \\
GRB120404A     		                            &        2012-04-05         &   9.6/9.6/9.6 	& 1.0/0.9/0.9JH 	& 1.7--1.3 		& 1.3  	    & 15.7      &  21.3    	& 2.876  		& \ref{120404} \\
GRB120422A   		                            &        2012-04-22         &   4.8/4.8/4.8 	& 1.0/0.9/0.9		& 1.3--1.3  	& 0.7   	& 16.5      &  22    	& 0.283  		& \ref{120422} \\
GRB120712A     		                            &        2012-07-13         &   4.8/4.8/4.8 	& 1.0/0.9/0.9 	    & 1.5--2.5 		& 1.5   	& 10.4      &  21.5    	& 4.175  		& \ref{120712} \\
GRB120714B     		                            &        2012-07-15         &   4.8/4.8/4.8 	& 1.0/0.9/0.9JH 	& 1.5--1.2 		& 1.2   	& 7.8     	&  22.1    	& 0.398  		& \ref{120714}\\
GRB120716A\tablefootmark{a}  		            &        2012-07-19         &   3.6/3.6/3.6 	& 1.0/0.9/0.9JH 	& 1.8--2.6 		& 1.1   	& 62      	&  20.9    	& 2.486  		& \ref{120716} \\
GRB120722A\tablefootmark{b} 		            &        2012-07-22         &   4.8/4.8/4.8 	& 1.0/0.9/0.9 	    & 1.3--1.3 		& 1.2   	& 10.3      &  23.6    	& 0.959  		& \ref{120722} \\
GRB120805A\tablefootmark{b} 		            &        2012-08-14         &   3.6/3.6/3.6 	& 1.0/0.9/0.9JH 	& 1.3--1.7 		& 0.9   	& 218      	&   >24    	& 3.9   		& \ref{120805} \\
GRB120815A\tablefootmark{a}     		        &        2012-08-15         &   2.4/2.4/2.4 	& 1.0/0.9/0.9 	    & 1.3--1.4 		& 0.7   	& 1.69      &  18.9     & 2.358  		& \ref{120815} \\
GRB120909A\tablefootmark{d}                     &        2012-09-09         &   1.2/1.2/1.2 	& 1.0/0.9/0.9 	    & 1.6--1.6 		& 1.6   	& 1.7     	&  21    	& 3.929  		& \ref{120909} \\
GRB120923A     		                            &        2012-09-23         &   9.6/9.6/9.6 	& 1.0/0.9/0.9JH 	& 1.2--1.4 		& 1.0   	& 18.5      &   >24    	& 7.84      	& \ref{120923} \\
GRB121024A     		                            &        2012-10-24         &   2.4/2.4/2.4 	& 1.0/0.9/0.9 	    & 1.2--1.1 		& 0.6   	& 1.8     	&  20    	& 2.300  		& \ref{121024} \\
GRB121027A     		                            &        2012-10-30         &   8.4/8.4/8.4 	& 1.0/0.9/0.9 	    & 1.3--1.3 		& 1.3   	& 69.4      &  21.15    & 1.773  		& \ref{121027} \\
GRB121201A     		                            &        2012-12-02         &   4.8/4.8/4.8 	& 1.0/0.9/0.9JH 	& 1.1--1.1 		& 1.1   	& 12.9      &  23    	& 3.385  		& \ref{121201} \\
GRB121229A     		                            &        2012-12-29         &   4.8/4.8/4.8 	& 1.0/0.9/0.9JH 	& 1.4--1.2 		& 1.5   	& 2     	&  21.5    	& 2.707  		& \ref{121229} \\
GRB130131B\tablefootmark{c} 		            &        2013-03-09         &   7.2/7.2/7.2 	& 1.0/0.9/0.9JH 	& 1.3--1.6 		& 1.1   	& 874       &   >24    	& 2.539  		& \ref{130131} \\
GRB130408A\tablefootmark{a}     		        &        2013-04-08         &   1.2/1.2/1.2 	& 1.0/0.9/0.9 	    & 1.0--1.0 		& 0.9   	& 1.9     	&  20    	& 3.758  		& \ref{130408} \\
GRB130418A     		                            &        2013-04-18         &   1.2/1.2/1.2 	& 1.0/0.9/0.9 	    & 1.4--1.3 		& 1.2   	& 4.6     	&  18.5    	& 1.222  		& \ref{130418} \\
GRB130427A     		                            &        2013-04-28         &   1.2/1.2/1.2 	& 1.0/0.9/0.9JH 	& 1.8--1.8 		& 0.8   	& 16.5      &  19    	& 0.340  		& \ref{130427A} \\
GRB130427B     		                            &        2013-04-28         &   1.2/1.2/1.2 	& 1.0/0.9/0.9JH 	& 1.2--1.0 		& 1.0   	& 20.3      &  22.7    	& 2.780   		& \ref{130427B} \\
GRB130603B\tablefootmark{f}		                &        2013-06-04         &   2.4/2.4/2.4 	& 1.0/0.9/0.9 	    & 1.4--1.4 		& 1.1   	& 8.2     	&  21.5    	& 0.356  		& \ref{130603} \\
GRB130606A     		                            &        2013-06-07         &   4.2/4.2/4.2 	& 1.0/0.9/0.9JH 	& 1.7--1.9 		& 0.9   	& 7.1     	&  19    	& 5.91   		& \ref{130606} \\
GRB130612A     		                            &        2013-06-12         &   1.2/1.2/1.2 	& 1.0/0.9/0.9 	    & 1.3--1.3 		& 1.5   	& 1.1     	&  21.5    	& 2.006  		& \ref{130612} \\
GRB130615A     		                            &        2013-06-15         &   1.2/1.2/1.2 	& 1.0/0.9/0.9 	    & 2.1--2.2 		& 1.0   	& 0.8     	&  21    	& 2.9  		& \ref{130615} \\
GRB130701A     		                            &        2013-07-01         &   1.2/1.2/1.2 	& 1.0/0.9/0.9JH 	& 2.0--2.0 		& 1.4   	& 5.5     	&  19.9    	& 1.155  		& \ref{130701} \\
GRB130925A                                      &        2013-09-25         &   5.88/6.0/6.9     & 1.0/0.9/0.9JH     & 1.0--1.0      & 0.6       & 3.5       &  >24     & 0.347         & \ref{130925} \\
GRB131011A\tablefootmark{a}			            &        2013-10-13         &   4.5/4.5/4.5 	& 1.0/0.9/0.9		& 1.1--1.1		& 0.8 	    & 34.2     	&   >24   	& 1.874			& \ref{131011} \\
GRB131030A			                            &        2013-10-31         &   3.6/3.6/3.6 	& 1.0/0.9/0.9		& 1.1--1.1		& 1.1 	    & 3.4     	&  18.0	    & 1.296			& \ref{131030} \\
GRB131103A			                            &        2013-11-05         &   2.4/2.4/2.4 	& 1.0/0.9/0.9JH		& 1.1--1.1		& 1.0 	    & 5.8     	&  20.48   	& 0.599			& \ref{131103} \\
GRB131105A			                            &        2013-11-05         &   4.8/4.8/4.8 	& 1.0/0.9/0.9		& 1.3--1.4		& 0.8 	    & 1.3     	&  22.4  	& 1.686			& \ref{131105} \\
GRB131117A			                            &        2013-11-17         &   4.8/4.8/4.8 	& 1.0/0.9/0.9JH		& 1.3--1.2		& 1.7 	    & 1.1     	&  20	    & 4.042			& \ref{131117} \\
GRB131231A\tablefootmark{a}			            &        2014-01-01         &   2.4/2.4/2.4 	& 1.0/0.9/0.9JH		& 1.4--1.3		& 0.9 	    & 20.2     	&  18.5  	& 0.642			& \ref{131231} \\
GRB140114A\tablefootmark{c}                     &        2014-03-28         &   5.4/5.4/5.4 	& 1.0/0.9/0.9JH		& 1.7--1.7		& 1.2 	    & 1746      &  >24  	& 3.0			& \ref{140114} \\
GRB140213A\tablefootmark{a}                     &        2014-02-14         &   1.2/1.2/1.2 	& 1.0/0.9/0.9JH		& 1.5--1.5		& 0.7 	    & 5.8     	&  19.5  	& 1.208			& \ref{140213} \\
GRB140301A			                            &        2014-03-02         &   7.2/7.2/7.2 	& 1.0/0.9/0.9JH		& 1.1--1.1		& 0.9 	    & 9     	&  23.1  	& 1.416			& \ref{140301} \\
GRB140311A\tablefootmark{a}                     &        2014-03-13         &   7.6/6.3/8.4 	& 1.0/0.9/0.9JH		& 1.2--1.2		& 0.6 	    & 32.5     	&  >24  	& 4.954			& \ref{140311} \\
GRB140430A\tablefootmark{a}                     &        2014-04-30         &   1.2/1.2/1.2 	& 1.0/0.9/0.9		& 2.0--1.8		& 1.6 	    & 2.5     	&  19	    & 1.601			& \ref{140430} \\
GRB140506A-1		                            &        2014-05-07         &   4.8/4.8/4.8 	& 1.0/0.9/0.9		& 1.3--1.4		& 0.7 	    & 8.8     	&   20.9  	& 0.889			& \ref{140506} \\
GRB140506A-2		                            &        2014-05-08         &   4.8/4.8/4.8 	& 1.0/0.9/0.9		& 1.2--1.3		& 0.7 	    & 32.9     	&      	    & 0.889			& \ref{140506} \\
GRB140515A			                            &        2014-05-16         &   4.8/4.8/4.8 	& 1.0/0.9/0.9		& 1.3--1.3		& 1.4 	    & 15.5     	&   >24  	& 6.327			& \ref{140515} \\
GRB140614A			                            &        2014-06-14         &   2.4/2.4/2.4 	& 1.0/0.9/0.9		& 1.8--1.8		& 0.7 	    & 3.8     	&   21.5   	& 4.233			& \ref{140614} \\
GRB140622A\tablefootmark{f}                     &        2014-06-22         &   1.2/1.2/1.2 	& 1.0/0.9/0.9		& 1.4--1.3		& 1.0 	    & 0.8     	&   >24     & 0.959			& \ref{140622} \\
GRB141028A\tablefootmark{a}                     &        2014-10-29         &   2.4/2.4/2.4 	& 1.0/0.9/0.9		& 1.5--1.4		& 1.0 	    & 15.4     	&   20 	    & 2.332			& \ref{141028} \\
GRB141031A\tablefootmark{a}\tablefootmark{c}    &        2015-01-29         &   2.4/2.4/2.4 	& 1.0/0.9/0.9		& 1.2--1.3		& 0.8 	    & 10912     &   >24   	& --			& \ref{141031} \\
GRB141109A-1		                            &        2014-11-09         &   2.4/2.4/2.4     & 1.0/0.9/0.9JH     & 1.5--1.7      & 0.8       & 1.9       &   19.2    & 2.993         & \ref{141109} \\
GRB141109A-2		                            &        2014-11-10         &   4.3/4.3/4.5     & 1.0/0.9/0.9JH     & 1.7--2.0      & 0.8       & 25.4      &           & 2.993         & \ref{141109} \\
GRB150206A\tablefootmark{a}                     &        2015-02-07         &   2.4/2.4/2.4     & 1.0/0.9/0.9       & 2.1--1.9      & 0.8       & 10        &  21.9     & 2.087         & \ref{150206} \\
GRB150301B			                            &        2015-03-02         &   3.6/3.6/3.6     & 1.0/0.9/0.9JH     & 1.2--1.2      & 1.1       & 5.1       &  21.0     & 1.517         & \ref{150301} \\
GRB150403A			                            &        2015-04-04         &   2.4/2.4/2.4     & 1.0/0.9/0.9       & 1.6--1.7      & 0.7       & 10.8      &  19.1     & 2.057         & \ref{150403} \\
GRB150423A\tablefootmark{d}\tablefootmark{f}    &        2015-04-23         &   4.8/4.8/4.8     & 1.0/0.9/0.9       & 2.7--2.4      & 1.4       & 0.4       &  >24      & 1.394         & \ref{150423} \\
GRB150428A			                            &        2015-04-28         &   2.4/2.4/2.4     & 1.0/0.9/0.9JH     & 1.6--1.5      & 0.8       & 3.7       &  >24      &  --           & \ref{150428} \\
GRB150514A\tablefootmark{a}                     &        2015-05-15         &   2.4/2.4/2.4     & 1.0/0.9/0.9       & 2.3--2.1      & 0.9       & 28.4      &  19.5     & 0.807         & \ref{150514} \\
GRB150518A\tablefootmark{a}                     &        2015-05-20         &   2.4/2.4/2.4     & 1.0/0.9/0.9JH     & 1.3--1.3      & 1.7       & 30.7      &  >24      & 0.256         & \ref{150518} \\
GRB150616A\tablefootmark{a}\tablefootmark{c}    &        2015-09-12         &   2.4/2.4/2.4     & 1.0/0.9/0.9JH     & 1.2--1.1      & 1.2       & 2092      &  >24      & 1.188         & \ref{150616} \\
GRB150727A			                            &        2015-07-28         &   3.6/3.6/3.6 	& 1.0/0.9/0.9JH		& 1.2--1.2		& 1.4 	    & 5.0     	&  20.5     & 0.313 		& \ref{150727} \\
GRB150821A\tablefootmark{d}                     &        2015-08-21         &   2.4/2.4/2.4 	& 1.0/0.9/0.9		& 2.0--1.8		& 1.3 	    & 0.2     	&  16     	& 0.755  		& \ref{150821} \\
GRB150910A			                            &        2015-09-11         &   1.8/1.8/1.8 	& 1.0/0.9/0.9JH		& 1.9--1.9		& 1.3 	    & 20.1     	&  21.2     & 1.359    		& \ref{150910} \\
GRB150915A			                            &        2015-09-16         &   4.8/4.8/4.8 	& 1.0/0.9/0.9JH		& 1.1--1.1		& 1.6 	    & 3.3     	& 23   	    & 1.968   		& \ref{150915} \\
GRB151021A\tablefootmark{d}                     &        2015-10-21         &   4.2/4.2/4.2 	& 1.0/0.9/0.9		& 1.0--1.1		& 1.4 	    & 0.75     	& 18.2     	& 2.330    		& \ref{151021} \\
GRB151027B			                            &        2015-10-28         &   2.4/2.4/2.4 	& 1.0/0.9/0.9JH		& 1.5--1.7		& 1.2 	    & 5     	& 20.5     	& 4.063   		& \ref{151027} \\
GRB151029A			                            &        2015-10-29         &   1.2/1.2/1.2 	& 1.0/0.9/0.9JH		& 1.9--1.7		& 1.1 	    & 1     	& 20   	    & 1.423 		& \ref{151029} \\
GRB151031A\tablefootmark{d}			                            &        2015-10-31         &   4.2/4.2/4.2 	& 1.0/0.9/0.9		& 1.1--1.1		& 1.1 	    & 0.3      & 20.4     	& 1.167   		& \ref{151031} \\
GRB160117B			                            &        2016-01-18         &   4.8/4.8/4.8 	& 1.0/0.9/0.9JH		& 1.1--1.2		& 1.1 	    & 13.5     	& 20.8   	& 0.870   		& \ref{160117} \\
GRB160203A\tablefootmark{d}                     &        2016-02-03         &   6.6/6.6/6.6 	& 1.0/0.9/0.9		& 1.0--1.8		& 1.0 	    & 0.3     	& 18   	    & 3.518   		& \ref{160203} \\
GRB160228A\tablefootmark{c}                     &        2016-03-12         &   4.8/4.8/4.8 	& 1.0/0.9/0.9JH		& 1.7--1.7		& 1.0 	    & 296       & >24    	& 1.640  		& \ref{160228} \\
GRB160303A\tablefootmark{f}                     &        2016-03-04         &   4.8/4.8/4.8 	& 1.0/0.9/0.9JH		& 1.6--1.5		& 0.8 	    & 19.1     	&  >24      &  --     		& \ref{160303} \\
GRB160314A			                            &        2016-03-15         &   4.8/4.8/4.8 	& 1.0/0.9/0.9JH		& 1.3--1.3		& 0.8 	    & 13.0     	&  21.7     & 0.726    		& \ref{160314} \\
GRB160410A\tablefootmark{d}\tablefootmark{f}    &        2016-04-10         &   1.8/1.8/1.8 	& 1.0/0.9/0.9		& 2.5--2.3		& 0.5 	    & 0.15     	& 20.3    	& 1.717   		& \ref{160410} \\
GRB160425A			                            &        2016-04-26         &   4.8/4.8/4.8 	& 1.0/0.9/0.9JH		& 1.3--1.3		& 0.5 	    & 7.2     	& 21.1      & 0.555 		& \ref{160425} \\
GRB160625B\tablefootmark{a}                     &        2016-06-27         &   2.4/2.4/2.4 	& 1.0/0.9/0.9JH		& 1.3--1.3		& 0.7 	    & 30     	& 19.1   	& 1.406 		& \ref{160625} \\
GRB160804A-1\tablefootmark{a}   		        &        2016-08-04         &   2.4/2.4/2.4  	& 1.0/0.9/0.9JH 	& 1.4--1.3 		& 0.6  	    & 22.4      & 21.2   	& 0.736  		& \ref{160804} \\
GRB160804A-2\tablefootmark{a}\tablefootmark{c}  &        2016-08-27         &   3.6/3.6/3.6  	& 1.0/0.9/0.9JH 	& 1.9--1.8 		& 0.6  	    & 574       &   >24   	& 0.736   		& \ref{160804} \\
GRB161001A			                            &        2016-10-01         &   2.4/2.4/2.4 	& 1.0/0.9/0.9JH		& 1.2--1.3		& 0.5 	    & 6.1     	&   >24     & 0.891  		& \ref{161001} \\
GRB161007A\tablefootmark{c}  	                &        2016-10-14         &   2.4/2.4/2.4  	& 1.0/0.9/0.9JH 	& 1.6--1.6 		& 0.7   	& 323      	&   >24   	&  --     		& \ref{161007} \\
GRB161014A   		                            &        2016-10-15         &   4.8/4.8/4.8  	& 1.0/0.9/0.9JH 	& 1.1--1.2 		& 0.5   	& 11.6      &  21.4   	& 2.823   		& \ref{161014} \\
GRB161023A\tablefootmark{a}			            &        2016-10-24         &   1.2/1.2/1.2 	& 1.0/0.9/0.9JH		& 1.2--1.2		& 0.9 	    & 3     	&  17.5   	& 2.710 		& \ref{161023} \\
GRB161117A                                      &        2016-11-17         &   2.4/2.4/2.4     & 1.0/0.9/0.9       & 1.8--1.6      & 2.6       & 0.73      &  19       & 1.549         & \ref{161117} \\
GRB161219B                                      &        2016-12-21         &   2.4/2.4/2.4     & 1.0/0.9/0.9JH     & 1.1--1.1      & 0.9       & 35.7      &  19.5     & 0.146        & \ref{161219} \\
GRB170113A                                      &        2017-01-14         &   4.8/4.8/4.8     & 1.0/0.9/0.9JH     & 1.5--1.4      & 0.9       & 15.23     &  21.7     & 1.968        & \ref{170113} \\
GRB170202A                                      &        2017-02-03         &   2.4/2.4/2.4     & 1.0/0.9/0.9JH     & 1.3--1.2      & 0.7       &  9.7      &  20.8     & 3.645        & \ref{170202} \\

%% file: tables/sample_properties.tex
\begin{table*}[!ht]
	\centering
	\begin{tabular}{cccc}
		\hline
		\hline\noalign{\smallskip}
		{} & {Full \textit{Swift} sample} & {Statistical sample} &  {Followed up bursts} \\
		\hline\noalign{\smallskip}
		N$_{BAT}$ & 981 & 163 & 92\\
		$\log(15-150$ keV fluence)  & $-5.9_{-0.6}^{+0.7}$ &  $-5.9_{-0.6}^{+0.7}$ &  $-5.9_{-0.7}^{+0.7}$   \\
		N$_{XRT}$ & 902 & 160 & 90\\
		$\log(0.3-10$ keV flux) & $-12.3_{-0.8}^{+0.7}$ &  $-12.4_{-0.8}^{+0.7}$ &  $-12.4_{-0.7}^{+0.9}$  \\
		N$_{\mathrm{HI_x}}$ & 248 & 99 & 79\\
		$\log(\mathrm{N}_{\mathrm{HI_x}})$ & $21.7_{-0.9}^{+0.6}$ &  $21.5_{-3.4}^{+0.7}$ &  $21.6_{-4.5}^{+0.7}$  \\
		\hline\noalign{\smallskip}

\end{tabular} 

\caption{
	Population properties (median and 16th and 84th percentiles as the
	error intervals) for the \textit{Swift sample} and the subset of bursts
	fulfilling the sample criteria. The population characteristics of the three
	samples are very similar, which shows that our selection criteria effectively
	conserve the statistical properties of the underlying population, as least for
	these parameters.  Notice that not all bursts have measurements of the
	quantities we compare. \label{tab:sample_properties}
	}

\end{table*}

%% file: tables/redshift_comparison.tex
\begin{table}[H]
	\centering
	\begin{tabular}{ccccc}
		\hline
		\hline\noalign{\smallskip}
		{Sample} & {N$_{\mathrm{bursts}}$} & {$z_{\mathrm{completeness}}$} &  {$z_{\mathrm{mean}}$} &  {$z_{\mathrm{med}}$} \\
		\hline\noalign{\smallskip}
		{\smallskip}
		XS-GRB & 129 & 88 \% & 1.89 & $1.52_{-0.91}^{+1.83}$ \\
		{\smallskip}
		SHOALS  & 119 &  99 \% &  2.18  & $2.06_{-1.20}^{+1.27}$ \\
		{\smallskip}
		BAT6 & 58 & 97 \% &  1.90 &  $1.71_{-1.04}^{+1.31}$ \\
		{\smallskip}
		TOUGH & 69 &  87 \% & 2.20 & $2.11_{-1.36}^{+1.42}$ \\
		{\smallskip}
		Fynbo09 & 146 &  49 \% &  2.2 & $2.1_{-1.23}^{+1.28}$ \\
		\hline\noalign{\smallskip}

\end{tabular} 

\caption{Comparison between the redshift distributions of previous complete
	samples. The SHOALS redshift characteristics are taken from \citet{Perley2016a},
	the BAT6 redshifts are from \citet{Salvaterra2012} with the update from
	\citet{Pescalli2016}, TOUGH is from \cite{Hjorth2012} with the update from
	\citet{Schulze2015}, and Fynbo09 are from \citet{Fynbo2009}. The errors shown on
	the median redshift contain 68 per cent of the probability mass.
	\label{tab:redshift_comparison}}

\end{table}

%% file: tables/HI_columns.tex
\begin{table}[!ht]
\caption{Hydrogen column densities for the 41 bursts exhibiting \lya~absorption in the spectral coverage of X-shooter. The corresponding measurements are shown in Fig. \ref{fig:HI1}. \label{tab:HI}}
\centering
\begin{tabular}{ll}
\hline
\hline\noalign{\smallskip}
{GRB} & {Hydrogen Column} \\
\hline\noalign{\smallskip}
{} & {\nh} \\
\hline\noalign{\smallskip}
GRB~090809A\tablefootmark{c}  & 21.7 $\pm$ 0.2    \\
GRB~090926A & 21.55 $\pm$ 0.10  \\
GRB~100219A\tablefootmark{c} & 21.2 $\pm$ 0.2  \\
GRB~100425A\tablefootmark{c} & 21.0 $\pm$ 0.2  \\
GRB~100728B & 21.2 $\pm$ 0.5  \\
GRB~110128A & 21.90 $\pm$ 0.15  \\
GRB~110818A & 21.9 $\pm$ 0.4    \\
GRB~111008A\tablefootmark{c} & 22.40 $\pm$ 0.10  \\
GRB~111107A\tablefootmark{c} & 21.0 $\pm$ 0.2    \\
GRB~120119A & 22.6 $\pm$ 0.2    \\
GRB~120327A\tablefootmark{c} & 22.00 $\pm$ 0.05   \\
GRB~120404A & 20.7 $\pm$ 0.3    \\
GRB~120712A & 19.95 $\pm$ 0.15  \\
GRB~120716A\tablefootmark{c} & 21.80 $\pm$ 0.15  \\
GRB~120815A\tablefootmark{c} & 22.10 $\pm$ 0.10  \\
GRB~120909A\tablefootmark{c} & 21.75 $\pm$ 0.10  \\
GRB~121024A\tablefootmark{c} & 21.85 $\pm$ 0.10  \\
GRB~121027A & 22.8 $\pm$ 0.3    \\
GRB~121201A\tablefootmark{a}\tablefootmark{c} & 22.0 $\pm$ 0.3  \\
GRB~121229A & 21.8 $\pm$ 0.2    \\
GRB~130408A\tablefootmark{c} & 21.80 $\pm$ 0.10    \\
GRB~130427B & 21.9 $\pm$ 0.3    \\
GRB~130606A & 19.91 $\pm$ 0.05  \\
GRB~130612A & 22.1 $\pm$ 0.2    \\
GRB~131011A & 22.0 $\pm$ 0.3    \\
GRB~131117A & 20.0 $\pm$ 0.3    \\
GRB~140311A & 22.40 $\pm$  0.15 \\
GRB~140430A & 21.8 $\pm$ 0.3    \\
GRB~140515A & 19.0 $\pm$ 0.5     \\
GRB~140614A & 21.6 $\pm$ 0.3    \\
GRB~141028A & 20.60 $\pm$ 0.15   \\
GRB~141109A & 22.10 $\pm$ 0.10    \\
GRB~150206A & 21.7 $\pm$ 0.4    \\
GRB~150403A & 21.8 $\pm$ 0.2    \\
GRB~150915A\tablefootmark{a} & 21.2 $\pm$ 0.3     \\
GRB~151021A\tablefootmark{a} & 22.3 $\pm$ 0.2    \\
GRB~151027B & 20.5 $\pm$ 0.2    \\
GRB~160203A & 21.75 $\pm$ 0.10  \\
GRB~160410A\tablefootmark{b} & 21.2 $\pm$ 0.2 \\
GRB~161014A & 21.4 $\pm$ 0.3    \\
GRB~161023A & 20.96 $\pm$ 0.05  \\
GRB~170202A\tablefootmark{a} & 21.55 $\pm$ 0.10  \\

\hline\noalign{\smallskip}

\end{tabular}
\tablefoot{
\tablefoottext{a}{Has \lya~emission in the trough.}
\tablefoottext{b}{Short burst.}
\tablefoottext{c}{Previously published in \citet{Cucchiara2015}.}
} 
\end{table}